\documentclass[pdftex]{pasj01}
\usepackage{color}
\usepackage{graphicx}
\usepackage{bm}

\begin{document}
\title{\REVIEWLABEL Introduction to Faraday tomography and its future prospects}
\author{Keitaro \textsc{Takahashi}$^{1,2}$}
\altaffiltext{1}{Kumamoto University, Graduate School of Science and Technology, 2-39-1 Kurokami, Chuo-ku, Kumamoto 860-8555, Japan}
\altaffiltext{2}{Kumamoto University, International Research Organization for Advanced Science and Technology, 2-39-1 Kurokami, Chuo-ku, Kumamoto 860-8555, Japan}
\email{keitaro@kumamoto-u.ac.jp}


\maketitle

\begin{abstract}
Faraday tomography is a new method of the study of cosmic magnetic fields enabled by broadband low-frequency radio observations. By Faraday tomography, it is possible to obtain the Faraday dispersion function which contains information on the line-of-sight distributions of magnetic fields, thermal electron density, and cosmic-ray electron density by measuring the polarization spectrum from a source of synchrotron radiation over a wide band. Furthermore, by combining it with 2-dimensional imaging, Faraday tomography allows us to explore the 3-dimensional structure of polarization sources. The application of Faraday tomography has been active in the last 20 years, when broadband observation has become technically feasible, and polarization sources such as interstellar space, supernova remnants and galaxies have been investigated. However, the Faraday dispersion function is mathematically the Fourier transform of the polarization spectrum, and since the observable band is finite, it is impossible to obtain a complete Faraday dispersion function by performing Fourier transform. For this purpose, various methods have been developed to accurately estimate the Faraday dispersion function from the observed polarization spectrum. In addition, the Faraday dispersion function does not directly reflect the distribution  of magnetic field, thermal-electron density, and cosmic-ray electron density in the physical space, and its physical interpretation is not straightforward. Despite these two difficult problems, Faraday tomography is attracting much attention because it has great potential as a new method for studying cosmic magnetic fields and magnetized plasmas. In particular, the next-generation radio telescope SKA (Square Kilometre Array) is capable of polarization observation with unprecedented sensitivity and broad bands, and the application of Faraday tomography is expected to make dramatic progress in the field of cosmic magnetic fields. In this review, we explain the basics of Faraday tomography with simple and instructive examples. Then representative algorithms to realize Faraday tomography are introduced and finally some applications are shown.
\end{abstract}
\tableofcontents

\section{Introduction}
\label{section:introduction}

Radio observations, especially polarization observations, have played an extremely important role in the study of cosmic magnetic fields. This is because it is possible to obtain information on the magnetic fields of radio sources and interstellar space through synchrotron radiation, which is caused by interactions between high-energy electrons and magnetic fields, and Faraday rotation in magnetized plasma. In recent years, the developments in broadband observation technology for radio waves have made it possible to expect revolutionary advances in the research method of cosmic magnetic fields. This is a technique called Faraday tomography and the idea itself dates back to \citet{1966MNRAS.133...67B}. In astronomy, it is generally difficult to obtain information on the distribution of physical quantities along the line of sight. As we will see later, contrastingly, Faraday tomography allows us to probe the line-of-sight distribution of magnetic fields, thermal-electron density and cosmic-ray electron density. Thus, combining it with 2-dimensional imaging in the sky, we can study 3-dimensional structure of sources. This will be a milestone for astronomy as a whole.

As mentioned above, the idea of Faraday tomography itself dates back to 1966, but it was not actually implemented until recently because it requires wideband polarization observations. As will be explained later, Faraday tomography is mathematically Fourier transform, and without broadband observation data, Fourier transform for obtaining line-of-sight information cannot be effectively performed. Therefore, Faraday tomography has received wide attention and has been applied aggressively since the 2000s, when broadband radio observation became practical \citep{2018Galax...6..140I}. In particular, the Square Kilometer Array (SKA), which will appear in the latter half of the 2020s, will enable unprecedented broadband and high-sensitivity observations, and is expected to further advance the study of cosmic magnetic fields through Faraday tomography \citep{2018PASJ...70R...2A}.

The targets of Faraday tomography are polarization sources with magnetic fields, and one of the most important is galaxies, especially spiral galaxies. Spiral galaxies have both turbulent magnetic fields and global magnetic fields with correlation lengths comparable to their size, and these magnetic fields have a significant influence on the dynamics of the galactic gas. In addition, the magnetic fields are considered to be maintained and amplified by the dynamo mechanism in galaxies, and both the global and turbulent magnetic fields play important roles there. However, many things about galactic dynamo and dynamics are still not well understood, both qualitatively and quantitatively \citep{2002RvMP...74..775W}. If Faraday tomography can provide information on the 3-dimensional structure of galaxies, such as the magnetic fields, thermal electrons, and cosmic-ray electrons, it will clarify the detailed mechanism of dynamos, the properties of turbulence, and the dynamics of gas.

One of the biggest problems about the cosmic magnetic fields is their origin \citep{2008RPPh...71d6901K,2012SSRv..166...37W}. As mentioned above, galactic magnetic fields are maintained and amplified by the dynamo, but the dynamo cannot create the magnetic fields from zero. Therefore, it needs the seed fields at the galaxy formation. Concerning the seed fields, a variety of hypotheses has been proposed such as cosmological generation in early universe \citep{2005PhRvL..95l1301T,2006Sci...311..827I} and magnetogenesis associated with structure formation and cosmic reionization \citep{2012SSRv..166....1R}, but none of them have been observationally confirmed yet. In any case, the seed magnetic fields are amplified by nonlinear processes in galaxies, and most of the information about the initial condition would have been lost. Nevertheress, it has been pointed out that traces of the seed fields may remain in the shape of large-scale magnetic fields in galaxies. Also, if seed fields are cosmologically generated, they should exist also in intergalactic space, and the detection of weak intergalactic magnetic fields has been reported by some authors \citep{2010Sci...328...73N,2012ApJ...744L...7T,2013ApJ...771L..42T}. Faraday tomography can be expected to play an important role here as well. As mentioned above, Faraday tomography can provide information on the distribution of the magnetic field in the line-of-sight direction, which means that, in principle, it is possible to measure the galactic and intergalactic magnetic fields separately. Therefore, Faraday tomography offers a new approach to the origin of the cosmic magnetic fields.

In this review, the basic principle of Faraday tomography will be explained with instructive examples, and then algorithms to realize it and some applications are given. This review is organized as follows. In section \ref{section:basic}, basic knowledge necessary to understand Faraday tomography, such as Stokes parameters, synchrotron radiation and Faraday rotation, are summarized. The principle of Faraday tomography is explained in section \ref{section:principle}. The Faraday dispersion function is the most important quantity in Faraday tomography, but its physical meaning and interpretation are not straightforward. Then, in section \ref{section:interpretation}, we will give simple examples to deepen our understanding of the Faraday dispersion function. In section \ref{section:implementation}, some algorithms to realize Faraday tomography are introduced. We will see some application of Faraday tomography in section \ref{section:applications}. Finally, section \ref{section:conclusion} is devoted to the conclusion.

\section{Basics}
\label{section:basic}

In this section, we briefly summarize the basic knowledge necessary to understand and formulate Faraday tomography. First, we introduce Stokes parameters, which are basic quantities to describe polarization states of radio waves and appear throughout this article. Synchrotron radiation is a major mechanism of polarized emission, which is caused by magnetic fields and high-energy charged particles. The targets of Faraday tomography are basically astronomical objects with synchrotron radiation. Polarized emission is not directly observed in general but affected by Faraday rotation, where polarization angle is rotated by magnetized media during the propagation of  the radio waves. This is a fundamental ingredient of Faraday tomography and Faraday rotation itself has been utilized for the study of cosmic magnetic fields. Further, polarized waves can be further affected and attenuated by various reasons and they should be taken into account properly to interpret observation data. More detailed explanations on these matters can be found, for example, in \citet{1986rpa..book.....R}.

\subsection{Stokes parameters}
\label{subsection:Stokes}

Here, we introduce Stokes parameters, which are convenient to express the polarization state of electromagnetic waves. We consider a plane wave propagating along $z$ axis and take ($x$,$y$) plane perpendicular to it. Complex electric field at a certain spatial point can be written as follows.
\begin{eqnarray}
&& E_x(t) = \epsilon_x e^{- i \omega t + i \delta_1}, \\
&& E_y(t) =  \epsilon_y e^{- i \omega t + i \delta_2},
\end{eqnarray}
where $\epsilon, \omega, \delta$ are the amplitude, frequency and phase. Here, let us consider a quasi-monochromatic wave, which is a superposition of plane waves having slightly different frequencies. If the amplitude and phase are totally random with respect to the frequency, it is not polarized as a whole. However, if they are correlated at different frequencies, polarization remains to some extent. The state of polarization is described by the following four Stokes parameters.
\begin{eqnarray}
&& I = \langle |E_x|^2 + |E_y|^2 \rangle = \langle \epsilon_x^2 \rangle + \langle \epsilon_y^2 \rangle \\
&& Q = \langle |E_x|^2 - |E_y|^2 \rangle = \langle \epsilon_x^2 \rangle - \langle \epsilon_y^2 \rangle \\
&& U = \langle 2 {\rm Re}(E_x E_y) \rangle = 2 \langle \epsilon_x \epsilon_y \cos{(\delta_1 - \delta_2)} \rangle \\
&& V = - \langle 2 {\rm Im}(E_x E_y) \rangle = 2 \langle \epsilon_x \epsilon_y \sin{(\delta_1 - \delta_2)} \rangle
\end{eqnarray}
Here, $\langle \cdots \rangle$ is an average with respect to frequency\footnote{We can also take a time average for a timescale much longer than the period of the radio waves to define the Stokes parameters for a specific frequency.}. $I$ represents the total intensity, $Q$ and $U$ are linear polarization and $V$ is circular polarization. While $I$ is non-negative, $Q$, $U$ and $V$ can have negative values. In general, we have,
\begin{equation}
I^2 \geq Q^2 + U^2 + V^2.
\end{equation}
The equality holds for purely polarized waves but partially polarized waves lead to the inequality. Then, we define total polarization fraction, linear polarization fraction and circular polarization fraction as,
\begin{eqnarray}
&& p \equiv \frac{\sqrt{Q^2 + U^2 + V^2}}{I} \\
&& p_L \equiv \frac{\sqrt{Q^2 + U^2}}{I} \\
&& p_V \equiv \frac{|V|}{I}.
\end{eqnarray}
For linearly polarized waves, polarization angle is defied as
\begin{equation}
\chi \equiv \frac{1}{2} \arctan{\frac{U}{Q}}.
\end{equation}
Here, we take a range of $-\pi/2 \leq \chi \leq \pi/2$, where it is understood that $-\pi/2 \leq \chi \leq -\pi/4$ for $Q \leq 0$ and $U \leq 0$, $-\pi/4 \leq \chi \leq \pi/4$ for $Q \geq 0$, and $\pi/4 < \chi < \pi/2$ for $Q \leq 0$ and $U \geq 0$.

For a propagation direction along $z$ axis, there is a rotational freedom for the choice of ($x$,$y$) plane. For a rotation in ($x$,$y$) plane by an angle $\theta$, electric field transforms as,
\begin{equation}
\left(  \begin{array}{c}
E'_x \\ E'_y
\end{array} \right)
=
\left( \begin{array}{cc}
\cos{\theta} & - \sin{\theta} \\
\sin{\theta} & \cos{\theta} 
\end{array} \right)
\left( \begin{array}{c}
E_x \\ E_y
\end{array} \right).
\end{equation}
The, Stokes parameters transform as,
\begin{equation}
\left(  \begin{array}{c}
I' \\ Q' \\ U' \\ V'
\end{array} \right)
=
\left( \begin{array}{cccc}
1 & 0 & 0 & 0 \\
0 & \cos{2\theta} & - \sin{2\theta} & 0 \\
0 & \sin{2\theta} & \cos{2\theta} & 0 \\
0 & 0 & 0 & 1
\end{array} \right)
\left( \begin{array}{c}
I \\ Q \\ U \\ V
\end{array} \right).
\end{equation}
Thus, while $I$ and $V$ are invariant under a rotation, $Q$ and $U$ are not invariant but a combination $Q^2 + U^2$ is still invariant. Therefore, $p$, $p_L$ and $p_V$ are also invariant. Further, it is seen that a rotation in ($x$,$y$) plane by an angle $\theta$ corresponds to a rotation in ($Q$,$U$) plane by an angle $2 \theta$. The polarization angle is transformed as $\chi' = \chi + \theta$, which shows that it has a fixed direction in the sky irrespective of the choice of ($x$,$y$) plane. Thus, if we define
\begin{equation}
\bm{P} \equiv \sqrt{Q^2 + U^2} (\cos{\chi}, \sin{\chi})
\end{equation}
in ($x$,$y$) plane, this behaves as a vector under rotation. This vector is called the polarization vector. On the other hand, we define the complex polarization intensity as,
\begin{equation}
P = Q + i U = |P| e^{2 i \chi}.
\end{equation}
It should be noted that the phase is $2 \chi$, not $\chi$.

Hereafter, we will consider only linear polarization with $\delta_1 - \delta_2 = 0, \pi$ and circular polarization is omitted.

\subsection{Synchrotron radiation}
\label{subsection:synchrotron}

Synchrotron radiation is emitted by relativistic charged particles accelerated by magnetic fields. High energy cosmic-ray electrons are widely distributed in galaxies and they emit synchrotron radiation through the interaction with galactic magnetic fields. If magnetic fields are coherent, the radiation has a high linear-polarization fraction with a polarization plane perpendicular to magnetic fields. Thus, synchrotron radiation is useful for the study of cosmic magnetism.

First, the intensity of synchrotron radiation per unit frequency by a single charged particle is expressed as a function of frequency $\nu$,
\begin{equation}
P(\nu) = \frac{\sqrt{3} q^3 B \sin{\varphi}}{m c^2} F\left( \frac{\nu}{\nu_c} \right),
\end{equation}
where $q$ and $m$ are electric charge and mass of the particle. Here, $\varphi$ is the angle between the particle velocity and magnetic field, and there is no emission if the particle moves along the magnetic field line ($\varphi = 0$). The critical frequency is denoted as $\nu_c$ and expressed, using the Lorentz factor of the particle $\gamma = E / m c^2$, as
\begin{equation}
\nu_c = \frac{3 q B \sin{\varphi}}{4 \pi m c} \gamma^2.
\end{equation}
Here, $F(x)$ is a function which represents the spectral shape and has a peak at $x \sim 0.29$. Further, it behaves as $F(x) \propto x^{1/3}$ and $F(x) \propto x^{1/2} e^{-x}$ for $x \ll 1$ and $x \gg 1$, respectively. Thus, most of the radiation energy is emitted at around the critical frequency $\nu_c$. In the case of an electron, we have,
\begin{equation}
\nu_c = 16~{\rm MHz} \left( \frac{B \sin{\varphi}}{1~{\rm \mu G}} \right) \left( \frac{E}{1~{\rm GeV}} \right)^2.
\end{equation}

Next, we consider synchrotron radiation from a ensemble of particles. Galactic cosmic rays often have a power-law energy spectrum and we assume the number density of the form,
\begin{equation}
N(\gamma) d \gamma = N_0 \gamma^{-\alpha} d \gamma
\end{equation}
for a range of $\gamma_1 \leq \gamma \leq \gamma_2$. Here the spectral index $\alpha$ takes a value around $\sim 2.6 - 3.0$. The intensity spectrum of synchrotron radiation from such particles is as follows.
\begin{equation}
J_\nu
= \int_{\gamma_1}^{\gamma_2} P(\nu) N(\gamma) d \gamma
\propto \nu^{-(\alpha-1)/2} B^{(\alpha+1)/2}
\end{equation}
Therefore, the radiation spectrum is also power-law and the spectral index and dependence on the magnetic field are determined by $\alpha$. For a nominal value of $\alpha = 3.0$, we have $J_\nu \propto \nu^{-1} B^{2}$. The value of $\alpha$ cannot be measured directly but can be estimated from observation of synchrotron radiation.

As we saw above, the intensity of synchrotron radiation is determined by the magnetic field strength and energy density of charged particles. We cannot separate them only from observation of synchrotron radiation. However, it is possible if we assume the equipartition of energy between magnetic fields and charged particles or minimize the total energy density.

While synchrotron radiation from each particle is elliptically polarized, the ensemble average leaves only linear polarization and the polarization fraction is given by,
\begin{equation}
p = \frac{\alpha+1}{\alpha+7/3}.
\label{eq:p-fraction}
\end{equation}
It should be noted that this is independent of the wavelength. A nominal value of $\alpha = 3.0$ gives $p = 0.75$. Such a high polarization fraction cannot be seen except for pulsars. In the discussion so far, it was assumed that the magnetic field is uniform, but in many systems, there is a turbulence in magnetized media and the global magnetic field is also curved. In such cases, the polarization is canceled (depolarization) and the polarization fraction decreases substantially.

For example, when magnetic fields have a coherent component with strength $B_{\rm c}$ and a turbulent component with mean of zero and variance of $\sigma_B^2$, the polarization fraction is given by \citep{1966MNRAS.133...67B},
\begin{equation}
p' = p \frac{B_{\rm c}^2}{B_{\rm c}^2 + \sigma_B^2}.
\end{equation}
As we will see in section \ref{subsection:depolarization}, there are many mechanisms which reduce the polarization fraction. The degree of depolarzation is dependent on wavelength for some mechanisms and independent for others.

\begin{figure}[t]
\begin{center}
\includegraphics[width=6cm]{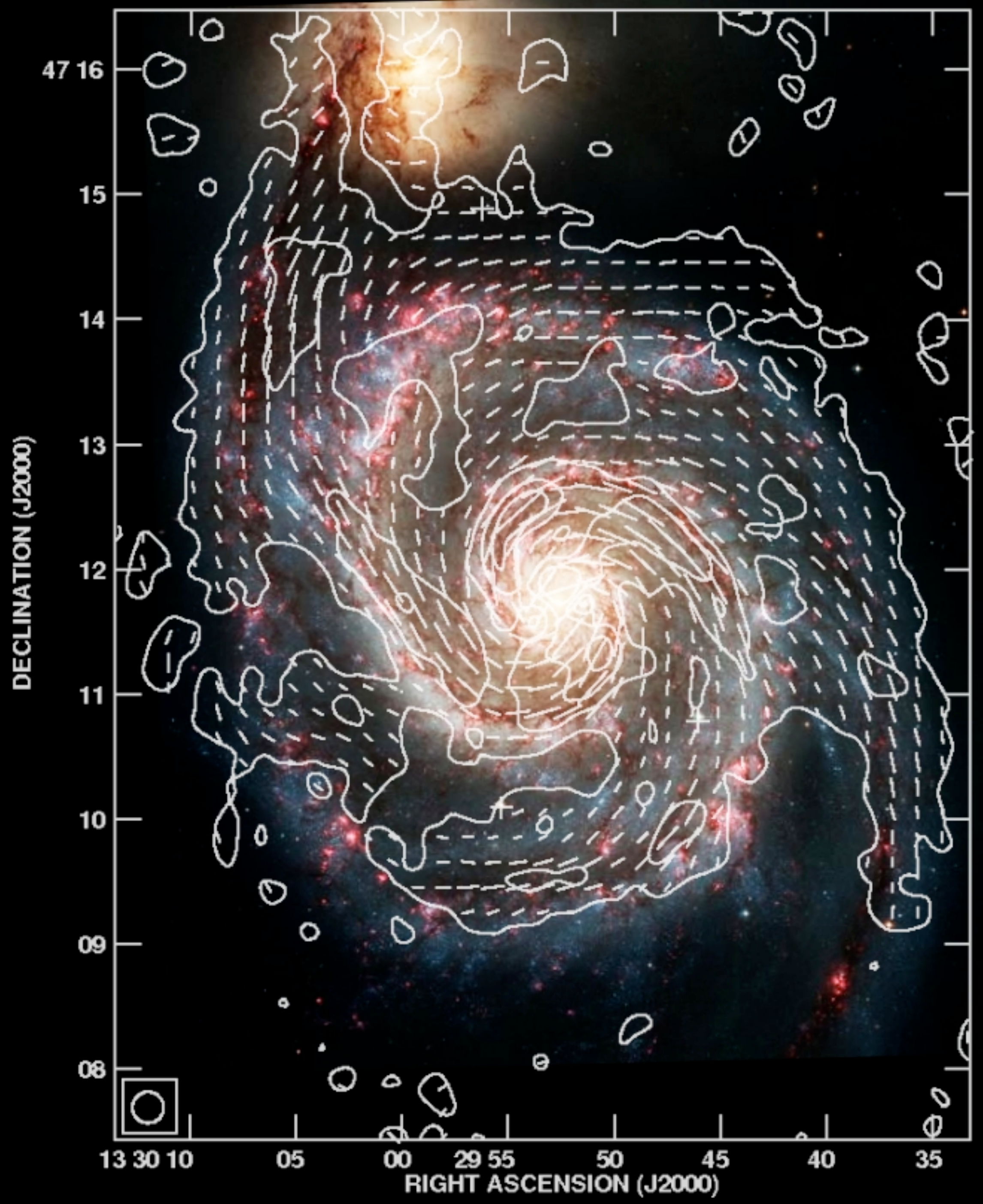}
\end{center}
\begin{center}
\includegraphics[width=6cm]{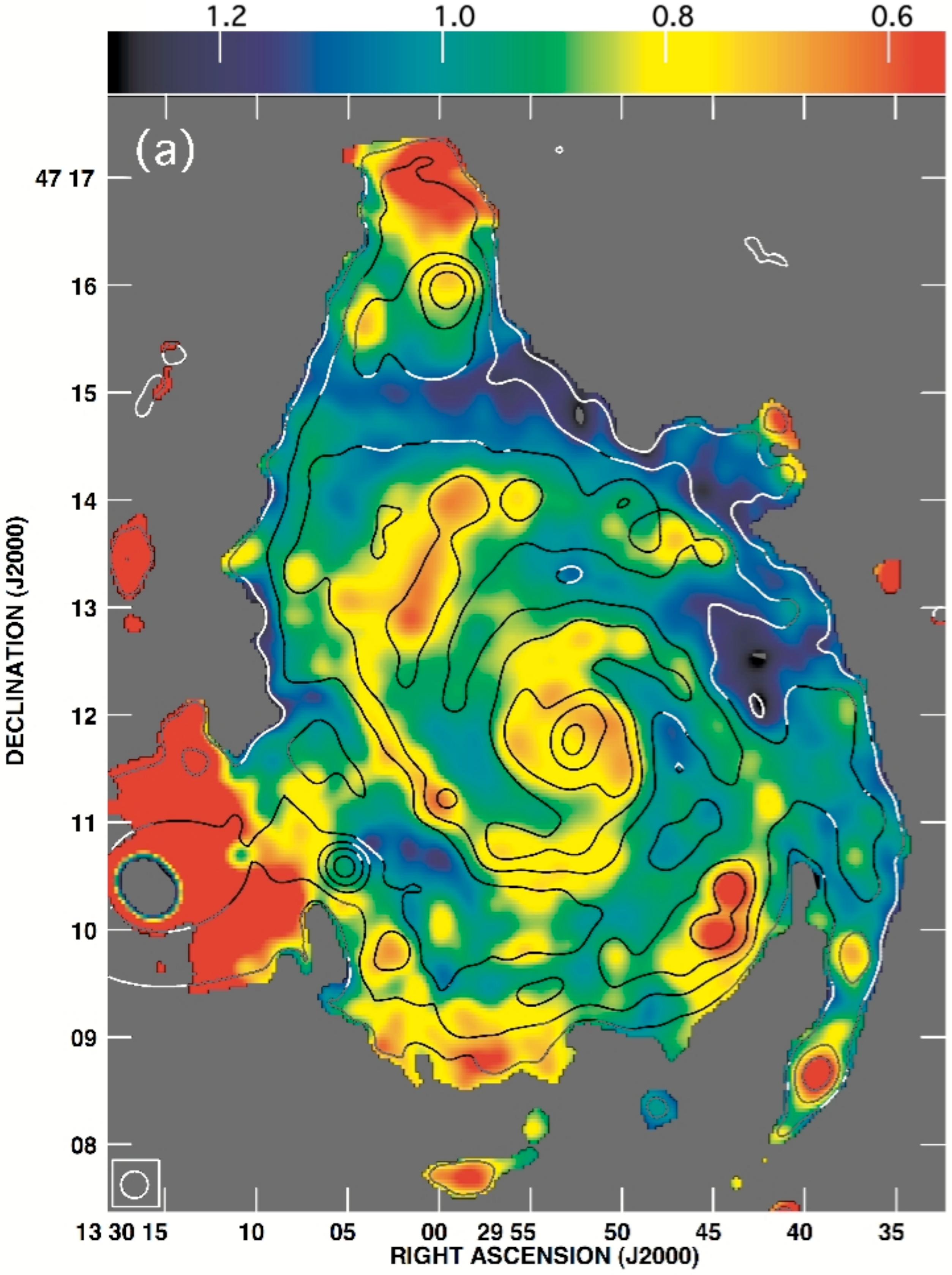}
\end{center}
\caption{Radio images of M51 \citep{2011MNRAS.412.2396F}. Top: contours of polarization intensity at $3~{\rm cm}$ are overlaid on an optical image by Hubble Space Telescope. Magnetic field lines estimated from the polarization angle are also shown. Bottom: map of spectral index of total intensity obtained from observations at $3~{\rm cm}$ and $20~{\rm cm}$. Contours of total intensity are also shown.}
\label{fig:Fletcher-2011}
\end{figure}

There have been many studies to explore the structure of the galactic magnetic fields by observing synchrotron radiation. Fig. \ref{fig:Fletcher-2011} is a radio image of M51 in \citet{2011MNRAS.412.2396F}. This galaxy is face-on so that it is suitable for studying the structure of magnetic field along the galactic plane. The top panel of Fig. \ref{fig:Fletcher-2011} shows contours of polarization intensity at $3~{\rm cm}$ overlaid on an optical image. Magnetic field lines estimated from the polarization angle are also shown. The angular resolution of the contours is $15''$, where $1''$ corresponds to $37~{\rm pc}$ at the distance of M51 ($7.6~{\rm Mpc}$). Therefore, the contours has a spatial resolution of about $560~{\rm pc}$. The spread of polarization intensity is wider than that of the arm structure of the optical image, suggesting that the magnetic field and cosmic ray electrons also exist in the space between the arms. Further, magnetic field lines are along the arms in the arm regions.

The bottom panel of Fig. \ref{fig:Fletcher-2011} represents a map of spectral index of total intensity obtained from observations at $3~{\rm cm}$ and $20~{\rm cm}$. The contours in this panel show the total intensity and it is seen that the contours have similar structure to that of optical image, compared with the polarization map. The spectral index is different between arm and inter-arm regions, and $-0.9 \lesssim \beta \lesssim -0.6$ for arms and $-1.2 \lesssim \beta \lesssim -0.9$ for inter-arm. The index in arms is larger than expected from synchrotron radiation, which suggests that thermal bremsstrahlung is significantly contributing in these region. Contrastingly, synchrotron radiation is dominant in inter-arm region.

By assuming energy equipartition, \citet{2011MNRAS.412.2396F} estimated magnetic field strength in galactic center, arm and inter-arm to be $30~{\rm \mu G}$, $20-25~{\rm \mu G}$ and $15-20~{\rm \mu G}$, respectively. As mentioned above, because incoherence of magnetic field below the scale of spatial resolution ($570~{\rm pc}$ in this case) causes depolarization, coherent component of magnetic field can be estimated from the degree of depolarization. As a result, strengths of $11-13~{\rm \mu G}$, $8-10~{\rm \mu G}$ and $10-12~{\rm \mu G}$ were obtained in inner arm region ($1-2~{\rm kpc}$ from the center), outer arm region and inter-arm region, respectively. Thus, they estimated that turbulent magnetic fields are stronger than these coherent fields by a factor of about 1.5 in inner and outer arm regions and about 1.2 in inter-arm region.

\subsection{Faraday rotation}
\label{subsection:faradayrotation}

Faraday rotation is a phenomenon in which the plane of polarization of an electromagnetic wave rotates in a magnetized plasma, and occurs because the dispersion relation between right- and left-circular polarizations differs in the magnetized plasma. Denoting the polarization angle at the emission as $\chi_0$, the polarization angle after the propagation through magnetized media is dependent on wavelength $\lambda$ and expressed as,
\begin{equation}
\chi = \chi_0 + RM ~ \lambda^2,
\label{eq:FR}
\end{equation}
which is linear with respect to $\lambda^2$. Here the coefficient is called rotation measure and have the following value,
\begin{eqnarray}
&& k \int n_e B_{||} dx \label{eq:RM} \nonumber \\
&& ~~~~~ \approx  811.9~[{\rm rad~m^{-2}}] \int \left(\frac{n_e}{\rm cm^{-3}}\right) \left(\frac{B_{||}}{\rm \mu G}\right) \left(\frac{dx}{\rm kpc}\right), \\
&& k \equiv \frac{e^3}{8 \pi^2 \epsilon_0 m_e^2 c^3}, \label{eq:RM-coeff}
\end{eqnarray}
where $\epsilon_0$ is vacuum permittivity and $m_e$ is electron mass. Here, $x$ is the spatial coordinate along the line of sight and $n_e$ is the density of thermal electrons. $B_{||}$ is the line-of-sight component of magnetic field and defined to be positive for magnetic field in the direction from the source to the observer. This simple linear relation between $\lambda^2$ and $\chi$, Eq. (\ref{eq:FR}), holds if only a single source exists along the line of sight. To be more precise, it holds if only a single Faraday-thin source, which will be defined later, exists along the line of sight. If the slope of the relation is observationally measured, we obtain the integration of the product, $n_e B_{||}$, from the observer to the source. Further, by assuming the electron density distribution along the line of sight, magnetic field strength can be estimated.

\begin{figure}
\begin{center}
\includegraphics[width=8cm]{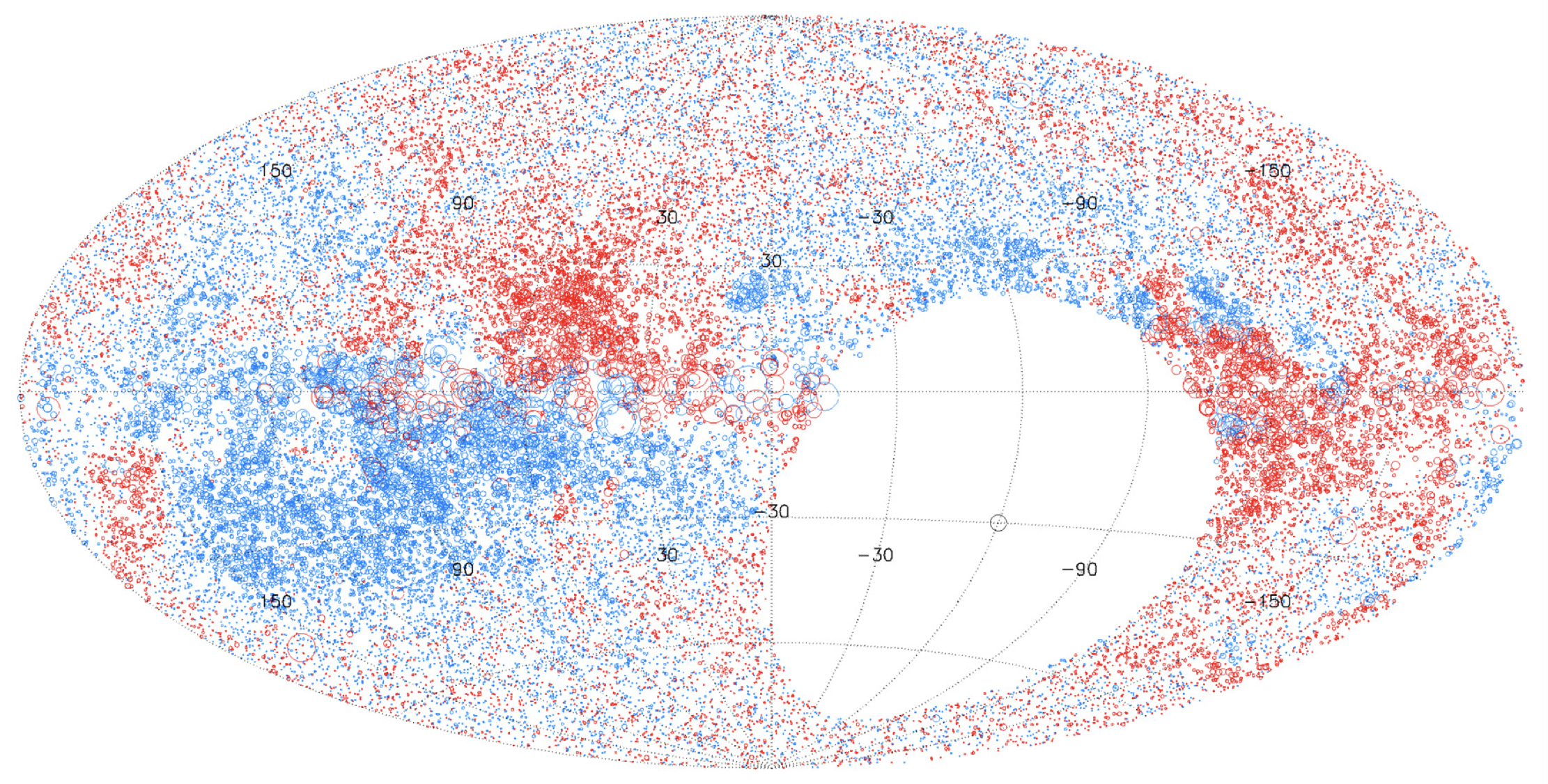}
\end{center}
\begin{center}
\includegraphics[width=8cm]{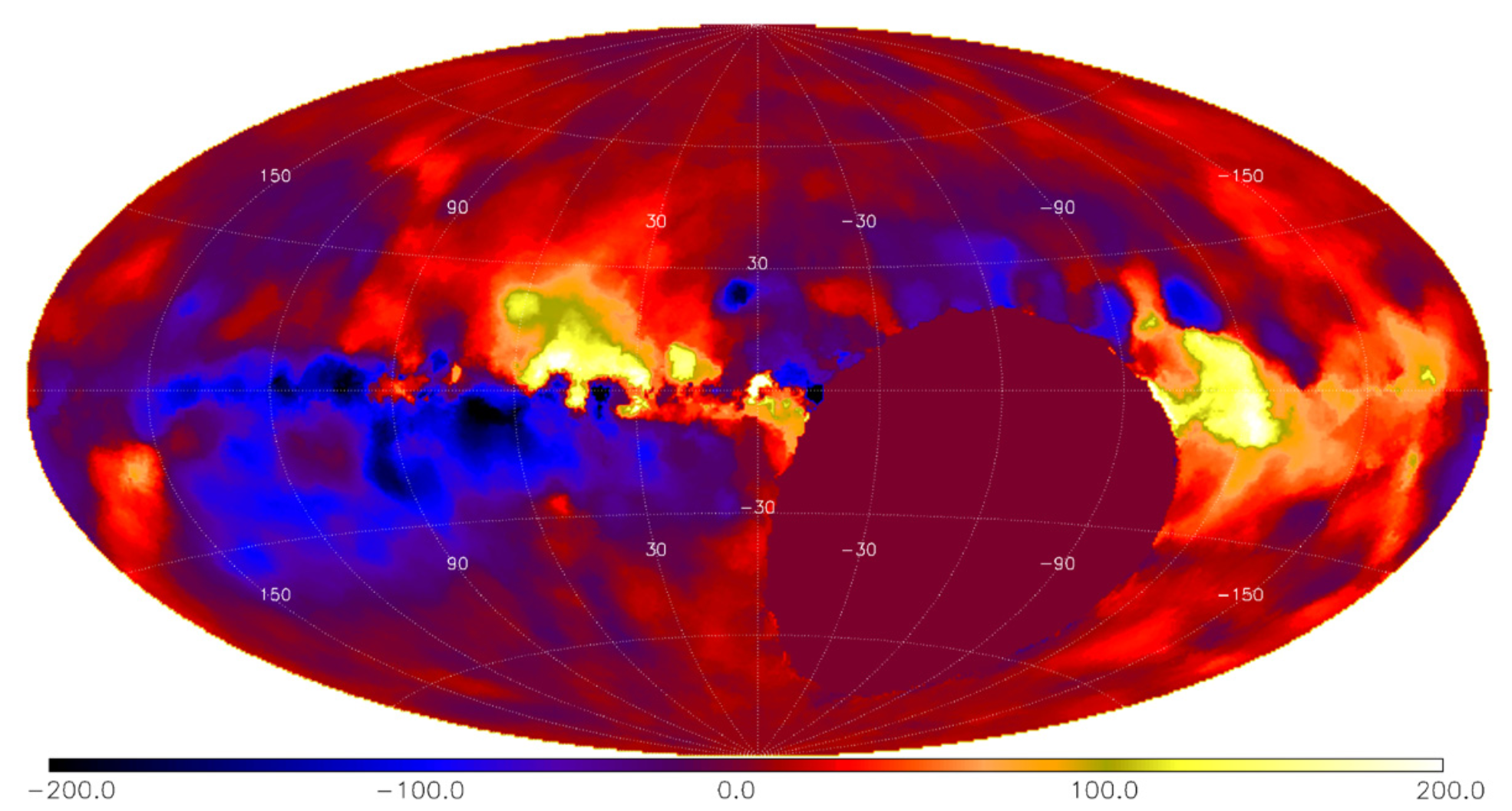}
\end{center}
\caption{Top: rotation measures of 37,543 polarized sources of NVSS \citep{2009ApJ...702.1230T}. Red and blue points indicate positive and negative values. Bottom: map of median value of rotation measure of sources within a circle with radius of $4^{\circ}$. \copyright AAS. Reproduced with permission.}
\label{fig:Taylor-2009}
\end{figure}

Faraday rotation, along with synchrotron radiation, has also been used extensively as a means of studying cosmic magnetic fields. In \citet{2009ApJ...702.1230T}, the large-scale structure of Galactic magnetic fields was studied using a large data set of NVSS (NRAO VLA Sky Survey). They derived rotation measure of 37,543 polarized sources observed at $1.4~{\rm GHz}$ with a typical error of $1-2{\rm rad/m^2}$. Fig. \ref{fig:Taylor-2009} is the obtained rotation measure map, which covers about $82\%$ of the sky area north of declination $-40^{\circ}$. The source density is about one per square degree. The bottom panel is a smoothed image, taking the median value of rotation measure of sources within a circle with radius of $4^{\circ}$. It is seen that rotation measure reach as high as $200~{\rm rad/m^2}$ at the Galactic plane, while it is $O(10)~{\rm rad/m^2}$ in the polar directions. Fluctuations of rotation measure over various scales are also seen. Extension and further detailed analyses of this map have been done in \citet{2012A&A...542A..93O,2015A&A...575A.118O}.

As we saw in Eq. (\ref{eq:RM}), rotation measure is essentially an integration of magnetic field strength and thermal electron density from the source to the observer, and it is in principle impossible to know the distribution along the line of sight. Such values of rotation measure obtained in \citet{2009ApJ...702.1230T} are contributed not just from our galaxy but also from medium around the source and intergalactic medium. Further, there can be intrinsic rotation measure inside the source. Thus, we need to be careful to interpret the map of rotation measure. Typical values of rotation measure of our galaxy and intergalactic medium are as follows.
\begin{itemize}
\item our galaxy: \\$n_e \sim 0.1~{\rm cm^{-3}}$, $B \sim 1~{\rm \mu G}$, $x \sim 1~{\rm kpc}$ $~ \rightarrow ~ {\rm RM} \sim 100~{\rm rad/m^2}$
\item intergalactic medium: \\$n_e \sim 10^{-6}~{\rm cm^{-3}}$, $B \sim 1~{\rm nG}$, $x \sim 1~{\rm Gpc}$ $~ \rightarrow ~ {\rm RM} \sim 1~{\rm rad/m^2}$
\end{itemize}
In reality, both the magnetic field and the thermal electron density have spatial fluctuations so that rotation measure also has a large dispersion depending on the direction in the sky. In addition, the contributions inside sources are not correlated between different sources, and the contributions from intergalactic medium to different sources are not correlated unless they are very close to each other. Thus, these contributions are considered to have little effect on the large-scale pattern of the rotation measure map.

Because most of the polarized sources in \citet{2009ApJ...702.1230T} are located outside the Galaxy, it is expected that their rotation measures include the contribution of the Galaxy to a considerable extent and, therefor, the nature of  Galaxy magnetic field can be investigated from this rotation measure map. In particular, the pattern at low Galactic latitudes reflects the shape of the global magnetic fields in the Galactic disk, and that at high Galactic latitudes reflects magnetic fields and interstellar gas in the cylindrical region extending perpendicular to the Galactic plane from the vicinity of the solar system, especially the halo region. In the successive paper \citep{2011ApJ...726....4S}, they analyzed the spatial fluctuations of rotation measure with structure function and discussed the structure of magnetic fields in the vicinity of the solar system and the entire Galaxy.

\begin{figure}
\begin{center}
\includegraphics[width=7cm]{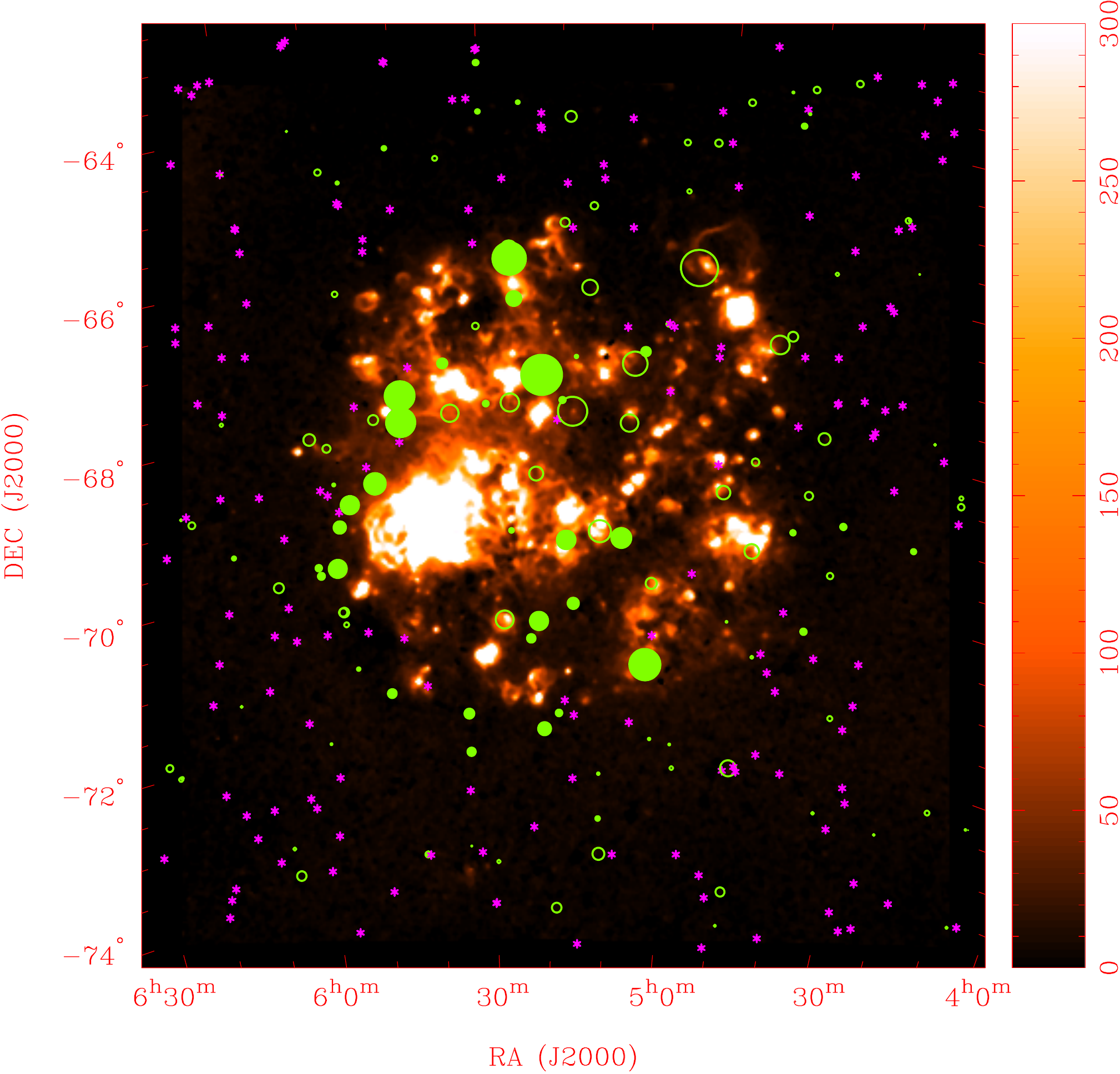}
\end{center}
\caption{Rotation measures of polarized sources behind Large Magellanic Cloud
\citep{2005Sci...307.1610G}. Filled and open green circles represent objects with positive and negative rotation measures and the size is proportional to the magnitude. Purple asterisks indicate rotation measures that are consistent with zero within their errors. The image shows the distribution of emission measure in units of ${\rm pc~cm^{-6}}$.}
\label{fig:Gaensler-2005}
\end{figure}

\citet{2005Sci...307.1610G} studied the strength and structure of magnetic fields in the Large Magellanic Cloud (LMC) with rotation measures of polarized sources behind it. They used rotation measures of 291 sources in sky area of $130~{\rm deg}^2$ around the LMC observed by ATCA (Australia Telescope Compact Array). 140 of the 291 objects are out of the LMC and their average rotation measure can be used to estimate the contribution from the Galaxy. Then Faraday rotation due to the LMC was estimated by subtracting the contribution of the Galaxy from the observed rotation measures of the sources behind the LMC. Fig. \ref{fig:Gaensler-2005} is the map of rotation measures due to the LMC. The distribution implies that the LMC has coherent axisymmetric spiral magnetic field of strength about $1~{\rm \mu G}$.

As described above, Faraday rotation has been used to investigate the structure of magnetic fields of galaxies. However, if there are few polarized objects behind the target object, much information cannot be obtained, so the target has been limited to the Galaxy and nearby galaxies. In the future, as the sensitivity of radio telescopes increases, the number of polarized objects that can be observed will increase substantially. Thus, it will be possible to understand in more detail the structure of magnetic field of not only the Galaxy and the LMC but also distant galaxies.

\subsection{Depolarization}
\label{subsection:depolarization}

As we saw in section \ref{subsection:synchrotron}, synchrotron radiation due to coherent magnetic field has a large linear polarization fraction. However, various processes can cause depolarization and reduce the polarization fraction of observed radio waves. Below, let us summarize some important processes of depolarization following \citet{1966MNRAS.133...67B}.

\begin{description}
\item[beam depolarization 1]
Spatial variation of polarization angle within a beam leads to depolarization. For example, we consider a case where there is a turbulent component in magnetic field which causes synchrotron radiation and assume that the polarization angle at a position $\bm{x}$, $\chi(\bm{x})$, follows Gaussian distribution with mean $\chi_0$ and variance $\sigma_\chi^2$.
\begin{eqnarray}
&& \chi(\bm{x}) = \chi_0 + \delta \chi(\bm{x}), \\
&& \langle \delta \chi(\bm{x}) \rangle = 0, \\
&& \langle ( \delta \chi(\bm{x}) )^2 \rangle = \sigma_\chi^2.
\end{eqnarray}
Here $\langle \cdots \rangle$ represents an average within a beam. Using the following formulas for Gaussian random variable,
\begin{eqnarray}
&& \langle (\delta \chi(\bm{x}))^{2n} \rangle = (2n-1)!! ~ \sigma_\chi^{2n}, \\
&& \langle (\delta \chi(\bm{x}))^{2n+1} \rangle = 0,
\end{eqnarray}
complex polarization intensity integrated within a beam is given by,
\begin{eqnarray}
\langle P(\bm{x}) \rangle
&=& \langle P_0 e^{2 i \chi(\bm{x})} \rangle \nonumber \\
&=& P_0 e^{2 i \chi_0} \Big\langle 1 + 2 i \delta \chi(\bm{x}) - \frac{2^2}{2!} (\delta \chi(\bm{x}))^2 - i \frac{2^3}{3!} (\delta \chi(\bm{x}))^3 \nonumber \\
& & ~~~~~~~~~~~~ + \frac{2^4}{4!} (\delta \chi(\bm{x}))^4 + \cdots \Big\rangle \nonumber \\
&=& P_0 e^{2 i \chi_0} \left( 1 - 2 \sigma_\chi^2 + \frac{2^2}{2!} \sigma_\chi^4 + \cdots \right) \nonumber \\
&=& P_0 e^{2 i \chi_0} e^{-2 \sigma_\chi^2}.
\label{eq:depolarization}
\end{eqnarray}
Here, $P_0$ is the intrinsic polarization fraction and we see the depolarization occurs by a factor of $e^{-2 \sigma_\chi^2}$. It should be noted that this effect is independent of wavelength.

\item[beam depolarization 2]
Let us consider a case where polarized radiation experiences Faraday rotation in a magnetized medium somewhere between the source and observer. Even if the initial polarization angle is uniform within a beam, observed polarization angle fluctuate within the beam if rotation measure varies depending on the location. As an example, we assume that  rotation measure at a position $\bm{x}$, $RM(\bm{x})$, follows Gaussian distribution with mean $RM_0$ and variance $\sigma_{RM}^2$.
\begin{eqnarray}
&& RM(\bm{x}) = RM_0 + \delta RM(\bm{x}) \\
&& \langle \delta RM(\bm{x}) \rangle = 0 \\
&& \langle ( \delta RM(\bm{x}) )^2 \rangle = \sigma_{RM}^2
\end{eqnarray}
Then, in a similar way as beam depolarization 1, complex polarization intensity integrated within a beam can be calculated as,
\begin{eqnarray}
\langle P(\bm{x}) \rangle
&=& \langle P_0 e^{2 i (RM(\bm{x}) \lambda^2 + \chi_0)} \rangle \nonumber \\
&=& P_0 e^{2 i (RM_0 \lambda^2 + \chi_0)} e^{-2 \sigma_{RM}^2 \lambda^4}
\label{eq:beam-depolarization2}
\end{eqnarray}
While the depolarization factor is again exponential with respect to the variance, it depends on wavelength in contrast to Eq. (\ref{eq:depolarization}).

\item[band-width depolarization]
Even if the intrinsic polarization angle is independent of wavelength, Faraday rotation induces the wavelength dependence. When rotation measure is relatively large and the frequency channel is wide, polarization angle varies significantly within a channel, which leads to depolarization. Within a channel with $\nu \sim \nu + \delta \nu$, which corresponds to squared wavelength difference of $\delta \lambda^2$, variation of polarization angle is given by,
\begin{eqnarray}
RM \times \delta \lambda^2
&=& 1.8 \times 10^{-2}~{\rm rad} \left( \frac{RM}{100~{\rm rad/m^2}} \right) \nonumber \\
& & \times \left( \frac{\nu}{1~{\rm GHz}} \right)^{-2} \left( \frac{\delta \nu/\nu}{10^{-3}} \right)
\label{eq:band-depolarization}
\end{eqnarray}
The depolarization is significant when this quantity is $O(1)$. The effect is larger for lower frequencies so that the channel width should be enough fine to avoid depolarization at $100~{\rm MHz}$ band.

\item[Faraday-depth depolarization]
When the line-of-sight component of magnetic fields is nonzero in a radiation region, radio waves experience different amount of Faraday rotation depending on the position along the line of sight. Therefore, even if perpendicular component of magnetic fields is coherent within an observation beam, superposition of polarized emission with different amount of Faraday rotation leads to depolarization.
\end{description}

As described above, depolarization occurs due to various processes and the observable polarization intensity is reduced. However, when the degree of depolarization depends on the frequency, the depolarization reflects the physical quantity related to that process. Therefore, it should be possible to obtain information by utilizing depolarization. Faraday tomography, the main theme of this paper, attempts to reproduce the distribution of various physical quantities related to Faraday rotation and synchrotron radiation along the line of sight by using the Faraday-depth depolarization.

\section{Principle of Faraday Tomography}
\label{section:principle}

\subsection{Faraday tomography}
\label{subsection:faradaytomography}

Let us consider a general case with multiple polarized sources with spatial thickness along the line of sight. Observed polarization spectrum, $P(\lambda^2)$, is the integration of polarization emissivity, $\varepsilon(x)$, along the line of sight and, noting that the amount of Faraday rotation depends on the position, it is given by,
\begin{equation}
P(\lambda^2) = \int \varepsilon(x) e^{2i(\chi_0(x) + \phi(x) \lambda^2)} dx.
\label{eq:P-x}
\end{equation}
Here $x$ is a coordinate along the line of sight and the fact that radiation with polarization angle $\chi_0(x)$ emitted at $x$ experiences Faraday rotation of $\phi(x) \lambda^2$ is taken into account in this equation. Further, $\phi(x)$ is an important quantity called Faraday depth and is defined as,
\begin{equation}
\phi(x) \equiv k \int n_e B_{||} dx.
\label{eq:Faraday-depth}
\end{equation}
This is the same expression as Eq. (\ref{eq:RM}), but the rotation measure actually represents the dependence of observed polarization angle on the wavelength and is defined as,
\begin{equation}
RM(\lambda^2) \equiv \frac{1}{2} \frac{d}{d\lambda^2} {\rm arg}[P(\lambda^2)] = \frac{d\chi}{d\lambda^2}.
\end{equation}
In fact, observed polarization angle is not linear with respect to $\lambda^2$ in general and, thererore, rotation measure itself is a function of $\lambda^2$ as we will see later. Thus, Faraday depth and rotation measure are totally independent quantities. They have the same value only in simple cases where there is only one Farady-thin source, which will be defined later, along the line of sight.

From Eq. (\ref{eq:Faraday-depth}), we can regard $x$ as a function of $\phi$ and change the integration variable of Eq. (\ref{eq:P-x}) to $\phi$.
\begin{equation}
P(\lambda^2) = \int^\infty_{-\infty} F(\phi) e^{2i \phi \lambda^2} d\phi
\label{eq:P}
\end{equation}
Here,
\begin{equation}
F(\phi) \equiv \int \varepsilon(x) e^{2i \chi_0(x)} \delta(\phi - \phi(x)) dx
\label{eq:F-epsilon}
\end{equation}
is called Faraday dispersion function (FDF) or Faraday spectrum and represents polarization intensity per unit Faraday depth at Faraday depth $\phi$. Eq. (\ref{eq:P}) shows that the relation between $P(\lambda^2)$ and $F(\phi)$ is mathematically Fourier transform. Then, we can formally write the inverse Fourier transform as follows.
\begin{equation}
F(\phi) = \frac{1}{\pi} \int^\infty_{-\infty} P(\lambda^2) e^{-2i \phi \lambda^2} d\lambda^2
\label{eq:F}
\end{equation}
It should be noted that the integration variable is not $\lambda$ but $\lambda^2$.

FDF $F(\phi)$ gives us information on the distribution of polarized intensity along the line of sight, because $\phi$ is a function of $x$. More specifically, Faraday depth $\phi$ is determined by the distribution of the line-of-sight component of magnetic field ($B_{||}$) and thermal electron density ($n_e$), while the polarized intensity is determined by the perpendicular component of magnetic field ($B_{\perp}$) and the density of high-energy charged particles ($n_{\rm cr}$). These are the physical quantities which we can probe with FDF.

In the conventional Faraday rotation method explained in the previous section, the information on the magnetic field and thermal electron density is confined to a single value of rotation measure, and only the integrated value from the observer to the polarization source can be probed. Therefore, if the FDF can be obtained, information along the line of sight, which has not been obtained so far, can be studied. Furthermore, if the image of the source is resolved and the FDF is obtained at each point, we can investigate the three-dimensional structure of the source. This methodology is called Faraday tomography and it can have a great impact on astronomy because it is generally difficult to obtain information in the line-of-sight direction in astronomy.

However, there are two major problems concerned with the use of the FDF.
\begin{enumerate}
\item As mentioned before, the polarization spectrum and the FDF are mathematically Fourier transform of each other and we need to perform integration of the range $-\infty < \lambda^2 < \infty$ in order to obtain the FDF from Eq. (\ref{eq:F}). However, negative values of $\lambda^2$ are not physical and we can obtain observation data only for a limited range of positive $\lambda^2$ depending on the frequency coverage of telescopes. Therefore, in practice, we cannot perform the integration perfectly. This is the problem of reconstruction of the FDF. We will discuss strategies to reconstruct the FDF as precisely as possible from finite data in section \ref{section:implementation}.
\item As we saw in Eq. (\ref{eq:Faraday-depth}), Faraday depth is an integration of magnetic field and thermal electron density and there is no one-to-one correspondence between Faraday depth $\phi$ and physical distance $x$ if there is a field reversal along the line of sight. Then, in general, different multiple spatial positions can have the same value of $\phi$ and the distribution of physical quantities in physical space cannot be determined uniquely from the FDF. Thus, even if we could succeed in the precise reconstruction of the FDF from polarization observation, its physical interpretation is not straightforward. This is the problem of physical interpretation of the FDF. We will discuss how the distribution of magnetic fields, thermal electrons and high-energy particles in physical space is reflected in the FDF in section \ref{section:interpretation}.
\end{enumerate}
These two problems are essential to Faraday tomography and it cannot be used effectively unless these are overcome.

\subsection{Dirty FDF and RMSF}
\label{subsection:RMSF}

As stated before, Fourier transform in Eq. (\ref{eq:F}) is impossible in principle, but if we put zeros outside observation band, which is called zero padding, the integration can be performed. Then, let us define a window function $W(\lambda^2)$, which is unity in observation band and zero outside, and we denote $\tilde{P}(\lambda^2) = W(\lambda^2) P(\lambda^2)$. We denote the FDF obtained by zero padding as $\tilde{F}(\phi)$:
\begin{eqnarray}
\tilde{F}(\phi)
&=& \frac{1}{\pi} \int^\infty_{-\infty} \tilde{P}(\lambda^2) e^{-2i \phi \lambda^2} d\lambda^2 \nonumber \\
&=& \frac{1}{\pi} \int^\infty_{-\infty} W(\lambda^2) P(\lambda^2) e^{-2i \phi \lambda^2} d\lambda^2
\label{eq:dirty-FDF}
\end{eqnarray}
This is called the dirty Faraday dispersion function (dirty FDF) and is generally a different function from the true FDF.

To see the difference between $F(\phi)$ and $\tilde{F}(\phi)$, let us consider a simple but important example where the FDF is a delta function:
\begin{equation}
F(\phi) = f e^{2i \chi_0} \delta(\phi - \phi_0).
\label{eq:delta}
\end{equation}
This represents a source with polarization amplitude $f$ and polarization angle $\chi_0$ at Faraday depth $\phi = \phi_0$. It should be noted that $\chi_0$ is the intrinsic polarization angle before experiencing Faraday rotation. In this case, polarization spectrum is given by,
\begin{eqnarray}
P(\lambda^2)
&=& \int^\infty_{-\infty} f e^{2i \chi_0} \delta(\phi - \phi_0) e^{2i \phi \lambda^2} d\phi
\nonumber \\
&=& f e^{2i (\phi_0 \lambda^2 + \chi_0)}.
\end{eqnarray}
Thus, the polarization spectrum is constant with respect to wavelength and the polarization angle varies linearly with $\lambda^2$. This is the expected spectrum when we consider the conventional Faraday rotation. The generalized rotation measure is given by,
\begin{equation}
RM = \frac{1}{2} \frac{d}{d\lambda^2} {\rm arg}[P(\lambda^2)] = \phi_0
\end{equation}
which is the same as the conventional Faraday rotation.

Let us assume we observe this source with a wavelength range of $\lambda^2_{\rm min} \leq \lambda^2 \leq \lambda^2_{\rm max}$. In this case, the dirty FDF can be calculated as,
\begin{eqnarray}
\tilde{F}(\phi)
&=& \frac{1}{\pi} \int_{-\infty}^{\infty} W(\lambda^2) f e^{2i (\phi_0 \lambda^2 + \chi_0)} e^{-2i \phi \lambda^2} d\lambda^2
\nonumber \\
&=& \frac{1}{\pi} \int_{\lambda_{\rm min}^2}^{\lambda_{\rm max}^2} f e^{2i (\phi_0 \lambda^2 + \chi_0)} e^{-2i \phi \lambda^2} d\lambda^2
\nonumber \\
&=& \frac{i f e^{2i \chi_0}}{2 \pi (\phi - \phi_0)} (e^{-2i (\phi - \phi_0) \lambda^2_{\rm max}} - e^{-2i (\phi - \phi_0) \lambda^2_{\rm min}}).
\label{eq:dirty-FDF_delta}
\end{eqnarray}
The absolute value and polarization angle can be written as,
\begin{equation}
\left| \tilde{F}(\phi) \right| = \frac{f}{\pi} \left| \frac{\sin{\{ (\phi - \phi_0) (\lambda_{\rm max}^2 - \lambda_{\rm min}^2) \}}}{\phi - \phi_0} \right|,
\end{equation}
\begin{eqnarray}
\tilde{\chi}(\phi)
&=& - \frac{1}{2} (\phi - \phi_0) (\lambda_{\rm max}^2 + \lambda_{\rm min}^2) + \chi_0.
\end{eqnarray}
First of all, at $\phi = \phi_0$, where the delta function is located, we have,
\begin{equation}
\tilde{F}(\phi_0) = \frac{f e^{2i \chi_0} (\lambda_{\rm max}^2 - \lambda_{\rm min}^2)}{\pi}.
\end{equation}
Thus, the polarization angle at $\phi = \phi_0$ coincides with the intrinsic polarization angle and the absolute value is proportional to $(\lambda_{\rm max}^2 - \lambda_{\rm min}^2)$.

\begin{figure}
\begin{center}
\includegraphics[width=6cm]{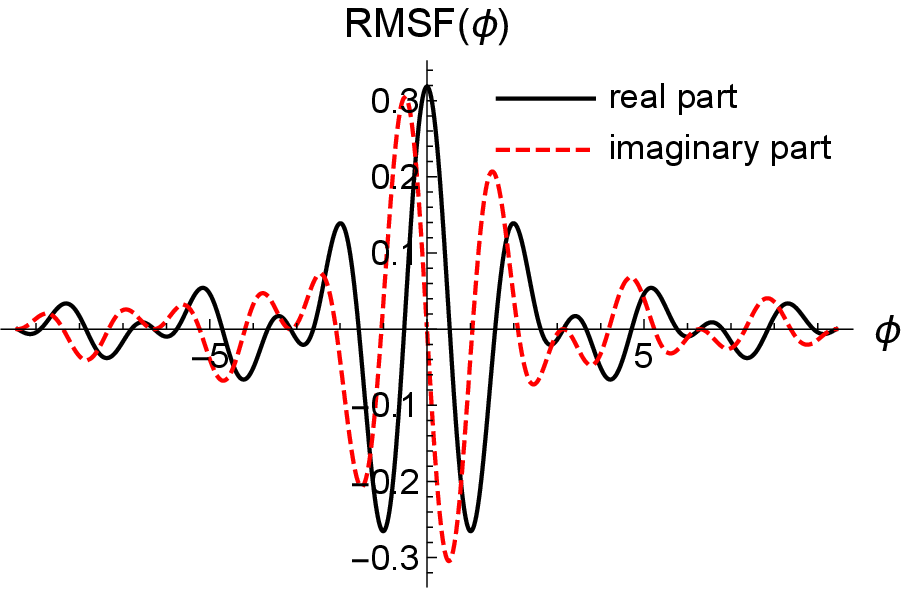} 
\end{center}
\begin{center}
\includegraphics[width=6cm]{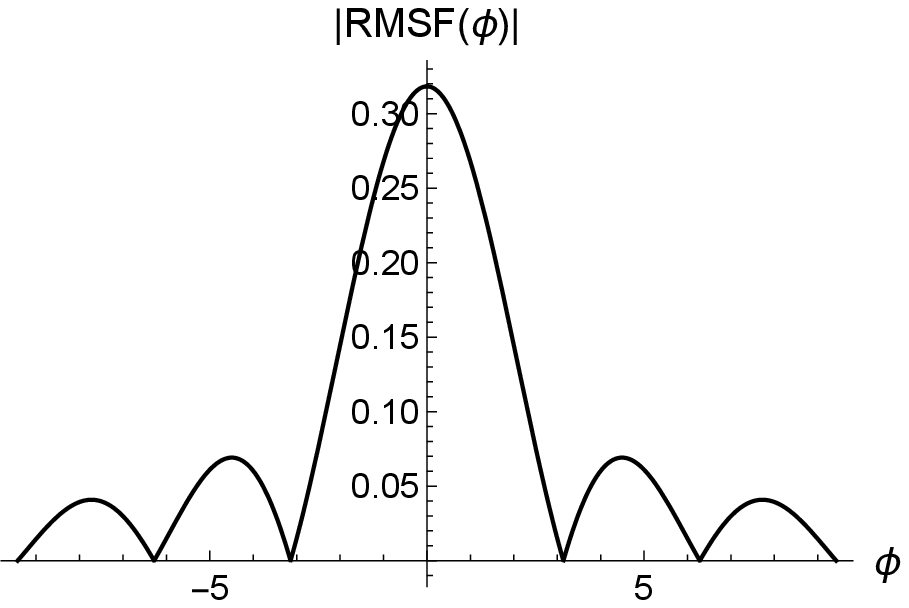} 
\end{center}
\begin{center}
\includegraphics[width=6cm]{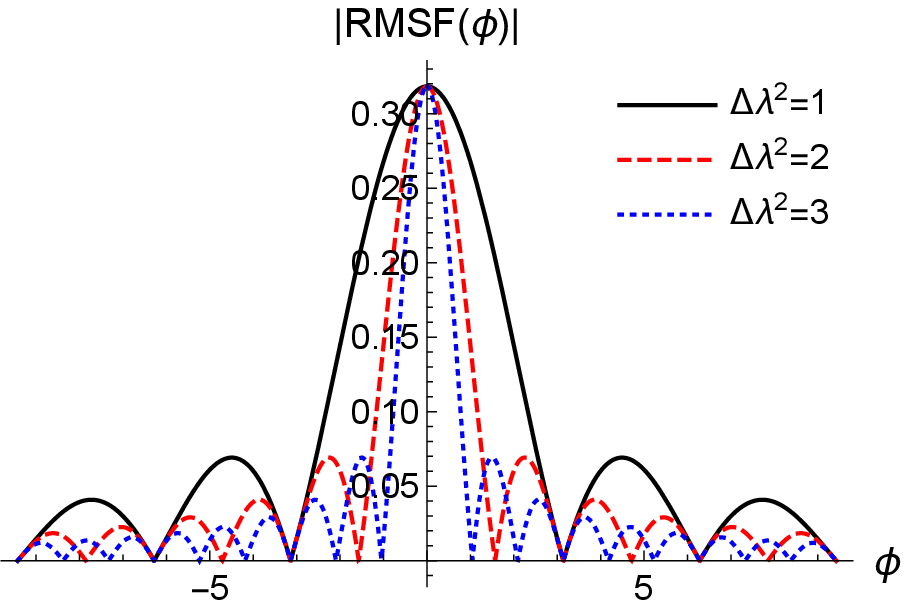} 
\end{center}
\caption{Top: Real and imaginary parts of the RMSF for $\Delta \lambda^2 = \lambda_{\rm max}^2 - \lambda_{\rm min}^2 = 1~{\rm m^2}$. Middle: Absolute value. Bottom: Comparison of absolute values of the RMSF for different values of $\Delta \lambda^2$}\label{fig:RMSF}
\end{figure}

In Fig.~\ref{fig:RMSF}, the RMSF, which is defined later and proportional to the dirty FDF, is plotted. Here, we set $f = 1, \phi_0 = 0, \chi_0 = 0$. The real and imaginary parts and absolute value are plotted in top and middle panels, respectively, setting $\lambda_{\rm max}^2 - \lambda_{\rm min}^2 = 1~{\rm m^2}$. As we can see, the dirty FDF has a peak at $\phi = \phi_0$ and decays slowly with oscillation in proportion to $1/(\phi - \phi_0)$. Aside from the main peak at $\phi = \phi_0$, there are multiple peaks on both sides, which are called sidelobes. It is understood that the delta function of the original FDF is broadened due to incomplete observation and the peak width is determined by the band width $\lambda_{\rm max}^2 - \lambda_{\rm min}^2$. The bottom panel of Fig.~\ref{fig:RMSF} shows a comparison of the absolute value of the dirty FDFs with different values of $\lambda_{\rm max}^2 - \lambda_{\rm min}^2$. It is seen that the main peak is narrower and higher for wider observation band. Thus, the effect of incompleteness of observation band is the broadening of the main peak and the existence of sidelobes.

If we put $\lambda^2 = \lambda_{\rm max}^2 = - \lambda_{\rm min}^2$ in Eq.~(\ref{eq:dirty-FDF_delta}), we have
\begin{equation}
\tilde{F}(\phi) =  \frac{f e^{2i \chi_0} \sin{\{ 2 (\phi - \phi_0) \lambda^2 \}}}{\pi (\phi - \phi_0)}.
\end{equation}
By taking a limit of $\lambda^2 \rightarrow \infty$, the dirty FDF reduces to the original FDF. Thus, the FDF could be completely reconstructed with a perfect observation including the negative region of $\lambda^2$, although it is practically impossible.

Here, we define the Rotation Measure Spread Function (RMSF) from the Fourier transform of the window function,
\begin{eqnarray}
&& R(\phi) \equiv \frac{K}{\pi} \int^\infty_{-\infty} W(\lambda^2) e^{-2i \phi \lambda^2} d\lambda^2
\label{eq:RMSF} \\
&& K \equiv \left( \int^\infty_{-\infty} W(\lambda^2) d\lambda^2 \right)^{-1}
\end{eqnarray}
This is the same as the quantity in Eq.~(\ref{eq:dirty-FDF_delta}) multiplied with $K$ setting $f = 1, \phi_0 = 0, \chi_0 = 0$ and is essentially the dirty FDF for a delta-function FDF. When the window function is unity only for a range $\lambda^2_{\rm min} \leq \lambda^2 \leq \lambda^2_{\rm max}$, we have $K = 1/(\lambda_{\rm max}^2 - \lambda_{\rm min}^2)$ and 
\begin{eqnarray}
R(\phi)
&=& \frac{i}{2 \pi \phi (\lambda^2_{\rm max} - \lambda^2_{\rm min})} (e^{-2i \phi \lambda^2_{\rm max}} - e^{-2i \phi \lambda^2_{\rm min}}) \nonumber \\
&=& \frac{i e^{-2i \phi \lambda^2_{\rm max}}}{2 \pi \phi (\lambda^2_{\rm max} - \lambda^2_{\rm min})} (1 - e^{2i \phi (\lambda^2_{\rm max} - \lambda^2_{\rm min})}), \\
|R(\phi)|
&=& \frac{\left| 1 - e^{2i \phi (\lambda^2_{\rm max} - \lambda^2_{\rm min})} \right|}{2 \pi \phi (\lambda^2_{\rm max} - \lambda^2_{\rm min})} \nonumber \\
&=& \frac{\left| \sin{\{ \phi (\lambda_{\rm max}^2 - \lambda_{\rm min}^2) \}} \right| }{\pi \phi (\lambda^2_{\rm max} - \lambda^2_{\rm min})}, \\
\chi(\phi)
&=& - \frac{1}{2} \phi (\lambda_{\rm max}^2 + \lambda_{\rm min}^2).
\label{eq:RMSF_chi}
\end{eqnarray}
Note that the absolute value $|R(\phi)|$ have a peak at $\phi = 0$ and the height is $|R(0)| = 1/\pi$, independent of the value of $(\lambda_{\rm max}^2 - \lambda_{\rm min}^2)$.

From the definition of the RMSF, Eq.~(\ref{eq:RMSF}), the window function can be written as,
\begin{equation}
W(\lambda^2) = K^{-1} \int^\infty_{-\infty} R(\phi) e^{2i \phi \lambda^2} d\phi.
\end{equation}
Then, the dirty FDF in Eq.~(\ref{eq:dirty-FDF}) can be rewritten as,
\begin{eqnarray}
\tilde{F}(\phi)
&=& \frac{1}{\pi} \int^\infty_{-\infty} W(\lambda^2) P(\lambda^2) e^{-2i \phi \lambda^2} d\lambda^2
\nonumber \\
&=& \frac{K^{-1}}{\pi} \int^\infty_{-\infty} d\lambda^2 \int^\infty_{-\infty} d\phi' \int^\infty_{-\infty} d\phi''
\nonumber \\
& & \times R(\phi') F(\phi'') e^{2i (\phi' + \phi'' - \phi) \lambda^2}
\nonumber \\
&=& K^{-1} \int^\infty_{-\infty} d\phi' \int^\infty_{-\infty} d\phi'' R(\phi') F(\phi'') \delta(\phi' + \phi'' - \phi)
\nonumber \\
&=& K^{-1} \int^\infty_{-\infty} d\phi' R(\phi') F(\phi - \phi')
\nonumber \\
&=& K^{-1} (R * F)(\phi)
\end{eqnarray}
Here, $(R * F)$ represents the convolution. This expression shows that the original FDF can be regarded as a collection of delta functions and that the dirty FDF is the superposition of the RMSFs with the weight of $F(\phi)$. If we write $\lambda^2 = \lambda_{\rm max}^2 = - \lambda_{\rm min}^2$ and take a limit of $\lambda^2 \rightarrow \infty$, the RMSF approaches to a delta function $\delta(\phi)$ and then the dirty FDF approaches to the original FDF. The more the RMSF deviates from a delta function, the more $\tilde{F}(\phi)$ deviates from $F(\phi)$. Because the RMSF has a finite width as we saw in Fig.~\ref{fig:RMSF}, the width is an important parameter which affects the quality of the reconstruction of the FDF. As a measure of the width, we adopt the Full Width at Half Maximum (FWHM). When the observation band is $\lambda^2_{\rm min} \leq \lambda^2 \leq \lambda^2_{\rm max}$, the FWHM is given by,
\begin{equation}
{\rm FWHM} = \frac{2 \sqrt{3}}{\lambda_{\rm max}^2 - \lambda_{\rm min}^2}.
\label{eq:FWHM}
\end{equation}
Here, the width of the RMSF depends only on the bandwidth in $\lambda^2$ domain.


Finally, from the Parseval's theorem, the following equation holds for a square-integrable FDF.
\begin{equation}
\int_{-\infty}^{\infty} |F(\phi)|^2 d\phi = \frac{1}{\pi} \int_{-\infty}^{\infty} |P(\lambda^2)|^2 d\lambda^2
\end{equation}
Further, the following relation holds between the dirty FDF and the observed polarization spectrum.
\begin{equation}
\int_{-\infty}^{\infty} |\tilde{F}(\phi)|^2 d\phi = \frac{1}{\pi} \int_{-\infty}^{\infty} W(\lambda^2) |P(\lambda^2)|^2 d\lambda^2
\end{equation}
Thus, the power of the dirty FDF for a finite observation band is smaller than that of the original FDF. For a specific case with a delta-function FDF and an observation band $\lambda^2_{\rm min} \leq \lambda^2 \leq \lambda^2_{\rm max}$, we have,
\begin{eqnarray}
\int_{-\infty}^{\infty} |\tilde{F}(\phi)|^2 d\phi
&=& \frac{1}{\pi} \int_{-\infty}^{\infty} W(\lambda^2) |P(\lambda^2)|^2 d\lambda^2 \nonumber \\
&=& \frac{f^2 (\lambda_{\rm max}^2 - \lambda_{\rm min}^2)}{\pi}.
\end{eqnarray}

\subsection{Interference in Faraday depth space}
\label{subsection:interference}

\begin{figure}[t]
\begin{center}
\includegraphics[width=6cm]{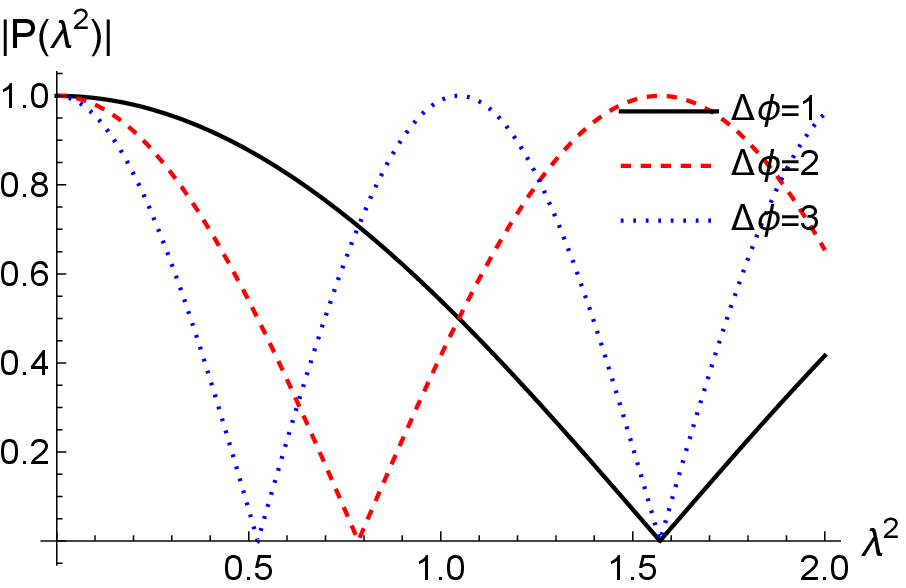} 
\end{center}
\begin{center}
\includegraphics[width=6cm]{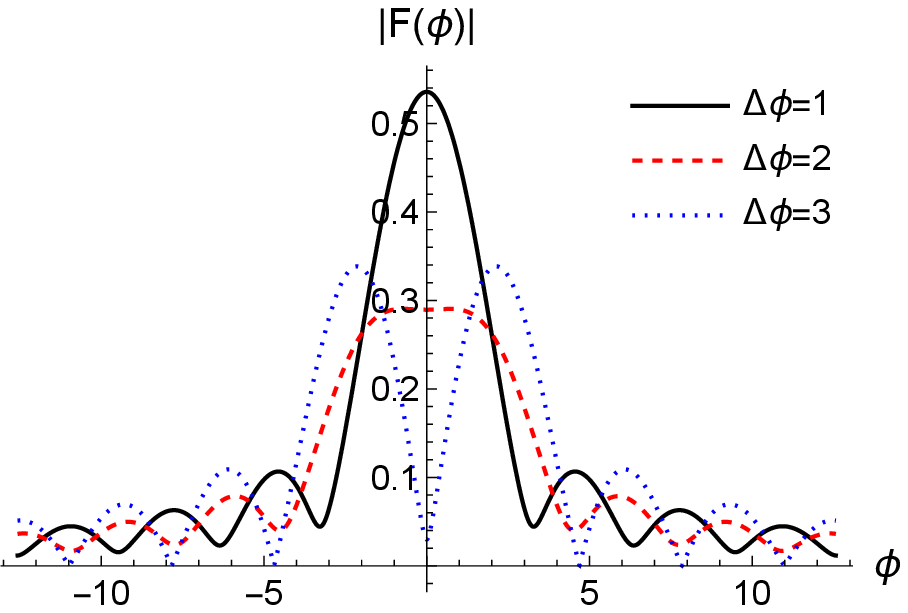} 
\end{center}
\caption{Polarization spectrum (top) and absolute value of the dirty FDF (bottom) in the case of two delta-function sources. The observation band is set to $0~{\rm m^2} \leq \lambda^2 \leq 1~{\rm m^2}$. The parameters for the FDF are set to $F_0 = 1~[{\rm Jy \cdot m^2/rad}]$, $\Delta F = 0~[{\rm Jy \cdot m^2/rad}]$, $\phi_0 = 0~[{\rm rad/m^2}]$, $\chi_0 = 0~[{\rm rad}]$ and $\Delta \chi = 0~[{\rm rad}]$. Three cases with $\Delta \phi = 1, 2, 3~[{\rm rad/m^2}]$ are compared.}
\label{fig:interference_phi}
\end{figure}

Because the FDF is a complex function, interference occurs when there are multiple sources along the line of sight or a source with a finite width in the Faraday depth space. Here, we see an example of interference considering an FDF with two delta functions. We denote the polarization amplitudes, polarization angles and Faraday depths of the two sources as $(F_0 \pm \Delta F)/2$, $\chi_0 \pm \Delta \chi/2$ and $\phi_0 \pm \Delta \phi/2$, respectively.
\begin{eqnarray}
F(\phi)
&=& \frac{F_0 + \Delta F}{2} e^{2i (\chi_0 + \Delta \chi/2)} \delta \left( \phi - (\phi_0 + \frac{\Delta \phi}{2}) \right)
\nonumber \\
& & + \frac{F_0 - \Delta F}{2} e^{2i (\chi_0 - \Delta \chi/2)} \delta \left( \phi - (\phi_0 - \frac{\Delta \phi}{2}) \right)
\end{eqnarray}
For this FDF, the polarization spectrum is given by,
\begin{eqnarray}
P(\lambda^2)
&=& e^{2i(\phi_0 \lambda^2 + \chi_0)}
\nonumber \\
& & \times \left( F_0 \cos{(\Delta\phi \lambda^2 + \Delta \chi)} + i \Delta F \sin{(\Delta\phi \lambda^2 + \Delta \chi)} \right).
\nonumber \\
\label{eq:P_interfere}
\end{eqnarray}
To understand the behavior of this spectrum, first let us set $\Delta F = 0$, that is, the polarization amplitudes of the two sources are the same. In this case, the polarization angle is given by $\chi = \phi_0 \lambda^2 + \chi_0$, which is linear with respect to $\lambda^2$ and equivalent to a single source with the polarization angle $\chi_0$ and Faraday depth $\phi = \phi_0$. In fact, such a single source does not exist and these values are average values of the two sources. On the other hand, the absolute value $|P(\lambda^2)|$ oscillates with $\lambda^2$. The upper panel of Fig.~\ref{fig:interference_phi} shows the comparison of the absolute value of the polarization spectrum for $\Delta \phi = 1, 2$, and $3~[{\rm rad/m^2}]$ fixing other values as $F_0 = 1~[{\rm Jy \cdot m^2/rad}]$, $\Delta F = 0~[{\rm Jy \cdot m^2/rad}]$, $\phi_0 = 0~[{\rm rad/m^2}]$, $\chi_0 = 0~[{\rm rad}]$ and $\Delta \chi = 0~[{\rm rad}]$. The oscillation in the absolute value is a typical symptom of interference. In particular, the polarization is perfectly canceled for wavelengths which satisfy $\Delta\phi \lambda^2 + \Delta \chi = (n + 1/2)\pi$ for a given value of $\Delta\phi$, because the polarization angles of the two sources differ by 90 degree. In contrast, the two sources interfere constructively when $\Delta\phi \lambda^2 + \Delta \chi = n \pi$.

\begin{figure}[t]
\begin{center}
\includegraphics[width=6cm]{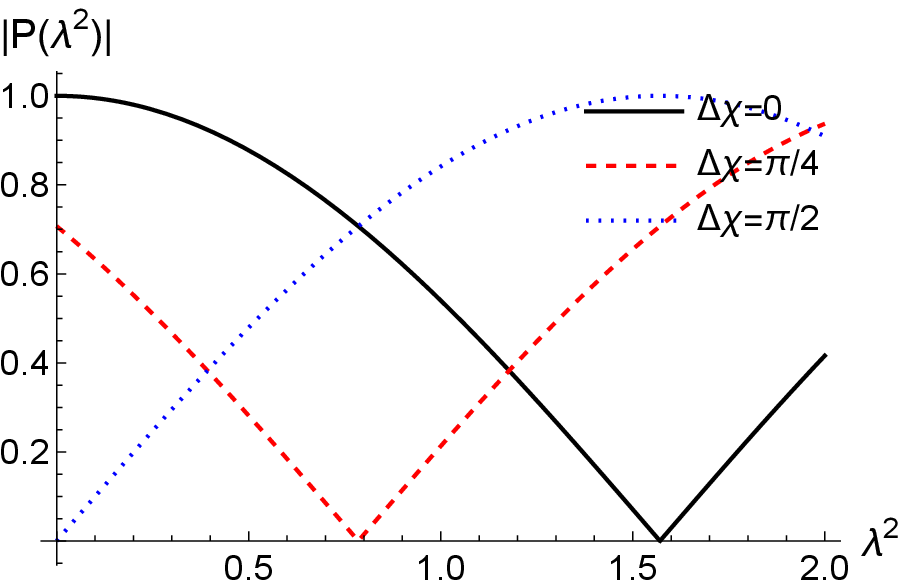} 
\end{center}
\begin{center}
\includegraphics[width=6cm]{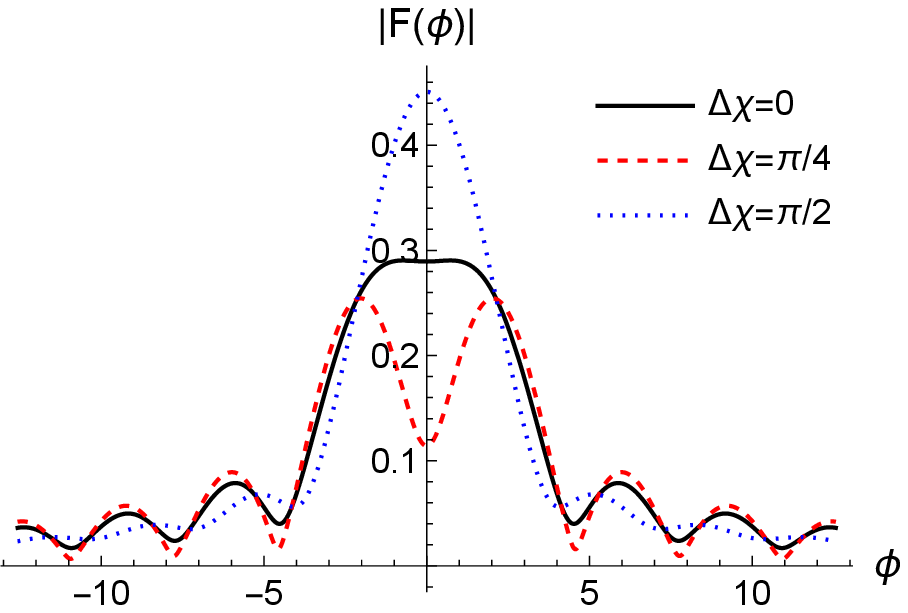} 
\end{center}
\caption{Same as Fig.~\ref{fig:interference_phi} but fixing $\Delta \phi = 2~[{\rm rad/m^2}]$ and comparing different values of $\Delta \chi = 0, \pi/4, \pi/2~[{\rm rad}]$.}
\label{fig:interference_chi}
\end{figure}

The dirty FDFs for the above polarization spectra are shown in the bottom panel of Fig.~\ref{fig:interference_phi}. Here the observation band is set to $0~{\rm m^2} \leq \lambda^2 \leq 1~{\rm m^2}$. The FWHM of the RMSF for this band is $2\sqrt{3}~[{\rm rad/m^2}]$ and it is seen that the two delta functions cannot be resolved when the difference in the Faraday depths is smaller than $\Delta \phi = 3~[{\rm rad/m^2}]$. As we can see in the top panel, within the observation band of $0~{\rm m^2} \leq \lambda^2 \leq 1~{\rm m^2}$, the polarization spectra are more different at longer wavelengths. Then, it is understandable that the case with $\Delta \phi = 1~[{\rm rad/m^2}]$ is less distinguishable with the case of a single delta function, where the absolute value of the polarization spectrum is constant with wavelength. Contrastingly, the spectral shape is significantly different for $\Delta \phi = 3~[{\rm rad/m^2}]$, which is why the two sources can be resolved.

\begin{figure}[t]
\begin{center}
\includegraphics[width=6cm]{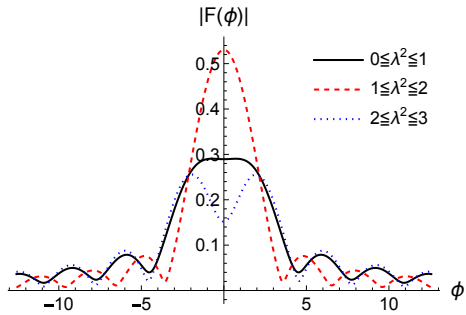} 
\end{center}
\caption{Same as Fig.~\ref{fig:interference_phi} but fixing $\Delta \phi = 2~[{\rm rad/m^2}]$ and comparing three cases with different observation bands: $0~{\rm m^2} \leq \lambda^2 \leq 1~{\rm m^2}$, $1~{\rm m^2} \leq \lambda^2 \leq 2~{\rm m^2}$ and $2~{\rm m^2} \leq \lambda^2 \leq 3~{\rm m^2}$.}
\label{fig:interference_lambda}
\end{figure}

In Fig.~\ref{fig:interference_chi}, three cases with $\Delta \chi = 0, \pi/4, \pi/2~[{\rm rad}]$ are compared by fixing $\Delta \phi = 2~[{\rm rad/m^2}]$. It is seen that the peak position of the polarization spectrum is shifted with $\Delta \phi$, as is also evident in Eq.~(\ref{eq:P_interfere}). As a result, the resolvability of the two sources also depends on $\Delta \phi$. Therefore, the FWHM of the RMSF should be considered as a rough measure of resolution in Faraday depth space.

As we saw in the previous section, the absolute value and FWHM of the RMSF depend only on $\Delta \lambda^2 = \lambda^2_{\rm max} - \lambda^2_{\rm min}$ but are independent of $\lambda^2_{\rm max}$ and $\lambda^2_{\rm min}$ themselves. On the other hand, its phase is dependent on $\lambda^2_{\rm max}$ and $\lambda^2_{\rm min}$ as in Eq.~(\ref{eq:RMSF_chi}). Because the interference pattern is affected by the difference in the polarization angles of the two sources, the shape of the dirty FDF varies depending on the observation band and is not determined just from the FWHM of the RMSF. Fig.~\ref{fig:interference_lambda} shows the comparison of the dirty FDF for three observation bands: $0~{\rm m^2} \leq \lambda^2 \leq 1~{\rm m^2}$, $1~{\rm m^2} \leq \lambda^2 \leq 2~{\rm m^2}$ and $2~{\rm m^2} \leq \lambda^2 \leq 3~{\rm m^2}$. Other parameters are the same as Fig.~\ref{fig:interference_phi} with $\Delta \phi = 2~[{\rm rad/m^2}]$. We can see that, for the same value of $\Delta \lambda^2$, two sources are resolved in the case of $2~{\rm m^2} \leq \lambda^2 \leq 3~{\rm m^2}$ but not resolved in the case of $1~{\rm m^2} \leq \lambda^2 \leq 2~{\rm m^2}$. From the top panel of Fig.~\ref{fig:interference_phi}, it is seen that, when the observation band is $1~{\rm m^2} \leq \lambda^2 \leq 2~{\rm m^2}$, $P(\lambda^2)$ does not fall to zero and is relatively similar to that of a single delta function.

\begin{figure}[t]
\begin{center}
\includegraphics[width=6cm]{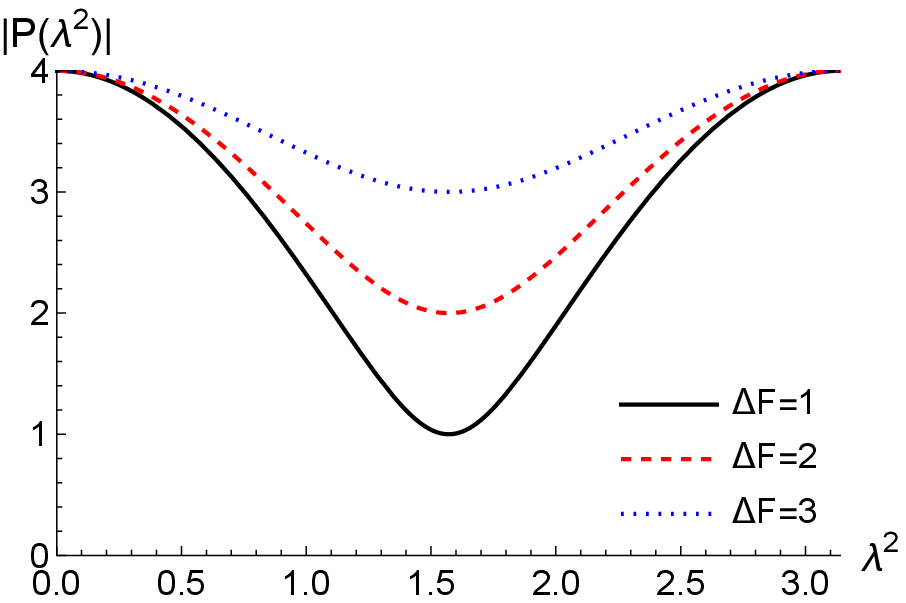} 
\end{center}
\begin{center}
\includegraphics[width=6cm]{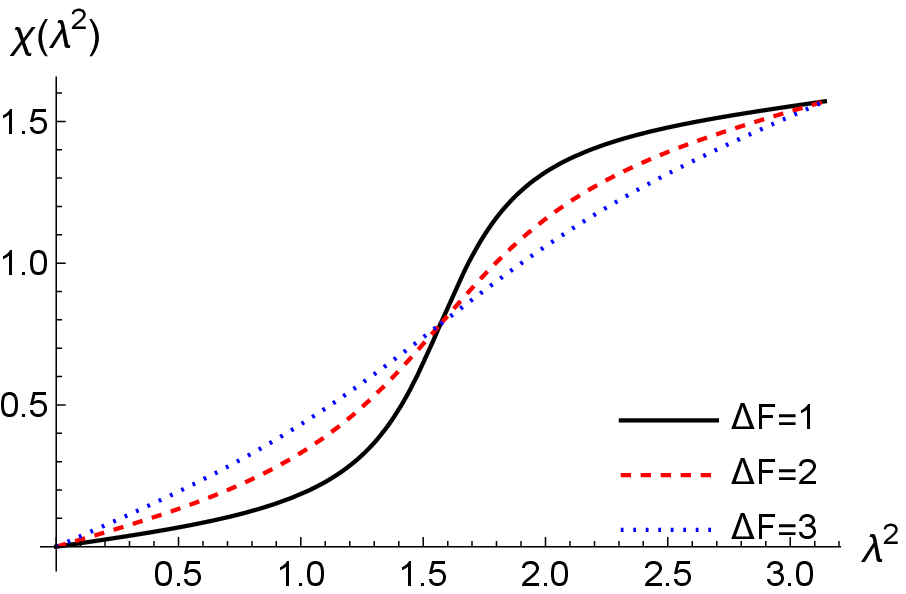} 
\end{center}
\caption{Comparison of absolute value of polarization intensity (top) and polarization angle (bottom) for $\Delta F = 1, 2, 3~[{\rm Jy \cdot m^2/rad}]$. Other parameters are set to $F_0 = 4~[{\rm Jy \cdot m^2/rad}]$, $\phi_0 = 0~[{\rm rad/m^2}]$, $\Delta \phi = 2~[{\rm rad/m^2}]$, $\chi_0 = 0~[{\rm rad}]$ and $\Delta \chi = 0~[{\rm rad}]$.}
\label{fig:interference_dF}
\end{figure}

A case with nonzero $\Delta F$ leads to more complicated behavior. Fig.~\ref{fig:interference_dF} shows the comparison of the absolute value and polarization angle for $\Delta F = 1, 2, 3~[{\rm Jy \cdot m^2/rad}]$. Other parameters are set to $F_0 = 4~[{\rm Jy \cdot m^2/rad}]$, $\phi_0 = 0~[{\rm rad/m^2}]$, $\Delta \phi = 2~[{\rm rad/m^2}]$, $\chi_0 = 0~[{\rm rad}]$ and $\Delta \chi = 0~[{\rm rad}]$. As in the case of $\Delta F = 0$, the interference of the polarization intensity can be constructive or destructive depending on the wavelength. However, since the brightness is different between the two sources when $\Delta F$ is not zero, the polarization is not completely canceled by the interference. Further, the polarization angle is not linear with respect to $\lambda^2$ and the RM depends on wavelength. Because two sources have different brightness in general, nonlinear behavior of the polarization angle $\lambda^2$ is a sign of interference of multiple sources. However, if the observation band is too narrow, the polarization angle is apparently linear and, thus, wideband observations are necessary to investigate the structure of the sources in Faraday depth space.

\subsection{Gaussian FDF}
\label{subsection:Gaussian}

Let us consider the Faraday dispersion function of Gaussian-function type, which is commonly used as well as delta-function type.
\begin{equation}
F(\phi) = \frac{f}{\sqrt{2 \pi} \sigma} \exp{\left[ - \frac{(\phi - \phi_0)^2}{2 \sigma^2} + 2i \chi_0 \right]}
\end{equation}
Here $F$ is the amplitude, $\phi_0$ is the central Faraday depth, $\sigma$ is the width and $\chi_0$ is the polarization angle. This can be regarded as a cluster of delta-function sources with a wide range of Faraday depth. The corresponding polarization spectrum is given by,
\begin{equation}
P(\lambda^2) = f \exp{\left[ -2 \sigma^2 \lambda^4 + 2i (\phi_0 \lambda^2 + \chi_0) \right]}.
\label{eq:P-Gauss}
\end{equation}
This is also a Gaussian function centered at $\lambda^2 = 0$ with the width $\sigma_P = 1/(2 \sigma)$. The behavior of the polarization angle and the RM are the same as the case with a delta function. This is nontrivial because Gaussian FDF is the sum of the polarization sources with various Faraday depth. Thus, the difference between Gaussian and delta-function FDF appears in the absolute value of the polarization spectrum. It falls rapidly for $\lambda^2 > \sigma_P$, which physically means the depolarization due to interference among sources with different Faraday depths at long wavelengths. Conversely, for a given observation band $\lambda^2_{\rm min} \leq \lambda^2 \leq \lambda^2_{\rm max}$, a Gaussian FDF broader than the following maximum width $\sigma_{\rm max}$ cannot be observed due to depolarization:
\begin{equation}
\sigma_{\rm max} = \frac{\pi}{\lambda_{\rm min}^2}.
\label{eq:max-width}
\end{equation}
It should be noted that this width represents the broadening in the Faraday-depth space rather than in physical space.

The polarization spectrum in Eq.~(\ref{eq:P-Gauss}) is equivalent to Eq.~(\ref{eq:beam-depolarization2}) for beam depolariztion. Therefore, when turbulent magnetized plasma exists in front of polarization sources which are delta-function type and have the same Faraday depth, the total FDF becomes Gaussian. In this way, the FDF includes information on non-emitting magnetized plasma as well as polarization sources along the line of sight.

\begin{figure}
\begin{center}
\includegraphics[width=6cm]{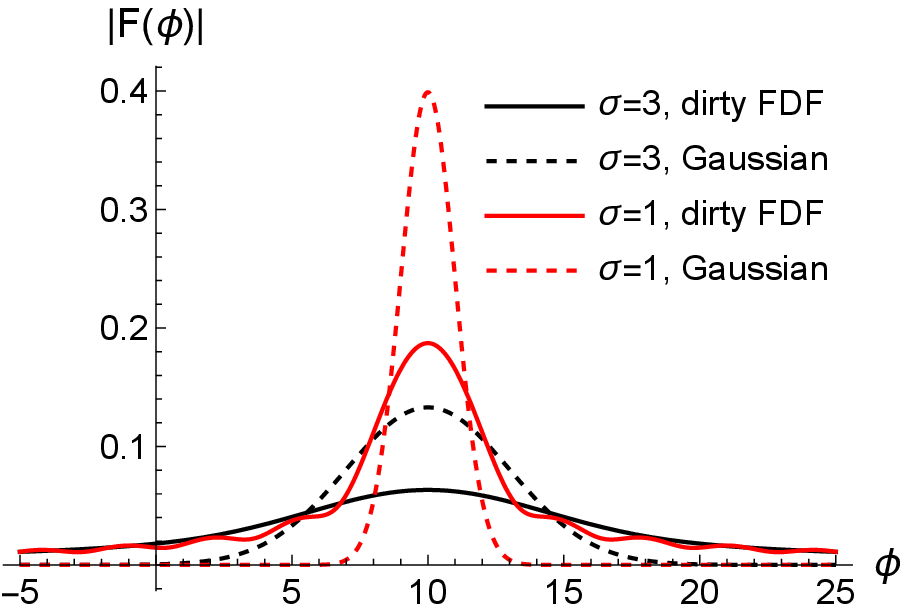} 
\end{center}
\begin{center}
\includegraphics[width=6cm]{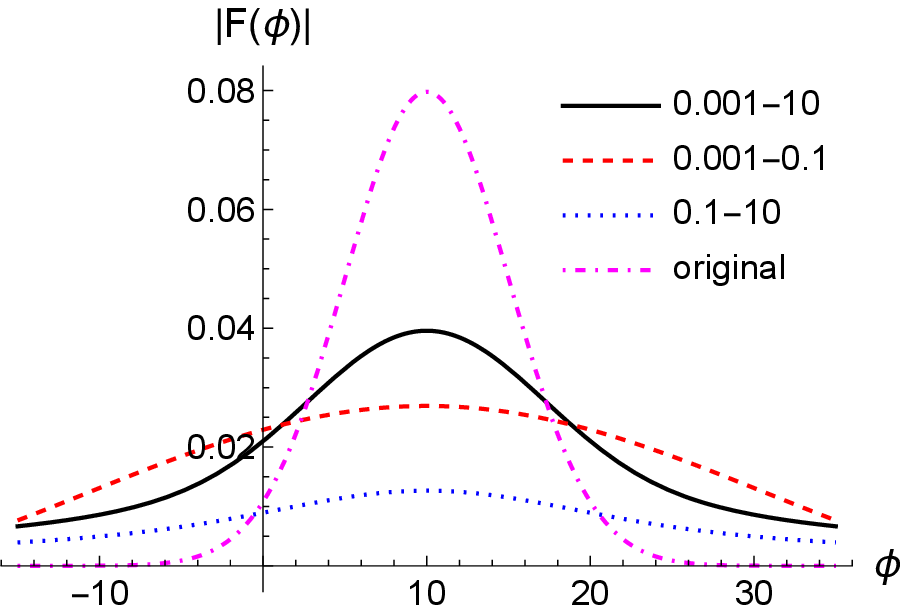} 
\end{center}
\caption{Dirty FDF for Gaussian FDFs. Top: original FDFs (dashed) and dirty FDFs (solid) for $\sigma = 1~{\rm rad/m^2}$ and $\sigma = 3~{\rm rad/m^2}$. Other parameters are set as $f = 1~{\rm Jy \cdot m^2/rad}$, $\phi_0 = 10~{\rm rad/m^2}$, $\lambda^2_{\rm min} = 0.01~{\rm m^2}$ and $\lambda^2_{\rm max} = 1~{\rm m^2}$. Bottom: Comparison of the dirty FDF for three observation bands ($0.001~{\rm m^2} \leq \lambda^2 \leq 10~{\rm m^2}$, $0.001~{\rm m^2} \leq \lambda^2 \leq 0.1~{\rm m^2}$ and $0.1~{\rm m^2} \leq \lambda^2 \leq 10~{\rm m^2}$) and the original Gaussian FDF. Other parameters are set as $f = 1~{\rm Jy \cdot m^2/rad}$, $\phi_0 = 10~{\rm rad/m^2}$ and $\sigma = 5~{\rm rad/m^2}$.}
\label{fig:Gauss}
\end{figure}

The dirty FDF for limited-band observation of the polarization spectrum, Eq.~(\ref{eq:P-Gauss}), is represented by error function and is shown in Fig.~\ref{fig:Gauss}. The top panel shows the dirty FDFs for $\sigma = 1~{\rm rad/m^2}$ and $\sigma = 3~{\rm rad/m^2}$. Other parameters are set as $f = 1~{\rm Jy \cdot m^2/rad}$, $\phi_0 = 10~{\rm rad/m^2}$, $\lambda^2_{\rm min} = 0.01~{\rm m^2}$ and $\lambda^2_{\rm max} = 1~{\rm m^2}$. Due to the incomplete observation, the dirty FDF is wider than the original Gaussian function, and the sidelobes which are seen in the delta-function case cannot be seen.

The bottom panel is comparison of the dirty FDF for three observation bands ($0.001~{\rm m^2} \leq \lambda^2 \leq 10~{\rm m^2}$, $0.001~{\rm m^2} \leq \lambda^2 \leq 0.1~{\rm m^2}$ and $0.1~{\rm m^2} \leq \lambda^2 \leq 10~{\rm m^2}$). Other parameters are set as $f = 1~{\rm Jy \cdot m^2/rad}$, $\phi_0 = 10~{\rm rad/m^2}$ and $\sigma = 5~{\rm rad/m^2}$. As we saw in Eq.~(\ref{eq:P-Gauss}), the polarization intensity is larger for shorter wavelengths so that the dirty FDF with shorter-wavelength observation has larger absolute value.

As we will see in section \ref{section:interpretation}, polarization sources have in general broad structure in Faraday depth space. When the width ($\sigma$ in the case of Gaussian function) is smaller and larger than the FWHM of the RMSF, it is called Faraday thin and Faraday thick, respectively. Gauss function and top-hat function discussed below are often used to characterize Faraday thick sources.

\subsection{Top-hat FDF}
\label{subsection:tophat}

A top-hat function is another commonly-used type of the FDF.
\begin{equation}
F(\phi)  =
\left\{
\begin{array}{ll}
F_0 & (0 \leq \phi \leq \phi_0) \\
0 & (\phi < 0,~ \phi > \phi_0)
\end{array}
\right.
\label{eq:F-tophat}
\end{equation}
In fact, this functional form is not realistic physically, but it is useful for understanding the relationship between the FDF and the polarization spectrum and the nature of Faraday tomography. Since the top-hat type is continuously distributed in Faraday depth space, interference occurs in a complicated form like the Gaussian-function type. First, the polarization spectrum is as follows.
\begin{equation}
P(\lambda^2)
= F_0 \phi_0 \frac{\sin{\phi_0 \lambda^2}}{\phi_0 \lambda^2} e^{2i (\phi_0/2) \lambda^2}
\end{equation}
Therefore, the behavior of the polarization angle is the same as the case with a single delta-function source at $\phi = \phi_0/2$. On the other hand, the polarization intensity decays with oscillation toward long wavelength. The phase of the oscillation is $\phi_0 \lambda^2$, which is a combination of squared wavelength and Faraday depth which characterize the FDF as is the case with two interfering delta-function sources. The intensity decays with $\lambda^{-2}$, which is much slower than the case of Gaussian source. Unlike the Gaussian function type, there is no characteristic wavelength that characterizes depolarization, but $\lambda^2 = \pi /  \phi_0$, where the intensity becomes zero for the first time, is one useful measure.

\begin{figure}
\begin{center}
\includegraphics[width=6cm]{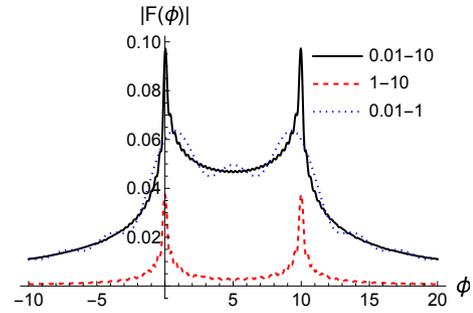} 
\end{center}
\caption{Dirty FDF of top-hat FDF for three observation bands, $0.01~[{\rm m^2}] \leq \lambda^2 \leq 10~[{\rm m^2}]$, $0.01~[{\rm m^2}] \leq \lambda^2 \leq 1~[{\rm m^2}]$ and $1~[{\rm m^2}] \leq \lambda^2 \leq 10~[{\rm m^2}]$. Other parameters are set as $\phi_0 = 10~[{\rm rad/m^2}]$ and $F_0 \phi_0 = 1~[{\rm Jy}]$.}
\label{fig:slab}
\end{figure}

Fig.~\ref{fig:slab} shows the dirty FDF of top-hat FDF. Here, the parameters are set as $\phi_0 = 10~[{\rm rad/m^2}]$ and $F_0 \phi_0 = 1~[{\rm Jy}]$, and three observation bands, $0.01~[{\rm m^2}] \leq \lambda^2 \leq 10~[{\rm m^2}]$, $0.01~[{\rm m^2}] \leq \lambda^2 \leq 1~[{\rm m^2}]$ and $1~[{\rm m^2}] \leq \lambda^2 \leq 10~[{\rm m^2}]$ are compared. Wider band observation results in better reconstruction of the FDF, but even for a very wide band $0.01~[{\rm m^2}] \leq \lambda^2 \leq 10~[{\rm m^2}]$ ($100~{\rm MHz} - 3~{\rm GHz}$) the shape of the dirty FDF is far from the top-hat shape. In fact, the widest physically-possible band ($\lambda^2 > 0$) does not improve the result compared with the case with $0.01~[{\rm m^2}] \leq \lambda^2 \leq 10~[{\rm m^2}]$. This is a limitation of reconstruction without information of unphysical band ($\lambda^2 < 0$), while the unnatural shape of the dirty FDF is due to the fact that the original FDF is not a continuous function.

By investigating the dirty FDFs for two bands, $0.01~[{\rm m^2}] \leq \lambda^2 \leq 1~[{\rm m^2}]$ and $1~[{\rm m^2}] \leq \lambda^2 \leq 10~[{\rm m^2}]$, which divides a wide band $0.01~[{\rm m^2}] \leq \lambda^2 \leq 10~[{\rm m^2}]$, we can see a generic characteristics of Faraday tomography. Because short-wavelength information of polarization spectrum reflects large-scale structure in Faraday-depth space, the dirty FDF for $0.01~[{\rm m^2}] \leq \lambda^2 \leq 1~[{\rm m^2}]$ reproduces the overall feature of that of $0.01~[{\rm m^2}] \leq \lambda^2 \leq 10~[{\rm m^2}]$. Contrastingly, long-wavelength information corresponds to small-scale structure in Faraday-depth space and its dirty FDF reproduces the spike structure at $\phi = 0, 10~{\rm rad/m^2}$. Thus, observation band determines what scale of Faraday-depth space structure can be reproduced.

\subsection{Indices for Faraday tomography}
\label{subsection:index}

In this section, we discussed the basic principle of Faraday tomography. Finally, we summarize some indices which represents the performance of Faraday tomography. Below, $\lambda_{\rm max}^2$, $\lambda_{\rm min}^2$ and $\delta \lambda^2$ denote maximum wavelength, minimum wavelength and channel width, respectively.
\begin{itemize}
\item RMSF FWHM: resolution in Faraday-depth space (see Eq.~(\ref{eq:FWHM}))
\begin{equation}
{\rm FWHM} = \frac{2 \sqrt{3}}{\lambda_{\rm max}^2 - \lambda_{\rm min}^2}
\end{equation}
\item max Faraday depth: maximum Faraday depth which can be probed (see Eq.~(\ref{eq:band-depolarization}))
\begin{equation}
\phi_{\rm max} = \frac{\sqrt{3}}{\delta \lambda^2}
\end{equation}
\item max Faraday scale: maximum width of polarization source in Faraday depth space which can be probed (see Eq.~(\ref{eq:max-width}))
\begin{equation}
\sigma_{\rm max} = \frac{\pi}{\lambda_{\rm min}^2}
\end{equation}
\end{itemize}

\section{Models and interpretation of Faraday dispersion function}
\label{section:interpretation}

In this section, we discuss the physical interpretation of the Faraday dispersion function and introduce some practical models of polarization sources. The FDF include information on the distribution of magnetic fields, thermal electrons and cosmic-ray electrons along the line of sight. However, as we saw in Eq.~(\ref{eq:Faraday-depth}), the interpretation of the FDF is not straightforward because there is in general no one-to-one correspondence between Faraday depth and physical distance. Here, we first consider physical models of the FDF which are simple but help our understanding of its nature in section \ref{subsection:coherent}-\ref{subsection:coherent-turbulent}. Then, in section \ref{subsection:galaxy}, we see some research examples of realistic polarization sources and discuss how they look like in Faraday space.

\subsection{Coherent magnetic field}
\label{subsection:coherent}

Let us consider a uniform slab \citep{1966MNRAS.133...67B,1998MNRAS.299..189S,2011MNRAS.414.2540F}. Magnetic field is uniform within the slab and has strength of $B$, and thermal-electron density, $n_e$, and cosmic-ray electron density, $n_{\rm CR}$, are also assumed to be uniform. Therefore, the distribution of polarization intensity in physical space is, setting the size of the slab along the line of sight as $L$,
\begin{equation}
\varepsilon(x)  =
\left\{
\begin{array}{ll}
\varepsilon_0 & (x_0 \leq x \leq x_0 + L) \\
0 & (x < x_0,~ x > x_0 + L)
\end{array}
\right. .
\end{equation}
Here, $\varepsilon_0$ is the polarization intensity per unit size. Faraday depth increases or decreases monotonically within the slab depending on the direction of magnetic field, and is constant outside the slab.
\begin{equation}
\phi(x)  =
\left\{
\begin{array}{ll}
0 & (x < x_0) \\
\frac{x - x_0}{L} \phi_0 & (x_0 \leq x \leq x_0 + L) \\
\phi_0 & (x > x_0 + L)
\end{array}
\right.
\end{equation}
Here, $\phi_0 = k n_e B_{||} L$. Because the polarized emission exist in a range of $x_0 \leq x \leq x_0 + L$ in physical space, the FDF is nonzero in a range of $0 \leq \phi \leq \phi_0$ in Faraday-depth space. The width of the emission in Faraday-depth space, $\phi_0$, is determined not only by the physical size of the slab but also by the strength of magnetic field along the line of sight and thermal-electron density. From Eq.~(\ref{eq:F-epsilon}), the FDF is expressed as,
\begin{equation}
F(\phi)  =
\left\{
\begin{array}{ll}
\frac{\varepsilon_0 L}{\phi_0} & (0 \leq \phi \leq \phi_0) \\
0 & (\phi < 0,~ \phi > \phi_0)
\end{array}
\right.
\label{eq:F-tophat2}
\end{equation}
Here, $\varepsilon_0 L/\phi_0$ is polarization intensity per unit Faraday depth. In this way, the FDF for a uniform slab is a top-hat function. There are 4 parameters which characterize a top-hat FDF: width, height, position (central Faraday depth) and polarization angle. On the other hand, a uniform slab can be characterized by 6 parameters in physical space: size, parallel and perpendicular components of coherent magnetic field, thermal-electron density and cosmic-ray electron density. Thus, even if we obtain such FDF from observation and assume that the source is a uniform slab, we cannot determine the latter 6 parameters from the former 4 parameters. In other words, we cannot obtain all parameters in physical space due to degeneracy.

Next, we consider a case where the direction of coherent magnetic field is reversed at the center of the slab, $x = x_0 + L/2$. Specifically, we assume the field points in the direction of the observer for $x < x_0 + L/2$. The field strength, thermal electrons and cosmic-ray electrons are again assumed to be uniform. In this case, Faraday depth increases and decreases monotonically for $x_0 < x < x_0 + L/2$ and $x_0 + L/2 < x < x_0 + L$, respectively, and reaches zero at $x = x_0 + L$.
\begin{equation}
\phi(x)  =
\left\{
\begin{array}{ll}
0 & (x < x_0) \\
\frac{x - x_0}{L} \phi_0 & (x_0 \leq x \leq x_0 + L/2) \\
\frac{L - (x - x_0)}{L} \phi_0 & (x_0 + L/2 \leq x \leq x_0 + L) \\
0 & (x > x_0 + L)
\end{array}
\right.
\end{equation}
It should be noted that one value of Faraday depth in the range of $0 < \phi < \phi_0/2$ corresponds to 2 positions in physical space. Thus, a range of polarized emission in Faraday-depth space becomes halves and the polarization intensity per unit Faraday depth doubles compared to the previous case.
\begin{equation}
F(\phi)  =
\left\{
\begin{array}{ll}
\frac{2 \varepsilon_0 L}{\phi_0} & (0 \leq \phi \leq \phi_0/2) \\
0 & (\phi < 0,~ \phi > \phi_0/2)
\end{array}
\right.
\label{eq:F-tophat}
\end{equation}
This is again a top-hat function and the shape is exactly the same as that of Eq.~(\ref{eq:F-tophat2}). Furthermore, in this case, the edges in Faraday-depth space ($\phi = 0$ and $\phi_0/2$) does not always correspond to the edges in physical space ($x = x_0, x_0 + L$). Therefore, when we obtain a top-hat FDF from observation, we suffer from not only the parameter degeneracy but also uncertainty in the configuration of magnetic field.

As we saw above, we cannot determine the distribution model in physical space from the FDF, even for such a very simple system as a uniform slab. This is a fundamental problem in Faraday tomography. Nevertheless, the FDF has much richer information compared with the conventional rotation measure and we can obtain various physical implication from it.

\subsection{Faraday caustics}
\label{subsection:inhomogeneous}

\begin{figure}
\begin{center}
\includegraphics[width=7cm]{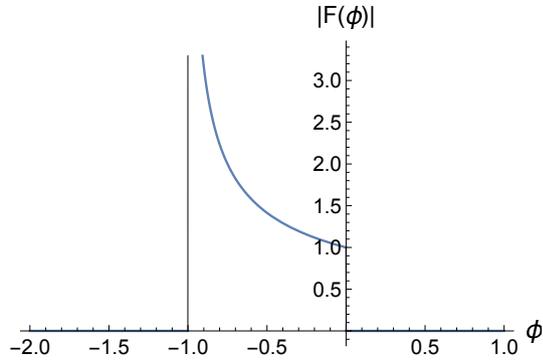} 
\end{center}
\caption{FDF with Faraday caustics expressed in Eq.~(\ref{eq:caustics}). Parameters are set as $\phi_0 = 1~{\rm rad/m^2}$ and $\epsilon_0 L = 1$}
\label{fig:caustics}
\end{figure}

\citet{2011A&A...535A..85B} proposed a simple configuration of magnetic field which leads to the FDF with a striking feature. Let us consider a emission region with uniform polarization emissivity $\epsilon_0$ and polarization angle $\chi_0 = 0~{\rm rad}$ within a range of $-L < x < L$. The line-of-sight component of magnetic field is assumed to vary as,
\begin{equation}
B_{||}(x) = B' x,
\end{equation}
where $B'$ is a constant. Then, Faraday depth can be calculated as,
\begin{eqnarray}
&& \phi(x) = \phi_0 \left( \frac{x^2}{L^2} - 1 \right) ~~~~~ (-L \leq x \leq L) \\
&& \phi_0 \equiv \frac{k n_e B' L^2}{2}
\end{eqnarray}
Here, thermal-electron density $n_e$ is assumed to be constant. Then, the resultant FDF is,
\begin{eqnarray}
F(\phi)
&=& \epsilon_0 \int_{-L}^{L} \delta \left( \phi -  \phi_0 \left( \frac{x^2}{L^2} - 1 \right) \right) dx \nonumber \\
&=& \frac{L}{\sqrt{\phi_0}} \frac{\epsilon_0}{\sqrt{\phi + \phi_0}} ~~~~~ \left( - \phi_0 \leq \phi \leq 0 \right)
\label{eq:caustics}
\end{eqnarray}
which is divergent at $\phi = - \phi_0$. This behavior was called "Faraday caustics" in analogy with optics in \citet{2011A&A...535A..85B} and shown in Fig.~\ref{fig:caustics} setting $\phi_0 = 1~{\rm rad/m^2}$ and $\epsilon_0 L = 1$. In this way, the FDF diverges at Faraday depth $\phi = -\phi_0$ which corresponds to the field reversal and has an asymmetric sharp peak. Which side of the peak has a non-zero polarization intensity depends on the direction of the magnetic field. This divergence appears because Faraday depth is almost constant at $x=0$ and polarization emissivity around $x=0$ contributes to the FDF of the very narrow range around $\phi = -\phi_0$. The FDF is polarization intensity per unit Faraday depth so that it diverges if finite amount of polarization intensity contributes to a very narrow range in Faraday-depth space. Although the peak will be smoothed on the scale of the FWHM of the RMSF in practical observations, such a feature will be helpful for the physical interpretation of the FDF.

Finally it should be noted that the total emissivity should be finite while the FDF itself is divergent. In fact, we have,
\begin{equation}
\int_{-\infty}^{\infty} |F(\phi)| d\phi = 2 \epsilon_0 L
\end{equation}
This depends only on the size of the slab and the emissivity per unit size and not on $\phi_0$.

\subsection{Helical magnetic field}
\label{subsection:helical}

\citet{2014MNRAS.441.2049H} considered an emitting region with helical magnetic field and derived the FDF to discuss the determination of parameters through observation. Helical magnetic fields are considered to play an important role in dynamo mechanism and also exist in galactic and protostar jets.

Helical magnetic field toward $z$ direction can be written as,
\begin{eqnarray}
&& B_x(z) = B_{\perp} \cos{(k_z (z - z_0) + \chi_0)} \\
&& B_y(z) = B_{\perp} \sin{(k_z (z - z_0) + \chi_0)} \\
&& B_z(z) = B_{||}.
\end{eqnarray}
Here, the strength of $z$ component, $B_{||}$, and the perpendicular components, $B_{\perp}$, are assumed to be constant and $k_z$ represents the wavenumber of helix in $z$ direction. When the line of sight is in $z$ direction, Faraday depth and polarization angle vary linearly with $z$.
\begin{eqnarray}
&& \phi(z) = k' B_{||} (z - z_0) \\
&& \chi(z) = k_z (z - z_0) + \chi'_0,
\end{eqnarray}
where $k' = k n_e$ using $k$ in Eq.~(\ref{eq:RM-coeff}) and thermal electron density $n_e$ is assumed to be constant. In this system, the phase of the FDF vary linearly with Faraday depth.
\begin{equation}
\chi(\phi) = \frac{k_z}{k' B_{||}} (\phi - \phi_0) + \chi'_0
\equiv \beta \phi + \chi_0
\end{equation}
Here, denoting the FDF in case of $\beta = 0$ as $F_0(\phi)$, the FDF with non-zero $\beta$ can be written as,
\begin{equation}
F(\phi) = F_0(\phi) e^{2 i \beta \phi}.
\end{equation}
Then, the polarization spectrum, $P(\lambda^2)$, is expressed by using that for $\beta = 0$, $P_0(\lambda^2)$, as,
\begin{eqnarray}
P(\lambda^2)
&=& \int_{-\infty}^{\infty} F(\phi) e^{2 i \phi \lambda^2} d\phi \nonumber \\
&=& \int_{-\infty}^{\infty} F_0(\phi) e^{2 i \phi (\lambda^2 + \beta)} d\phi \nonumber \\
&=& P_0(\lambda^2 + \beta).
\end{eqnarray}
This is $P_0(\lambda^2)$ shifted in $\lambda^2$ direction by $(-\beta)$.

Next, let us consider a case where the directions of line of sight and helix do not coincide. We take $z'$ axis as the new line-of-sight direction by rotation by $\theta$ in $(x,z)$ plane. The coordinate transformation is given by,
\begin{equation}
\left( \begin{array}{c} x \\ y \\ z \end{array} \right)
=
\left( \begin{array}{ccc}
\cos{\theta} & 0 & \sin{\theta} \\
0 & 1 & 0 \\
-\sin{\theta} & 0 & \cos{\theta}
\end{array} \right)
\left( \begin{array}{c} x' \\ y' \\ z' \end{array} \right),
\end{equation}
and magnetic field components in the new coordinate system is given by,
\begin{eqnarray}
&& \left( \begin{array}{c} B_{x'} \\ B_{y'} \\ B_{z'} \end{array} \right) \\
&& =
\left( \begin{array}{ccc}
\cos{\theta} & 0 & \sin{\theta} \\
0 & 1 & 0 \\
-\sin{\theta} & 0 & \cos{\theta}
\end{array} \right)
\left( \begin{array}{c}
B_{\perp} \cos{(k_z (z - z_0) + \chi_0)} \\ B_{\perp} \sin{(k_z (z - z_0) + \chi_0)} \\ B_{||}
\end{array} \right) \nonumber \\
&& =
\left(  \begin{array}{c}
B_{\perp} \cos{\theta} \cos{(k_z (z - z_0) + \chi_0)} + B_{||} \sin{\theta} \\
B_{\perp} \sin{(k_z (z - z_0) + \chi_0)} \\
- B_{\perp} \sin{\theta} \cos{(k_z (z - z_0) + \chi_0)} + B_{||} \cos{\theta}
\end{array} \right).
\label{eq:helical-B_new}
\end{eqnarray}
Then, polarization angle and Faraday depth depend on $x'$ and $z'$ through $z(x',z')$ and expressed as,
\begin{eqnarray}
\chi(x',z')
&=& \arctan{\left[ \frac{B_{\perp} \sin{(k_z (z - z_0) + \chi_0)}}{B_{\perp} \cos{\theta} \cos{(k_z (z - z_0) + \chi_0)} + B_{||} \sin{\theta}} \right]} \nonumber \\ \\
\phi(x',z')
&=& k' \int_{z'_0}^{z'} \left[ - B_{\perp} \sin{\theta} \cos{(k_z (z - z_0) + \chi_0)} + B_{||} \cos{\theta} \right] dz' \nonumber \\
&=& - \frac{k' B_{\perp} \sin{\theta}}{k_z \cos{\theta}} \left[ \sin{(k_z (z - z_0) + \chi_0)} - \sin{\chi_0} \right]
\nonumber \\
& & + k' B_{||} (z' - z'_0) \cos{\theta}
\end{eqnarray}
Faraday depth is generally not a monotonic function with respect to $z'$ because the line-of-sight component of the magnetic field can be inverted when observing helical magnetic field from an angle. In this case, $\chi$ and $\phi$ are not single-valued functions with respect to $z'$.

\begin{figure}[t]
\begin{center}
\includegraphics[width=7cm]{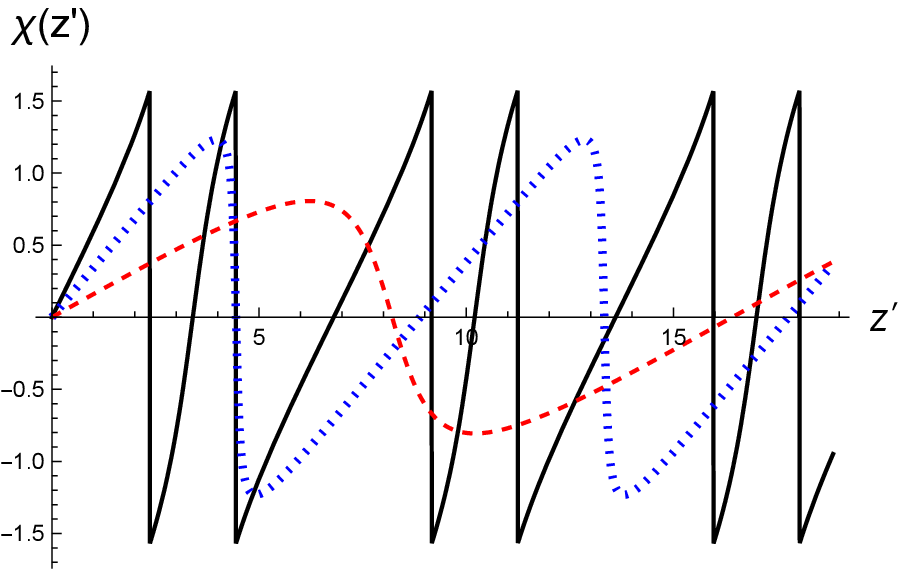} 
\end{center}
\begin{center}
\includegraphics[width=7cm]{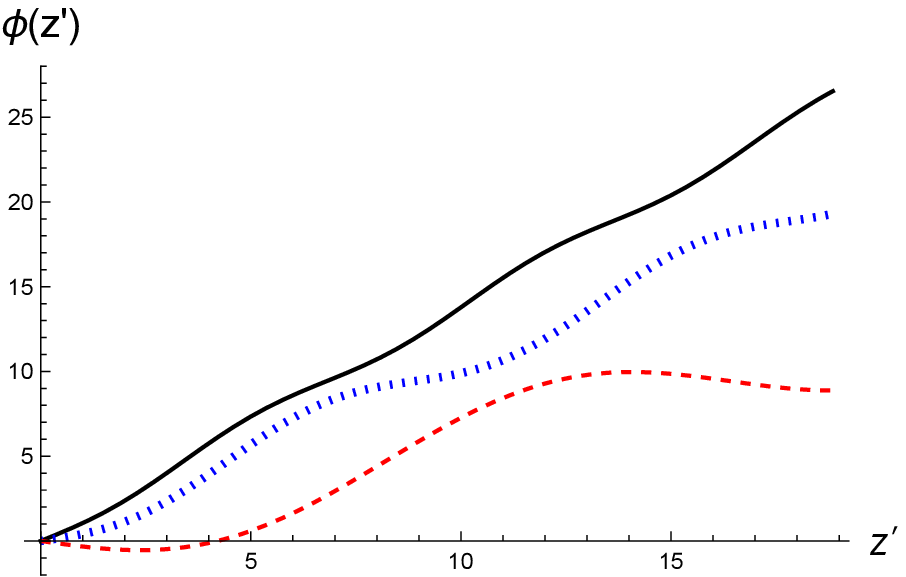} 
\end{center}
\begin{center}
\includegraphics[width=7cm]{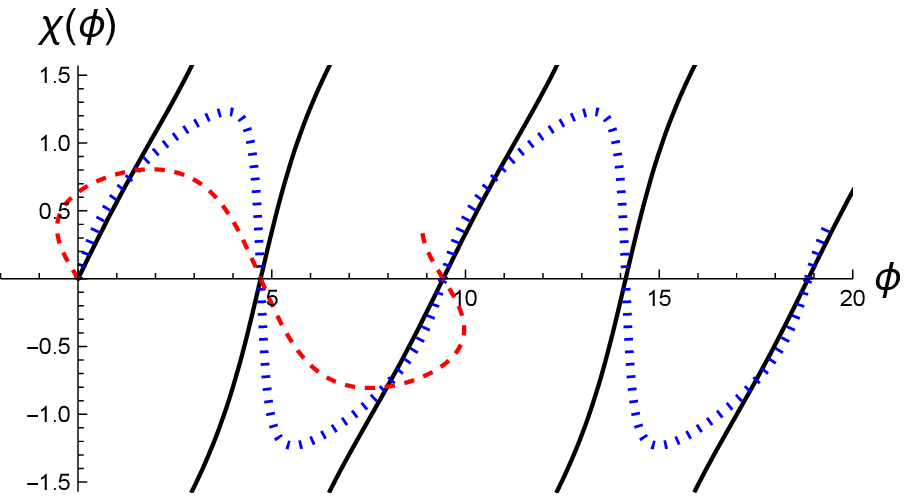} 
\end{center}
\caption{Behavior of $\chi(z')$ (top), $\phi(z')$ (middle) and $\chi(\phi)$ (bottom), when emitting region with helical magnetic field is observed from an angle. Here, parameters are set as $k' B_{\perp} = 1$, $k' B_{||} = 1.5$, $k_z = 1$, $z_0 = 0$, $\chi_0 = 0$ and $x' = 0$. Cases with $\theta = \pi/8, \pi/4$ and $3 \pi/8$ are shown with black solid, blue dotted and red dashed lines.}
\label{fig:helical}
\end{figure}

Fig.~\ref{fig:helical} shows an example of $\chi(z')$, $\phi(z')$ and $\chi(\phi)$. Here, 3 cases with $\theta = \pi/8$, $\pi/4$ and $3 \pi/8$ are compared and other parameters are set as $k' B_{\perp} = 1$, $k' B_{||} = 1.5$, $k_z = 1$, $z_0 = 0$, $\chi_0 = 0$ and $x' = 0$. In the case of $\theta = \pi/8$, $B_{||}$ is dominant in the line-of-sight component (see Eq.~(\ref{eq:helical-B_new})) so that Faraday depth is monotonic with respect to $z'$ and polarization angle is almost linear with respect to Faraday depth. However, in the case of $\theta = 3 \pi/8$, the line-of-sight component of the magnetic field is inverted along the line-of-sight direction and Faraday depth is not monotonic with respect to $z'$. Therefore, multiple positions in physical space have the same value of Faraday depth and polarization angle is not single-valued for some ranges of Faraday depth. In this case, polarization emission with different polarization angle contribute to the FDF of the same Faraday depth, which leads to depolarization.

As we saw above, the dependence of the FDF phase (polarization angle) on Faraday depth has in general information on the shape and configuration of magnetic fields.

\subsection{Turbulent magnetic fields}
\label{subsection:turbulent}

Turbulence is often developed in magnetic plasmas in interstellar space, and the presence of turbulence complicates the FDF \citep{2012A&A...543A.113B,2017ApJ...843..146I}. In order to discuss the FDF due to the polarization emission of turbulent plasma, we start from the following simple model of turbulence following \citet{2017ApJ...843..146I}.
\begin{itemize}
\item divide radiation region into regular lattice which consists of many cubic cells of size $L_{\rm cell}^3$
\item suppose $N$ cells are lined up in the line-of-sight direction
\item magnetic is coherent within each cell
\item direction of turbulent magnetic field is random and there is no correlation in the direction and strength between different cells
\item each cell can have coherent magnetic field common to all emission region
\item thermal-electron density and cosmic-ray density are uniform in the whole emitting region
\end{itemize}
Let us first consider the behavior of Faraday depth along the line of sight. Faraday depth of $n$-th cell is proportional to the sum of magnetic fields from the first cell to $n$-th cell. Denoting line-of-sight component of magnetic field and Faraday depth of $i$-th cell as $B_{||}^i$ and $\phi^i$,
\begin{equation}
\phi^n = k n_e L_{\rm cell} \sum_{i=1}^n B_{||}^i
\label{eq:phi-turbulent}
\end{equation}
Here, we assume $B_{||}^i$ is a Gaussian random variable with a mean of zero and a standard deviation of $\sigma_B$. Then, Faraday depth behaves as random walk.

To consider the FDF, it is necessary to consider the polarization intensity distribution in each cell. Here, for simplicity, let us assume that the component of the magnetic field perpendicular to the line of sight have the same strength and orientation in all cells. This may appear to contradict with the existence of turbulent magnetic fields. However, for example, in a case of a face-on galaxy which has strong coherent magnetic field along the galactic plane, it is possible that the perpendicular component is dominated by the coherent field while the line-of-sight component is dominated by the turbulent fields. In this case, all cells have the same polarization intensity and polarization angle, and the FDF is proportional to the distribution of Faraday depth because the complex polarization intensities of cells with the same Faraday depth are summed coherently.

\begin{figure}
\begin{center}
\includegraphics[width=6cm]{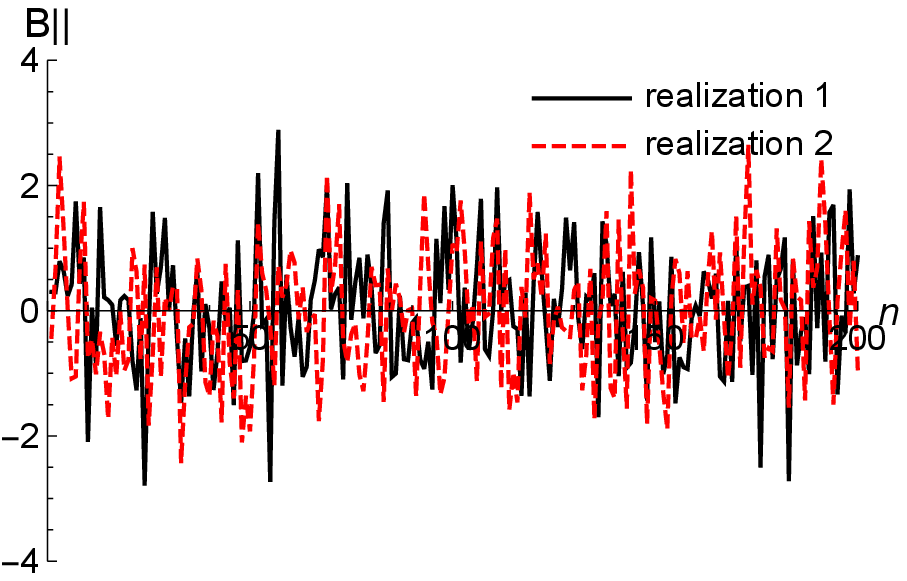} 
\end{center}
\begin{center}
\includegraphics[width=6cm]{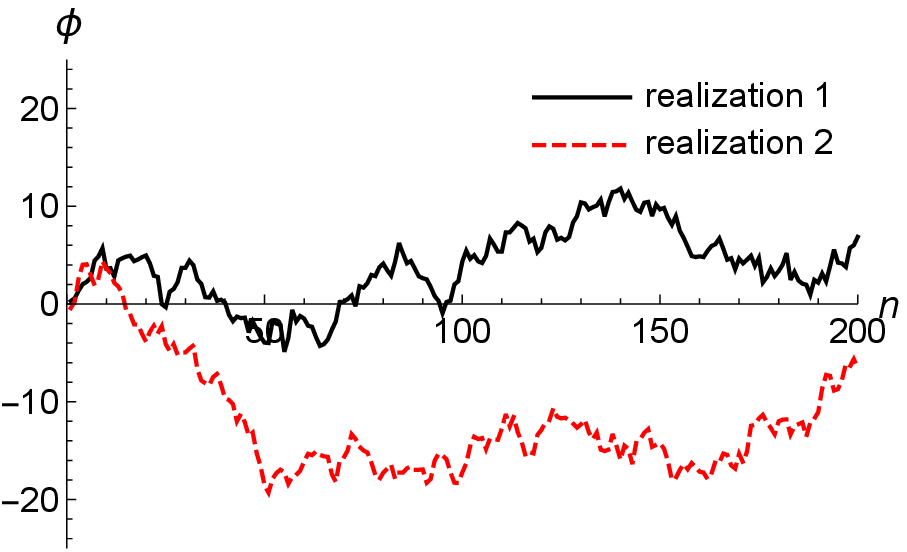} 
\end{center}
\begin{center}
\includegraphics[width=6cm]{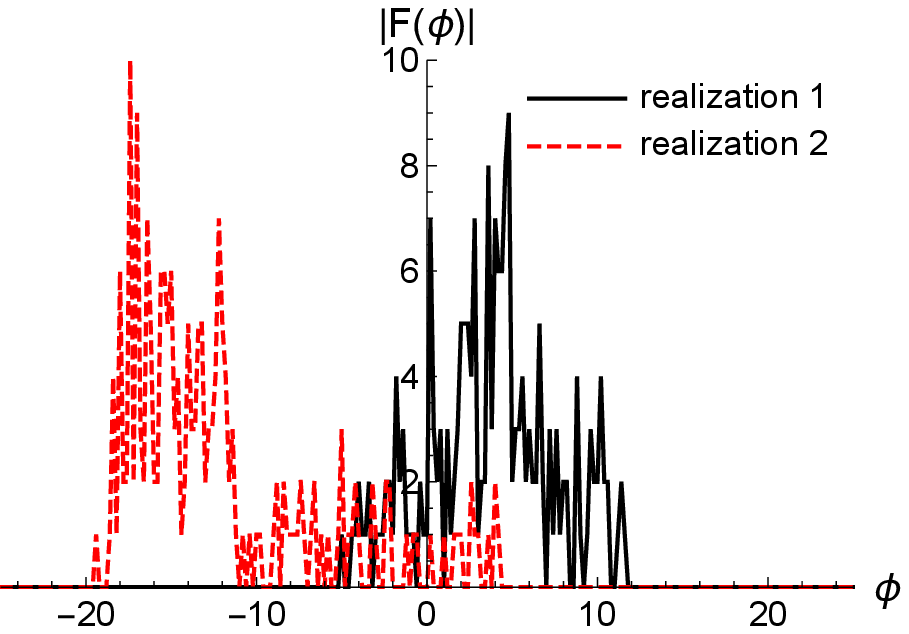} 
\end{center}
\caption{Simulation of turbulence based on the cell model. The standard deviation of $B_{||}^i$ is set to $\sigma_B = 1~{\rm \mu G}$ and the number of cells along a line of sight is $N = 200$. The top and middle panels show $B_{||}$ and the sum of $B_{||}$ from the first cell to $n$-th cell which is proportional to Faraday depth of $n$-th cell, taking the horizontal axis as the cell number $n$. The bottom panel shows the histogram of $\phi$ which is proportional to the absolute value of the FDF. Two independent realizations of turbulent fields are shown in black solid line and red dashed line.}
\label{fig:random}
\end{figure}

Fig.~\ref{fig:random} shows the result of simulations of turbulent magnetic fields. The standard deviation of $B_{||}^i$ is set to $\sigma_B = 1~{\rm \mu G}$ and the number of cells along a line of sight is $N = 200$. The top panel is two independent realizations of the line-of-sight components of random magnetic fields in the 200 cells. The middle panel is the sum of $B_{||}$ from the first cell to $n$-th cell, which is proportional to Faraday depth of $n$-th cell. As the two lines are based on two independent realizations of turbulent magnetic fields, they behave as two independent random walks. The bottom panel shows the histogram of $\phi$ which is proportional to the absolute value of the FDF. We can see that the two FDFs have significantly different shapes and average Faraday depths. This is due to the randomness of the turbulence, even if the two realizations have the same statistical property of turbulence and distribution of thermal electrons and cosmic-ray electron. Therefore, the turbulent magnetic field is one of the factors that make it difficult to extract physical information from the FDF.

\begin{figure}[t]
\begin{center}
\includegraphics[width=6cm]{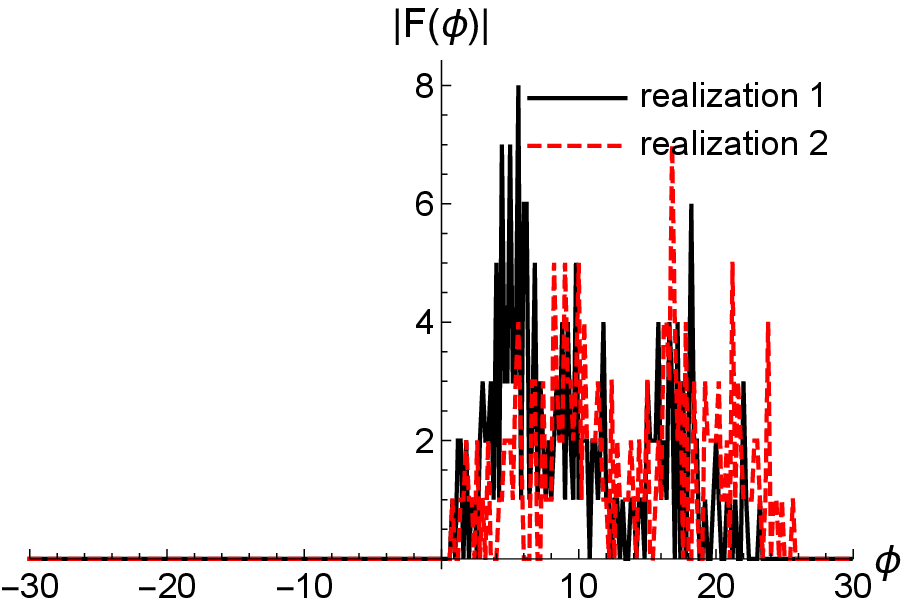} 
\end{center}
\begin{center}
\includegraphics[width=6cm]{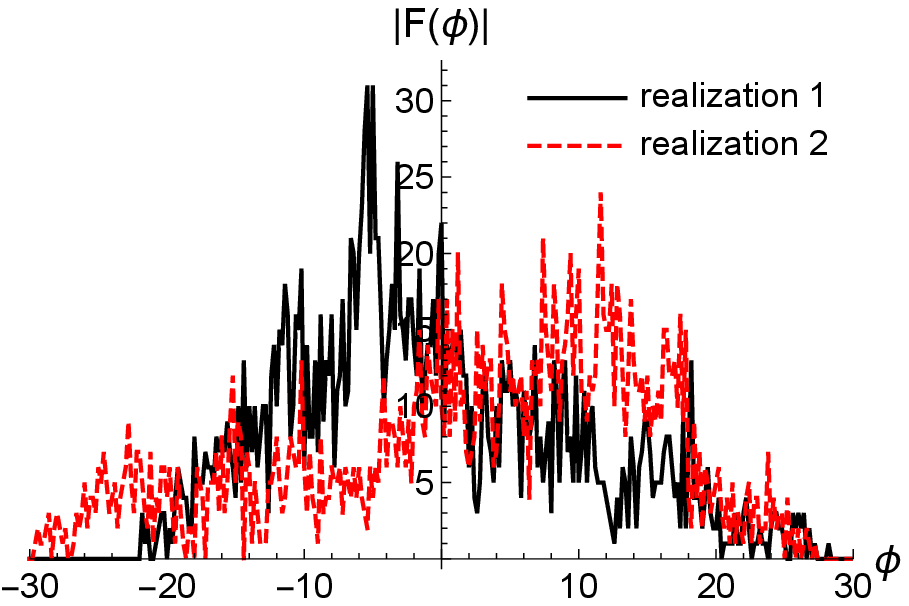} 
\end{center}
\begin{center}
\includegraphics[width=6cm]{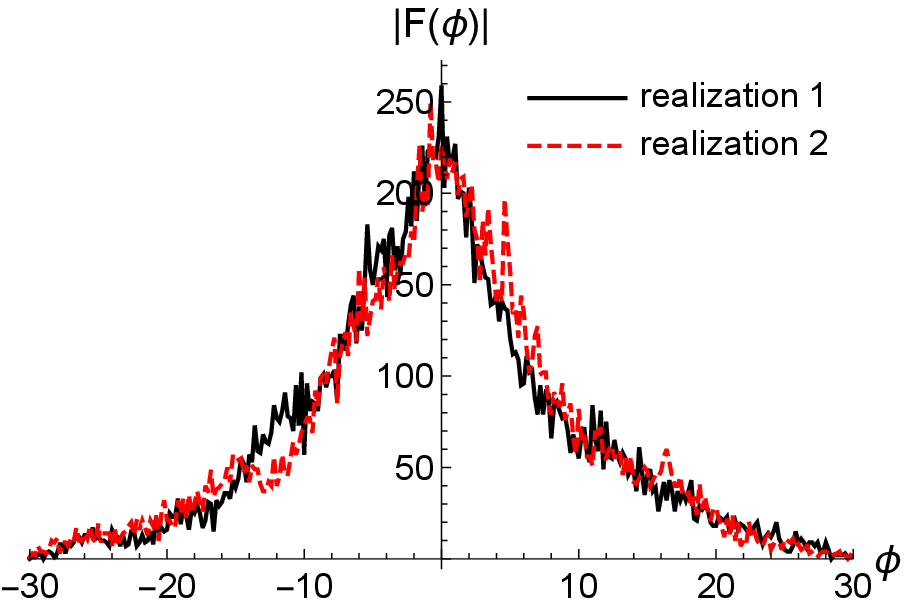} 
\end{center}
\begin{center}
\includegraphics[width=6cm]{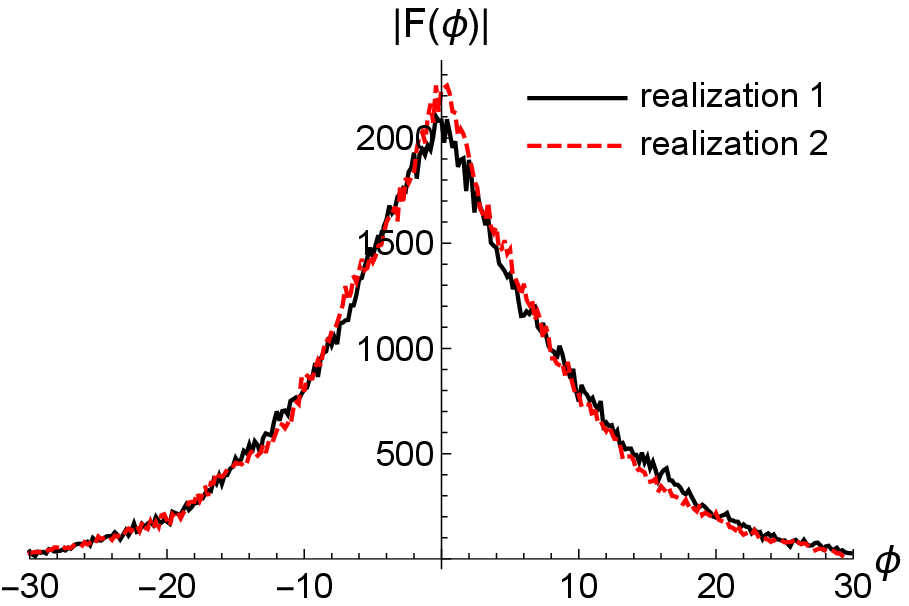} 
\end{center}
\caption{Simulation of the FDF due to turbulent plasma. While the parameters are the same as Fig.~\ref{fig:random}, independent FDFs are added together. The number of independent FDFs is $N = 1, 10, 100$, and $1000$, and the FDFs of two different realizations are shown. As $N$ increases, the difference between the two realizations becomes small.}
\label{fig:random2}
\end{figure}

\citet{2017ApJ...843..146I} proposed that adding many statistically independent FDFs will reduce the randomness and make it easier to extract physical information. This is automatically realized when the angular resolution is much larger than the coherence length of turbulence so that many turbulent cells are included within a beam. For example, the angular size of a cell of size $100~{\rm pc}$ located at $100~{\rm Mpc}$ is about $0.2~{\rm arcsec}$. If the angular resolution is $1~{\rm arcsec}$, about 25 independent cells are lined up perpendicular to the line of sight in a beam, and the observed FDF is a sum of those in the direction of these cells. Fig.~\ref{fig:random2} shows that the randomness reduces gradually as more statistically independent FDFs are added together. Approximately 100 or more additions reduce the difference between different realizations, and the resultant FDF is considered to represent the average behavior and statistical properties of turbulence.

Such an average behavior of turbulence can be understood analytically \citep{2017ApJ...843..146I}. First, we divide the emission region into cubic cells with size $L_{\rm cell}$. We assume that $N$ cells are arranged in the line-of-sight direction, and $N_{\perp} \times N_{\perp} $ cells are arranged in a square shape in the direction perpendicular to the line of sight. Therefore, the size of the emission region is $(N_{\perp} L_{\rm cell})^2 \times N L_{\rm cell}$. This can be thought of as a square layer of $N_{\perp} \times N_{\perp}$ cells stacked along the line of sight. The average and dispersion of Faraday depth of $n$-th cell are given by,
\begin{eqnarray}
&& \langle \phi^n \rangle =  k n_e L_{\rm cell} \sum_{i=1}^n \langle B_{||}^i \rangle = 0
\\
&& \langle (\phi^n - \langle \phi^n \rangle)^2 \rangle
= k^2 n_e^2 L_{\rm cell}^2 \sum_{i=1}^n \langle (B_{||}^i)^2 \rangle
= n k^2 n_e^2 L_{\rm cell}^2 \sigma_B^2
\end{eqnarray}
Here, $\langle \cdots \rangle$ represents the statistical average. Therefore, the expectation value of Faraday depth is zero and the dispersion is larger for farther layer and proportional to $n$.

\begin{figure}
\begin{center}
\includegraphics[width=6cm]{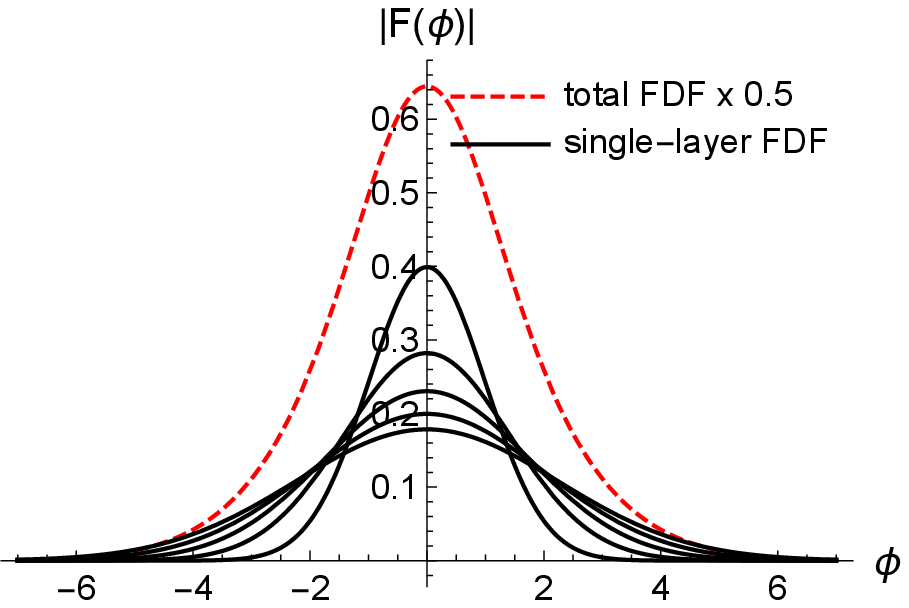} 
\end{center}
\begin{center}
\includegraphics[width=6cm]{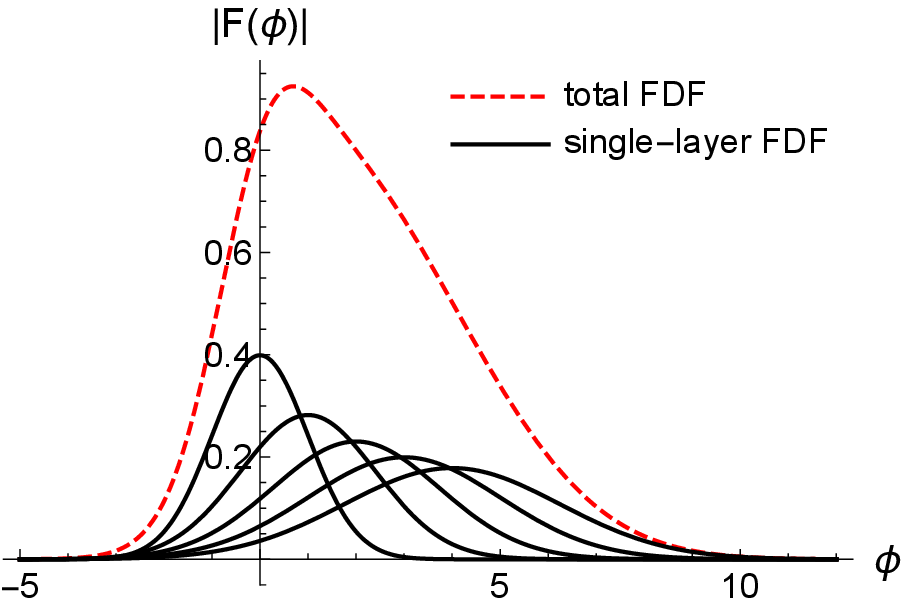} 
\end{center}
\caption{Summation of Gaussian functions. Top panel shows 5 Gaussian functions with a mean of zero and increasing width ($\sigma_{\phi}^2 = 1 \sim 5$) and the sum multiplied with $0.5$. Bottom panel shows the case with an increasing mean and width ($\phi_0 = \sigma_{\phi} = 1$).}
\label{fig:analytic}
\end{figure}

In general, the probability distribution of $\phi^n$ is determined by that of $B_{||}$. However, because $\phi^n$ is a sum of independent and identically distributed variables ($B_{||}^i, i=1,2,\cdots,n$), it obeys Gaussian distribution with a mean of zero and a dispersion of $n k^2 n_e^2 L_{\rm cell}^2 \sigma_B^2 \equiv n \sigma_{\phi}^2$ for $n \gg 1$ according to the central limit theorem. Then, the probability distribution of Faraday depth of a cell in $n$-th layer, $P_n(\phi)$, is,
\begin{equation}
P_n(\phi) = \frac{1}{\sqrt{2 \pi n \sigma_{\phi}^2}} \exp{\left[ - \frac{\phi^2}{2 n \sigma_{\phi}^2} \right]}
\label{eq:P_n}
\end{equation}
Assuming that the polarization intensity and polarization angle are the same for all cells, the FDF is proportional to the histogram of Faraday depth of $N_{\perp}^2 N$ cells. For a large $N_{\perp}$, the histogram of $n$-th layer approaches the probability distribution in Eq.~(\ref{eq:P_n}) multiplied with $N_{\perp}^2$ and the FDF is approximately given by,
\begin{equation}
F(\phi) \propto \sum_{n=1}^N P_n(\phi)
\label{eq:F_sum}
\end{equation}
Therefore, the FDF is a sum of many Gaussian functions with different widths. A demonstration of a sum of Gaussian functions with different widths is shown in Fig.~\ref{fig:analytic}. The resultant function has a heavier tail and a sharper peak than a Gaussian function. The two bottom panels in Fig.~\ref{fig:random2} have the same feature. On the other hand, the phase of the FDF is constant with respect to Faraday depth. It should be noted that $P_n(\phi)$ depends on the probability distribution of $B_{||}$ for small values of $n$ and the approximation of Gaussian function is not valid. However, if $N \gg 1$, the contribution of the layers of such small $n$ to the sum in Eq.~(\ref{eq:F_sum}) is not significant and the resultant FDF is well approximated by Eq.~(\ref{eq:F_sum}).

So far, the direction and strength of magnetic field perpendicular to the line of sight were assumed to be the same for all cells. If the perpendicular component is also assumed to be random, the complex polarization intensity is not coherently summed at the same Faraday depth, which causes depolarization. Because the polarization angle is angle, the polarization intensity at a specific value of Faraday depth is proportional to $\sqrt{N_\phi}$, where $N_\phi$ is the number of cells which has the value of Faraday depth. In this case, the FDF of each layer is proportional to a square root of a Gaussian function. Even then, the qualitative property remains the same in the sense that the total FDF is a sum of functions which are broader for deeper layers. Contrastingly, there is no correlation between the phases of the total FDF at different Faraday depths. The FDF of turbulent plasma was also calculated with numerical MHD simulations of turbulence in \citet{2019Galax...7...89B}.

\subsection{Coherent and turbulent magnetic fields}
\label{subsection:coherent-turbulent}

Let us consider a case where both coherent and turbulent magnetic fields exist in a emission region. The parallel component of each cell, $B_{||}$, is a sum of coherent ($B_{\rm coh}$) and turbulent ($B_{\rm tur}$) components.
\begin{equation}
B_{||} = B_{\rm coh} + B_{\rm tur}
\end{equation}
Here, $B_{\rm coh}$ is a constant common to all cells, while $B_{\rm tur}$ is a random variable. In this case, the expectation value of Faraday depth and its variance of a cell at $n$-th layer is given by,
\begin{eqnarray}
&& \langle \phi^n \rangle = n k n_e L_{\rm cell} B_{\rm coh} \equiv n \phi_0
\\
&& \langle (\phi^n - \langle \phi^n \rangle)^2 \rangle
= n \sigma_{\phi}^2
\end{eqnarray}
It should be noted that both are proportional to $n$. Therefore, for $n \gg 1$, the probability distribution of Faraday depth at $n$-th layer is a Gaussian function whose center is shifted by $ \langle \phi^n \rangle$.
\begin{equation}
P_n(\phi) = \frac{1}{\sqrt{2 \pi n \sigma_{\phi}^2}} \exp{\left[ - \frac{(\phi - n \phi_0)^2}{2 n \sigma_{\phi}^2} \right]}
\label{eq:P_n2}
\end{equation}
As discussed in section \ref{subsection:turbulent}, if the coherent field is dominant in the perpendicular component, the total FDF is again expressed by Eq.~(\ref{eq:F_sum}), that is, a sum of Gaussian functions. In the current case, not only the width but the central value of the Gaussian functions changes at a certain rate toward deeper layers. An example with $\phi_0 = \sigma_{\phi} = 1$ is shown in the bottom panel of Fig.~\ref{fig:analytic}. As we can see, while each Gaussian function is symmetric, the total FDF is not symmetric and has a finite skewness. The overall shape is determined by the relative strength of coherent and turbulent fields.
\citet{2017ApJ...843..146I} focused on the width, skewness and kurtosis to characterize the FDF and investigated the relation to the strength of the coherent field.

So far, in our simple model, we treated turbulence by dividing an emission region into cubic cells. In actual interstellar turbulence, fluctuations of various scales exist, and a turbulent magnetic field with a power-law power spectrum is often considered. One of the major models is the Kolmogorov turbulence. \citet{2017ApJ...843..146I} performed a series of numerical simulations of turbulent magnetic fields with a power-law power spectrum to make a comparison with the simple cell model and evaluate the effect of the spectral index on the FDF. As a result, it was shown that the difference between the two models is not significant and the dependence on the spectral index is also small.

As we saw in Eq.~(\ref{eq:phi-turbulent}), Faraday depth is a sum of many independent variables and the much information of the original statistical properties of turbulent fields is lost due to the central limit theorem. Thanks to this fact, while the FDF of an emission region with a turbulent magnetic field is universal and relatively easy to interpret physically, it makes it impossible to explore the statistical properties of turbulence.

Finally, in the simple cell model, we assumed the thermal-electron density and cosmic-ray electron density to be constant, but they are generally non-uniform in practice. However, the method of considering the uniform and turbulent fields separately is still effective, and the simple cell model provides a generic feature of the FDF of turbulent plasma and works as a benchmark to understand more realistic and complicated FDFs.

\subsection{Galactic models}
\label{subsection:galaxy}

There are many kinds of astronomical objects which have magnetic fields and emit synchrotron radiation, and a galaxy is one of the most interesting target of Faraday tomography. In previous studies, there have been some attempts to theoretically predict realistic galaxy FDFs to help the physical interpretation of observationally reconstructed FDFs.

\begin{figure}
\begin{center}
\includegraphics[width=7cm]{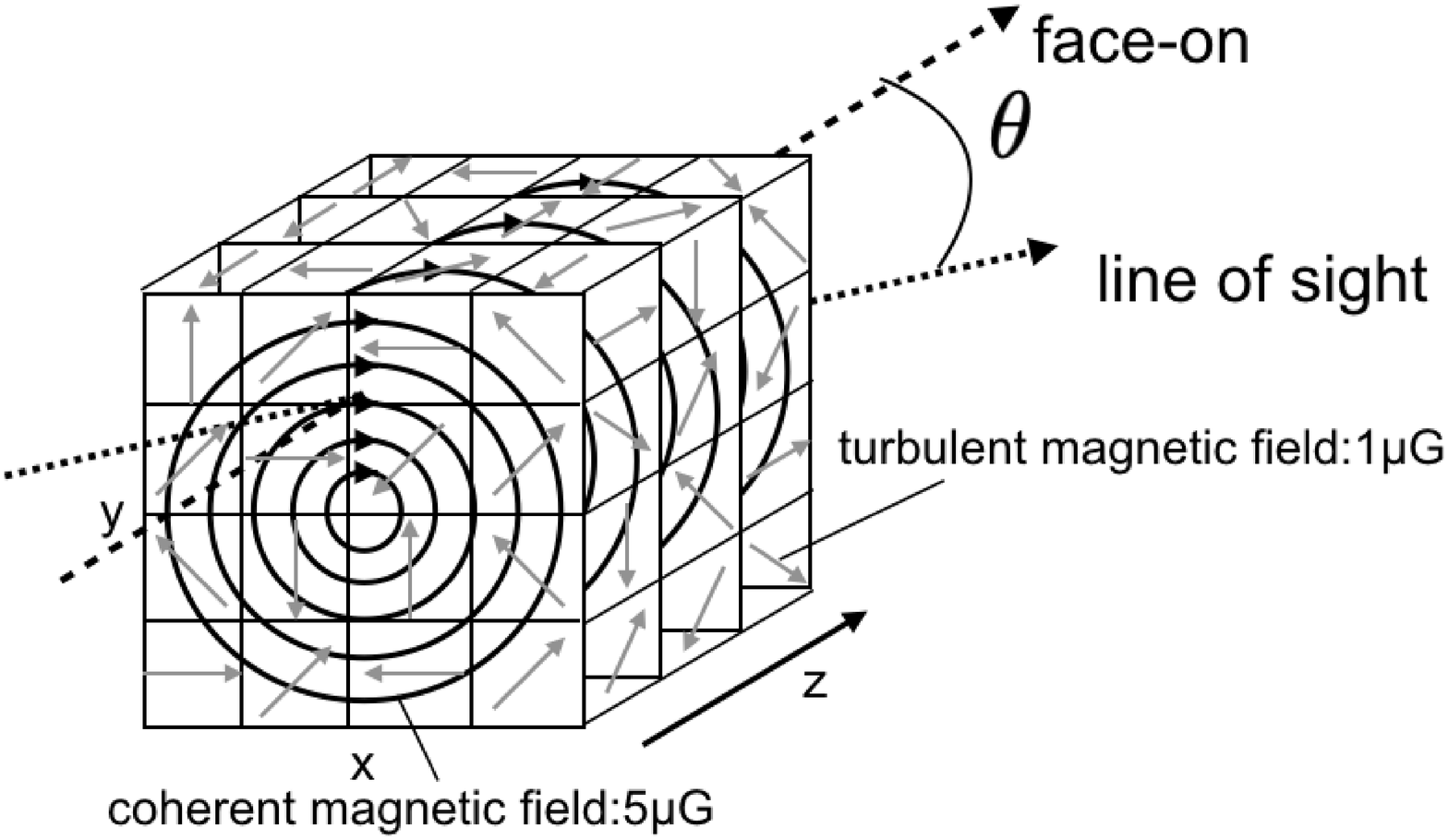} 
\end{center}
\begin{center}
\includegraphics[width=7cm]{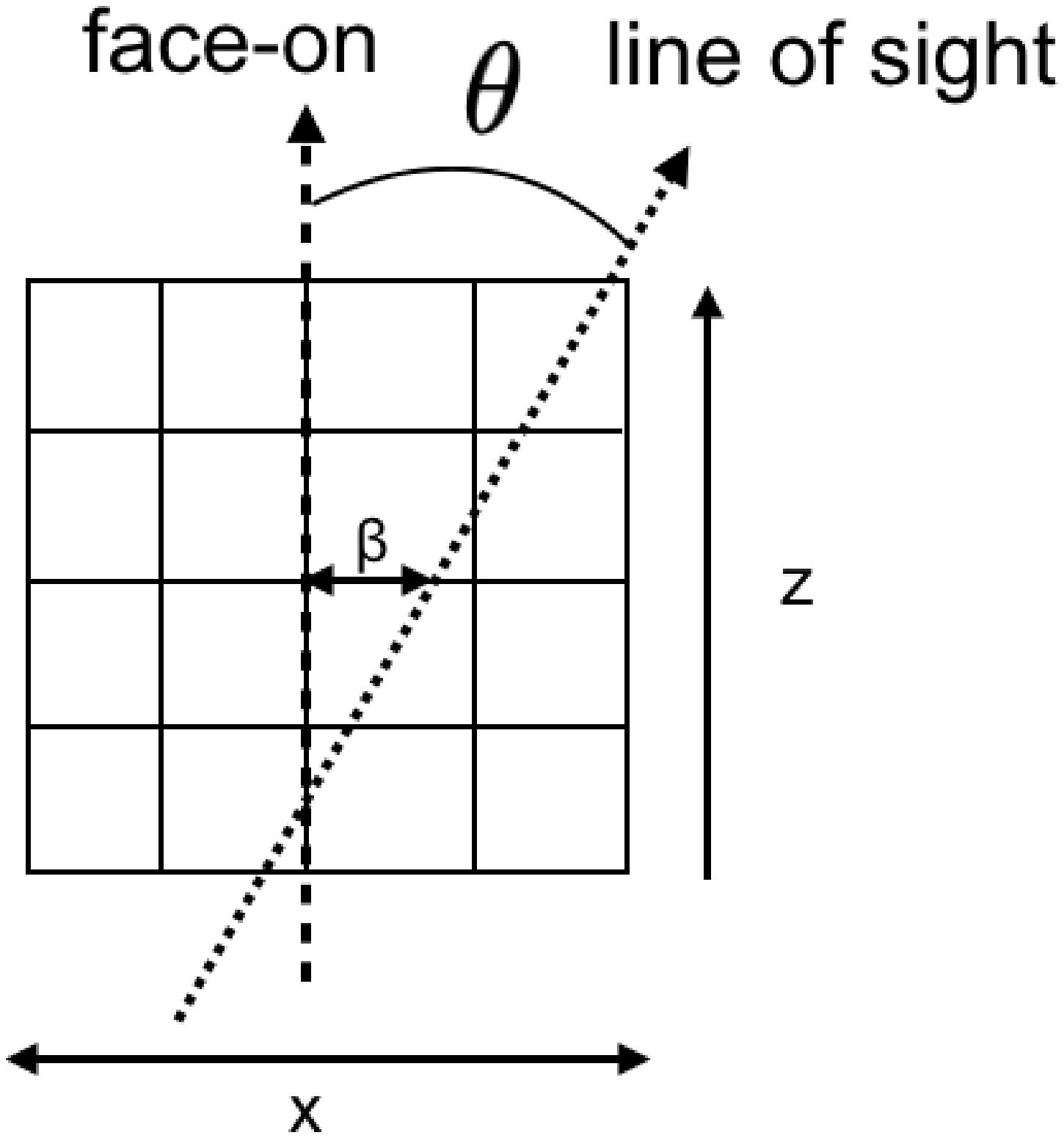} 
\end{center}
\caption{Schematic view of the galactic model by \citet{2020ApJ...899..122E}. Magnetic fields consist of global axisymmetric field and turbulent field similar to the cell model in section \ref{subsection:turbulent}. The $(x, y)$ plane is set along the galactic plane and the $z$ axis is perpendicular to it. The line of sight is assumed to be parallel to the $y$ axis and the inclination is denoted as $\theta$. The distance between the galactic center and the intersection of the line of sight and $z=0$ plane is denoted as $\beta~[{\rm pc}]$. \copyright AAS. Reproduced with permission.}
\label{fig:Eguchi-2020_model}
\end{figure}

\citet{2020ApJ...899..122E} introduced global axisymmetric magnetic fields to the simple cell model of the previous section in order to discuss the FDF of spiral galaxies. Fig.~\ref{fig:Eguchi-2020_model} is the schematic view of their galactic model. Global magnetic field is given to each cell depending on the position in the galaxy and turbulent field is randomly distributed. The thermal-electron density and cosmic-ray electron density are assumed to be uniform. Therefore, it is physically equivalent to the cell model except that an axisymmetric magnetic field along the galactic plane is considered. However, \citet{2020ApJ...899..122E} considered this galactic model to be observed from various directions, and as we will see later, the shape of the FDF changes significantly depending on the inclination angle because the relative relationship between the direction of the global magnetic field and the line-of-sight direction changes. Further, they took an effect of a finite beam size into account and the shape of the FDF also depends on the position in the galaxy.

To calculate the FDF, the $(x, y)$ plane is set along the galactic plane and the $z$ axis is perpendicular to it as in Fig.~\ref{fig:Eguchi-2020_model}. The line of sight is assumed to be parallel to the $y$ axis and the inclination is denoted as $\theta$. The distance between the galactic center and the intersection of the line of sight and $z=0$ plane is denoted as $\beta~[{\rm pc}]$. A cell is a cube with a size of $(10~{\rm pc})^3$ and the thickness of the galaxy is set to $1~{\rm kpc}$. The fiducial model has the thermal-electron density of $n_e = 0.02~{\rm cm^{-3}}$, the strength of global magnetic field of $5~{\rm \mu G}$ and the standard deviation of turbulent fields of $1~{\rm \mu G}$. In fact  thermal electrons and cosmic-ray electrons are not uniform and the density is often considered to decrease exponentially in the direction perpendicular to the galactic plane with the scale height of $1~{\rm kpc}$. Thus, a uniform galactic plane with a thickness of $1~{\rm kpc}$ can be regarded as a zeroth-order approximation. Another important model parameter is a pitch angle, which is the angle between the tangential direction of a circle centered on the galactic center and the global magnetic field line. The global field is ring-like when the pitch angle is zero, which is adopted as a fiducial value.

\begin{figure}[t]
\begin{center}
\includegraphics[width=7cm]{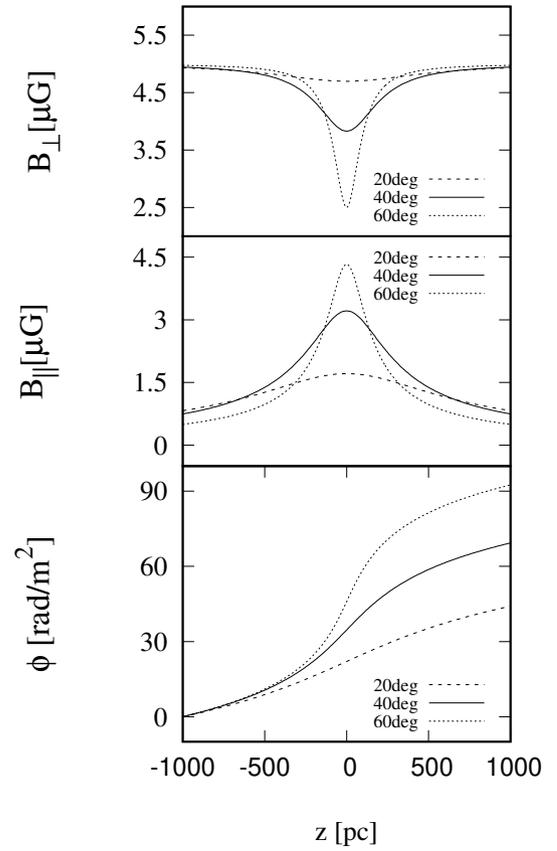}
\end{center}
\caption{Line-of-sight profile of the perpendicular and parallel components of the global field and Faraday depth taking only the global field into account from top to bottom \citep{2020ApJ...899..122E}. The position of the line of sight is set to $y = 200~{\rm pc}$ and $\beta = 0~{\rm pc}$ and three cases with different inclination, $\theta = 20, 40$ and $60~{\rm deg}$, are compared. \copyright AAS. Reproduced with permission.}
\label{fig:Eguchi-2020_profile}
\end{figure}

Fig.~\ref{fig:Eguchi-2020_profile} shows the line-of-sight profile of the perpendicular and parallel components of the global field, $B_{\perp}$ and $B_{||}$, respectively,  and Faraday depth taking only the global field into account. Here, the position of the line of sight is set to $y = 200~{\rm pc}$ and $\beta = 0~{\rm pc}$ and three cases with different inclination, $\theta = 20, 40$ and $60~{\rm deg}$, are compared. First, it should be noted that in the case of face-on observation ($\theta = 0$) the global field is perpendicular to the line of sight and does not contribute to Faraday depth. A larger inclination leads to a stronger $B_{||}$ and Faraday depth evolves more rapidly with $z$. Because the sign of $B_{||}$ doesn't change, Faraday depth monotonically increases. For a fixed inclination, $B_{||}$ is stronger around the plane of $z=0$ and weaker away from it so that the increase rate of Faraday depth changes accordingly. Contrastingly, $B_\perp$ is weaker around $z=0$ so that the polarization intensity is smaller there compared to other region.

\begin{figure}[t]
\begin{center}
\includegraphics[width=7cm]{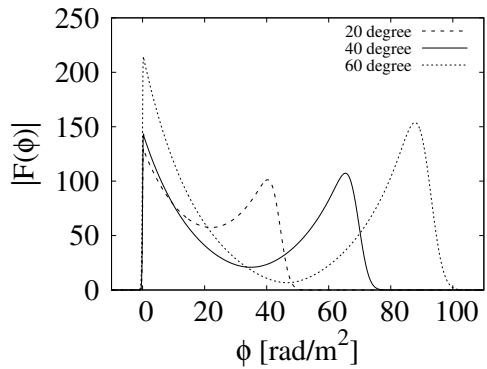} 
\end{center}
\caption{Absolute value of the FDFs corresponding to the three cases in Fig.~\ref{fig:Eguchi-2020_profile}. \copyright AAS. Reproduced with permission.}
\label{fig:Eguchi-2020_FDF}
\end{figure}

Fig.~\ref{fig:Eguchi-2020_FDF} shows the absolute value of the corresponding FDFs. It can be seen that for a larger inclination angle, the FDF has a non-zero value over a wider range of Faraday depth. There is a characteristic dip at the center of each FDF. Because Faraday depth is monotonically increasing along the line of sight, the center in Faraday depth space corresponds to the center in physical space ($z \sim 0$). The existence of the dip is attributed to the facts that the polarization intensity is relatively weak around $z \sim 0$ as we saw in Fig.~\ref{fig:Eguchi-2020_profile} and that the increase rate of $B_{||}$ is larger there. Thus, for the same galactic model with the same values of parameters, the FDFs are apparently different depending on the inclination.

In \citet{2020ApJ...899..122E}, taking the above parameters as fiducial values, they calculated the FDFs varying the position of the line of sight, the pitch angle and the relative strength of global and turbulent fields. It was found that the shape of the FDF changes variously depending on these parameters and that the FDF contains much information about galaxies. On the other hand, the diversity that exists even in such a simple model will make it difficult to physically interpret the FDFs. If some of the parameters such as the inclination angle could be measure by optical observation, it will be helpful for the physical interpretation.

\begin{figure}
\begin{center}
\includegraphics[width=8cm]{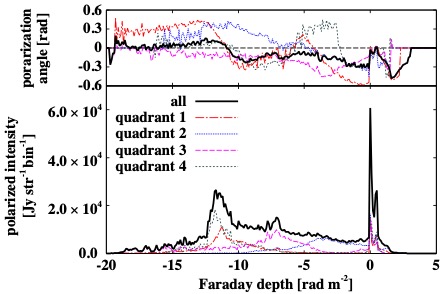} 
\end{center}
\caption{Phase (top) and absolute value (bottom) of the FDF calculated in \citet{2014ApJ...792...51I}. The FDFs of the whole region ($500~{\rm pc} \times 500~{\rm pc}$, black solid line) and 4 sub-regions of $250~{\rm pc} \times 250~{\rm pc}$ are plotted. \copyright AAS. Reproduced with permission.}
\label{fig:Ideguchi-2014}
\end{figure}

\citet{2014ApJ...792...51I} calculated the FDF of a realistic galaxy model constructed by \citet{2013ApJ...767..150A} with multi-wavelength observation data. The galaxy model consists of the following ingredients.
\begin{itemize}
\item thermal-electron density \citep{2002astro.ph..7156C}: a model of our Galaxy called NE2001 based on dispersion measure of pulsars
\item coherent magnetic field along the galactic plane \citep{2008A&A...477..573S}: axisymmetric and bisymmetric fields obtained from all-sky observations of radio intensity, polarization intensity and RMs
\item toroidal magnetic field in halo \citep{2010RAA....10.1287S}: large-scale patterns of all-sky Faraday rotation map suggest the existence of toroidal magnetic fields in opposite directions north and south of the galactic plane
\item poloidal magnetic field in the halo \citep{2010JCAP...08..036G}: suggested from the arrival direction distribution of ultra-high energy cosmic rays
\item turbulent fields \citep{1998ApJ...506L.139K}: estimation from MHD simulations of turbulence in interstellar medium
\item cosmic-ray electron density \citep{2008A&A...477..573S}: exponentially falls in the direction perpendicular to the galactic plane and in the radial direction, and the scale heights are $1~{\rm kpc}$ and $8.5~{\rm kpc}$, respectively
\end{itemize}
Incorporating these elements, \citet{2014ApJ...792...51I}  constructed a Galactic model of $500~{\rm pc} \times 500~{\rm pc}$ along the galactic plane near the Sun and a region of $-10~{\rm kpc} < z < 10~{\rm kpc}$ perpendicular to it. Then, the FDF was calculated assuming this region ($500~{\rm pc} \times 500~{\rm pc} \times 20~{\rm kpc}$) was placed outside our Galaxy and observed from the direction perpendicular to the plane. The angular size of a region of $500~{\rm pc} \times 500~{\rm pc}$ is $10'' \times 10''$ if it is located at $10~{\rm Mpc}$. Fig.~\ref{fig:Ideguchi-2014} is the phase and absolute value of the calculated FDF. The FDFs of the whole region and 4 sub-regions of $250~{\rm pc} \times 250~{\rm pc}$ are plotted. The FDF extends in the range of $-15~{\rm rad/m^2} \lesssim \phi \lesssim 2~{\rm rad/m^2}$ and has narrow peaks at $\phi \sim 0~{\rm rad/m^2}$ and a wide peak at $\phi \sim -12~{\rm rad/m^2}$. In addition, there are small fluctuations in the phase and absolute value of the FDF, which suggests that the effect of turbulence is not sufficiently smoothed out. In fact, it was shown that the FDF obtained form a different realization of turbulent magnetic fields have a significantly different shape, and it would be difficult to interpret such FDFs obtained by observations of an emitting region of this size. Nevertheless, it was pointed out that information such as the scale height of the density of thermal electrons and cosmic-ray electrons can be obtained from the characteristic quantities such as the width, skewness and kurtosis of the absolute value of the FDF. It is desirable for the FDF using a model of the entire galaxy to be calculated.

As described above, there are two approaches to understanding the FDF of galaxies: one is to incorporate various elements into a simple model, and the other is to construct a realistic model based on observational data. The former is relatively easy to understand qualitatively because the effect of each element on the FDF can be seen, but it is difficult to make a precise model to make a quantitative comparison with observational data. On the other hand, the latter can be directly compared with observational data, but it is difficult to make physical interpretations and estimate parameters due to the complexity of the model. Since the FDF of galaxies involves many factors and physical processes, it is necessary to deepen the qualitative and quantitative understanding through both approaches and compare it with observational data.


\subsection{Intergalactic magnetic field}
\label{subsection:IGMF}

\begin{figure}
\begin{center}
\includegraphics[width=7cm]{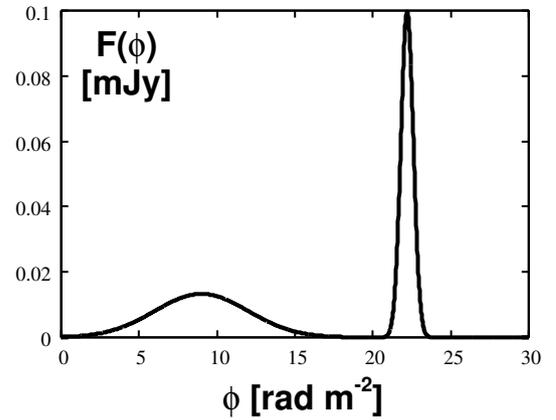} 
\end{center}
\caption{Schematic view of the FDF which reflects the intergalactic magnetic field appearing as a gap between two polarization sources \citep{2014PASJ...66....5I}.}
\label{fig:Ideguchi-2014_IGMF}
\end{figure}

While the FDF contains information on the magnetic field of polarization sources and the line-of-sight distribution of polarized radiation as we saw above, it has been proposed that it is also useful for exploring the intergalactic magnetic field (IGMF) \citep{2010ApJ...723..476A,2011ApJ...738..134A}. As an example, let us consider two polarization sources along the line of sight. Two sources can be our Galaxy and an external galaxy, or 2 external galaxies in the same direction. The intergalactic space is between the two sources, and, because the gas density is very small there, almost no radiation is emitted from the intergalactic matter. Nevertheless, the presence of a magnetic field of some strength can contributes to Faraday rotation. Therefore, the IGMF can create a gap between the two sources in Faraday-depth space. Fig.~\ref{fig:Ideguchi-2014_IGMF} represents this situation \citep{2014PASJ...66....5I}.

To identify a gap in Faraday-depth space, we need to measure precisely the width of two sources in Faraday-depth space. Because the sidelobes seen in the dirty FDF are much more widespread than the sources really are, they are a major obstacle to the gap identification. Further, as we saw in section \ref{subsection:faradayrotation}, the RM contributed from the IGMF is expected to be only $O(1)~{\rm rad/m^2}$, two sources will be very close to each other and high resolution observation in Faraday-depth space is necessary do resolve them. Therefore, study of the IGMF with Faraday tomography requires a wideband observation including long-wavelength bands.

\citet{2014PASJ...66....5I} estimated the measurement accuracy of rotation measure due to the IGMF by Fisher analysis, assuming a combination of observations with LOFAR, GMRT and ASKAP and the data analysis with QU fit method described later. As a result, it was shown that, if the brightness of two sources is more than $0.1~{\rm mJy}$, the IGMF about $3~{\rm rad/m^2}$ is measurable with 1-hour observation of each telescope. Practically, due to the presence of Radio Frequency Interference (RFI), some ranges of the observation bands are not available even if they are covered by the telescope. Based on the situation of the RFI, \citet{2014PASJ...66...65A,2018PASJ...70..115A} discussed which band is suitable for observing the IGMF.

Here, it should be noted that, even if the rotation measure due to the IGMF is larger than expected, it does not necessarily create a gap in Faraday depth space. As discussed in section \ref{subsection:coherent}, in a situation where the magnetic field is inverted inside the emitting region, the edges in real space generally does not correspond to the edges in Faraday depth space. Therefore, depending on the configuration of the magnetic field in the sources or in the intergalactic matter, Faraday rotation in the intergalactic matter may not appear as a gap. Contrastingly, the presence of a gap in Faraday-depth space is a sufficient condition to indicate the existence of the intergalactic magnetic field. In this sense, this method is effective to probe the IGMF.

\section{Algorithms of Faraday tomography}
\label{section:implementation}

In section \ref{section:interpretation}, we learned that the FDF contains a wealth of information on polarization sources with magnetic fields. However, the FDF cannot be directly observable but has to be reconstructed from observation of polarization spectrum. Let us remind the relation between the FDF $F(\phi)$ and polarization spectrum $P(\lambda^2)$:
\begin{equation}
P(\lambda^2) = \int^\infty_{-\infty} F(\phi) e^{2i \phi \lambda^2} d\phi
\label{eq:P2}
\end{equation}
Here, while the integration range is formally $-\infty < \lambda^2 < \infty$, we can practically obtain the polarization spectrum only for a limited range of positive $\lambda^2$ and the FDF cannot be reconstructed perfectly. In other words, there are an infinite number of FDFs which give the same polarization spectrum for a finite range and observed polarization spectrum cannot determine the FDF uniquely. In this situation, the purpose of Faraday tomography is to obtain a physically reasonable FDF by some means. Accurate reconstruction of the FDF not only provides the physical information of the polarization source, but also leads to the detection of the intergalactic magnetic field as seen in section \ref{subsection:IGMF}.

The simplest of Faraday tomography is to perform an inverse Fourier transform with only the obtained polarization data, that is, to obtain a dirty FDF, which is called Rotation Measure synthesis (RM synthesis). As we saw in section \ref{section:principle}, a dirty FDF has artificial features such as the spread in Faraday depth space and sidelobes due to the finite observation band, which makes it difficult to physically interpret the FDF. So far, many algorithms have been proposed to alleviate these problems. In this section, we review the outline of some of major algorithms.

\subsection{RM CLEAN}
\label{subsection:RMCLEAN}

RM CLEAN is a method developed based on the image synthesis algorithm CLEAN \citep{1974A&AS...15..417H} of radio interferometer, and attempts to reconstruct the FDF by regarding it as a collection of delta functions \citep{2007PhDT.......303B,2009A&A...503..409H}. As we saw in section \ref{subsection:RMSF}, when a delta-function-type FDF is observed with a finite frequency range, the dirty FDF has a finite width as shown in Fig.~\ref{fig:RMSF}, which is the RMSF $R(\phi)$. Therefore, RM CLEAN estimates a set of delta functions (CLEAN components) that produces the dirty FDF, assuming that the dirty FDF is a superposition of the RMSF pattern. The specific algorithm is as follows.
\begin{enumerate}
\item find a peak in the dirty FDF $\tilde{F}(\phi)$ and set the peak position as $\phi = \phi_0$
\item add a delta function $\gamma F(\phi_0) \delta(\phi - \phi_0)$ to CLEAN component, where $\gamma$ is a constant gain and is often taken as $\gamma \sim 0.1$
\item subtract $\gamma F(\phi_0) R(\phi-\phi_0)$ from $\tilde{F}(\phi)$
\item find a peak in the residual, $\tilde{F}(\phi) - \gamma F(\phi_0) R(\phi-\phi_0)$ and repeat 2 and 3 above
\item finish the iteration when the residual is below a constant $\epsilon$ at all $\phi$
\item multiply each of CLEAN component by a Gaussian function with the same width of the RMSF, considering the resolution in Faraday-depth space determined by the observation frequency range
\item add the residual to obtain the final result called the cleaned FDF
\end{enumerate}

\begin{figure}
\begin{center}
\includegraphics[width=7cm]{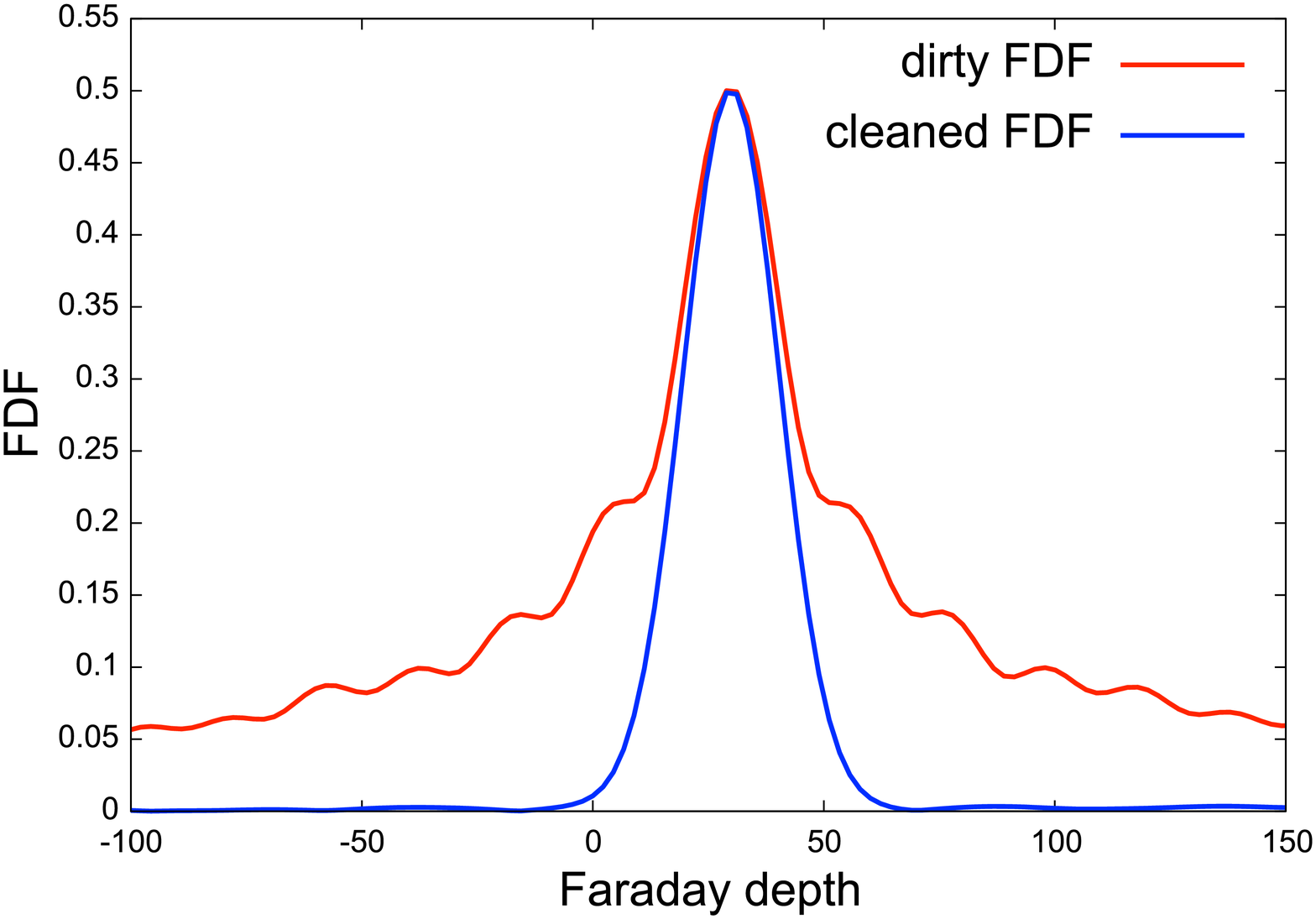} 
\end{center}
\caption{Example of RM CLEAN. A delta-function source is put at $\phi = 30~{\rm rad/m^2}$ and the dirty FDF (red) and cleaned FDF (blue) assuming an observation of a $700 \sim 1800~{\rm MHz}$ band are shown.}
\label{fig:RMCLEAN_delta}
\begin{center}
\includegraphics[width=7cm]{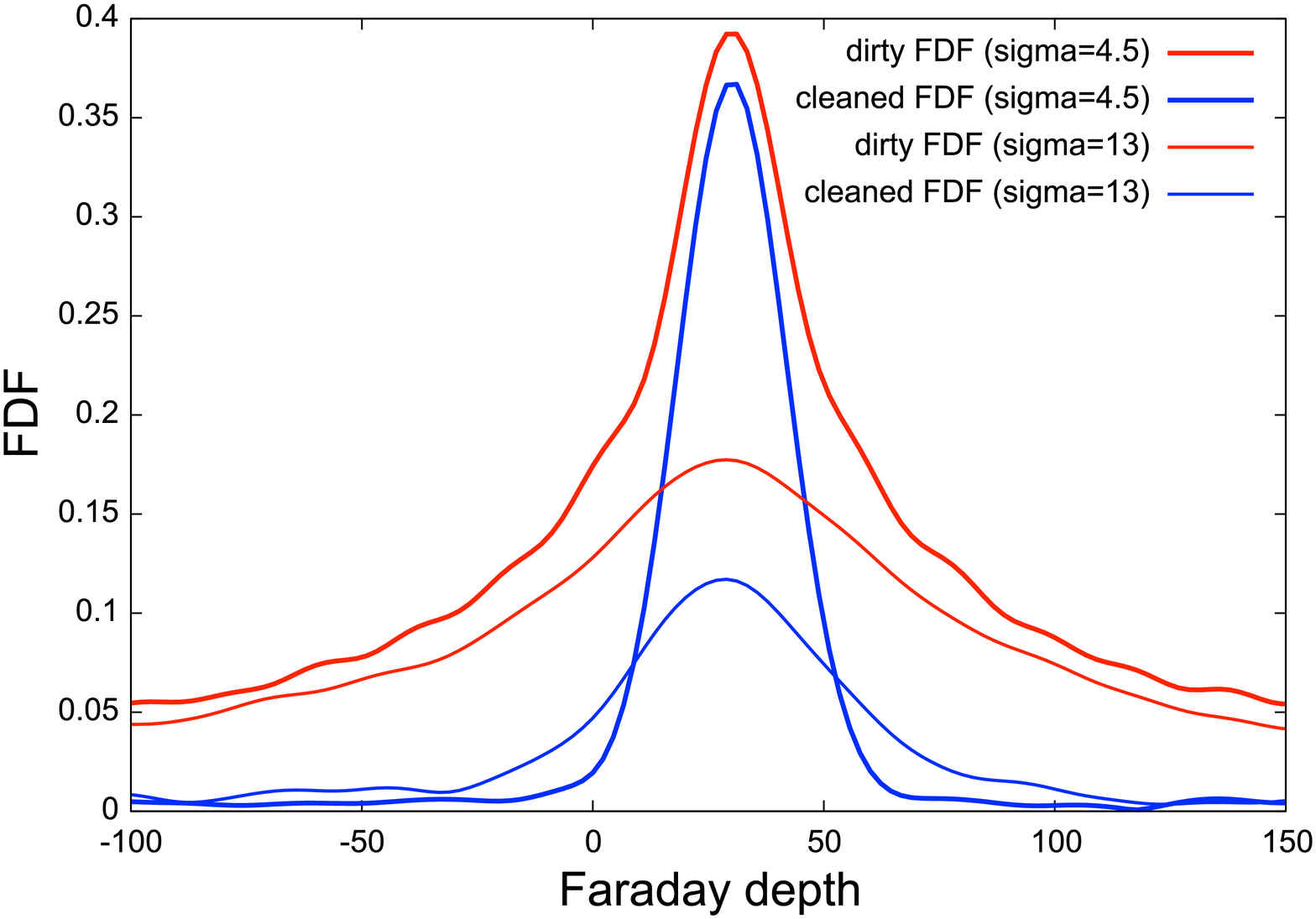} 
\end{center}
\caption{Example oRM CLEAN for a Gaussian source with $\phi = 30~{\rm rad/m^2}$. The dirty FDF (red) and cleaned FDF (blue) are plotted for $\sigma = 4.5~{\rm rad/m^2}$ (thick) and $\sigma = 13~{\rm rad/m^2}$ (thin). An observation with a $700 \sim 1800~{\rm MHz}$ band is assumed.}
\label{fig:RMCLEAN_gauss}
\end{figure}

Fig.~\ref{fig:RMCLEAN_delta} is an example of RM CLEAN. Here, a delta-function source is put at $\phi = 30~{\rm rad/m^2}$ and the dirty FDF and cleaned FDF assuming an observation of a $700 \sim 1800~{\rm MHz}$ band are shown. The dirty FDF is the RMSF itself because there is only one delta function. The cleaned FDF has a Gaussian function with the width of the RMSF (FWHM $\sim 22~{\rm rad / m^2}$) at the correct position and small residuals. Thus, the polarization source is originally a delta function, but it is reproduced with a finite width due to the finite observation band. Nevertheless, RM CLEAN eliminates the sidelobes and outputs an FDF close to the original FDF.

RM CLEAN is widely used in polarization analysis because it produces reasonable results for its low computational cost compared to other methods. However, RM CLEAN may not work well in some cases. That is, for example, when two delta-function FDFs are located close to each other in the $\phi$ space by about the FWHM of the RMSF or closer \citep{2011AJ....141..191F}. As an example of $\Delta \phi = 2$ in the bottle panel of Fig.~\ref{fig:interference_phi}, when two sources are close, they interfere and only one peak appears in the dirty FDF. Then, because RM CLEAN assumes that there is a delta-function source at the peak of the dirty FDF, in this example, a CLEAN component stands at $\phi = 0$ where there is actually no source. This phenomenon is called Rotation Measure ambiguity (RM ambiguity). As we saw in the bottom panel of Fig.~\ref{fig:interference_chi}, the interference of two sources depends on the difference in polarization angle as well.

\citet{2014PASJ...66...61K} performed a series of simulations to identify the condition of RM ambiguity. More specifically, they consider two delta-function sources and investigate the condition of RM ambiguity varying the gap in the $\phi$ space, difference in polarization angle and brightness ratio. As a result, it was discovered that RM ambiguity occurs even for a gap of $1.5 \times$ FWHM depending on the condition. \citet{2016PASJ...68...44M} performed more detailed simulations to find that the estimation of Faraday depth, polarization angle and polarization intensity of two sources is biased by several tens of percent even if RM ambiguity does not happen, when the gap between the two sources is less than $1.2 \times$ FWHM.

Further, it is difficult to reconstruct FDFs with extended structure in the $\phi$ space by RM CLEAN. Fig.~\ref{fig:RMCLEAN_gauss} shows an example of a Gaussian function. It can be seen that a wider Gaussian function leads to a cleaned FDF with a smaller peak value and the reproducibility is poorer. As we saw above, while RM CLEAN is simple and the result of reconstruction is often reasonable, it should be noted that it may return a poor result depending on the shape of the original FDF.

\subsection{QU fit}
\label{subsection:QUfit}

QU fit is another common method of Faraday tomography, which assumes a model (functional form) of the FDF and determine the model parameters by fitting the polarization spectrum calculated from it to the observed data. Some of the commonly used models are following:
\begin{itemize}
\item delta function (parameters: $\bm{\theta} = \{ f_0, \chi_0, \phi_0 \}$)
	\begin{equation}
	F(\phi) = f_0 e^{2 i \chi_0} \delta(\phi - \phi_0)
	\end{equation}
\item Gaussian function (parameters: $\bm{\theta} = \{ f_0, \chi_0, \phi_0, \sigma \}$)
	\begin{equation}
	F(\phi) = \frac{f_0}{\sqrt{2 \pi} \sigma} e^{-\frac{(\phi - \phi_0)^2}{2 \sigma^2} + 2 i \chi_0}
	\end{equation}
\item top-hat function (parameters: $\bm{\theta} = \{ f_0, \chi_0, \phi_0, \phi_1 \}$, where $\phi_1 > \phi_0$)
	\begin{equation}
	F(\phi) = f_0 e^{2 i \chi_0} \left[ \Theta(\phi - \phi_0) - \Theta(\phi - \phi_1) \right]
	\end{equation}
\end{itemize}
In practice, because there may be multiple sources along the line of sight or the FDF can have a complicated structure, a model with multiple of these functions is also used. The following chi-square is often used as an index to evaluate the fit between the observed data and the model.
\begin{equation}
\chi^2(\bm{\theta}) = \sum_j \frac{(P_{\rm model}(\lambda_j^2,\bm{\theta}) - P_{\rm obs}(\lambda_j^2))^2}{\sigma_j^2}
\end{equation}
Here $j$ represents the index of wavelength channel, $\sigma_j$ is an error of $P_{\rm obs}(\lambda_j^2)$ and $\bm{\theta}$ is a vector of model parameters. By finding the parameters that minimize this chi-square value, we can find an FDF that explains the observed data well.

In RM Synthesis, the inverse Fourier transform is performed on the polarization spectrum in the observed band, putting artificially $P(\lambda^2)=0$ in the unobserved band including negative $\lambda^2$. The same is true for RM CLEAN because it is based on dirty FDF, and such an operation causes the delta function source to spread in the $\phi$ space. On the other hand, QU fit does not require any explicit assumption on the unobserved bands, limiting the possibility for artifacts in the FDF.

In QU fit, the shape of FDFs is assumed in advance, but it is not obvious whether the true FDF can be accurately expressed by the assumed functional form. In fact, as we saw in section \ref{section:interpretation}, FDFs are not expressed in a simple functional form even for a simple galactic model. Therefore, it is necessary to try several models with different functional forms and number of sources and to select the model that best approximates the true FDF. The chi-square mentioned earlier is a common index of the goodness of data fitting, but it cannot simply be considered that the model with a smaller value of chi-square is better. In general, a model with more parameters is easier to fit to the data, but a complex model with too many parameters is difficult to interpret physically and is not a good model. Further, due to the presence of observational errors, even if the correct model is used, the data cannot be perfectly fitted, and an unnecessarily complex model will overfit the data. Therefore, the balance between the goodness of fit to the data and the simplicity of the model is important, and the information criterion is often used as an index for model selection. There are various information criteria, and the simple information criteria that are often used are as follows.
\begin{itemize}
\item AIC (Akaike information criterion)
\begin{equation}
{\rm AIC} = -2 \ln{(L)} + 2 k
\end{equation}
\item BIC, Bayesian information criterion)
\begin{equation}
{\rm BIC} = -2 \ln{(L)} + k \ln{(n)}
\end{equation}
\end{itemize}
Here, $L = e^{-\chi^2/2}$ is likelihood function, $n$ is the number of observation data and $k$ is the number of parameters. In the expression of these two information criteria, the first term represents the goodness of fit to the data and becomes smaller as it fits better. The second term is a penalty for complex models, which increases with more parameters. By selecting a model that minimizes the information criterion, the fit to the data and the simplicity of the model can be balanced. Comparing AIC and BIC, BIC gives a stronger penalty for the number of parameters unless the number of data is extremely small, so a simpler model tends to be selected using BIC. There are other types of information criteria, and the applicability varies depending on the size of the data and the nature of the error.

In estimating parameters and selecting models, it is necessary to search the parameter space to find the place where the chi-square is minimized. However, this is not easy for the following two reasons. One is that the number of parameters is generally large. As there can be multiple sources within an observation field or one source can have a complex FDF as seen in section \ref{section:interpretation}, a model with multiple delta functions and Gaussian functions is often used to fit data. For example, if a model with three Gaussian functions is adopted, the number of parameters is 12 and it is not practical to perform a grid search in the 12-dimensional parameter space considering the calculation time. Therefore, various methods for efficiently searching the parameter space have been developed such as the Markov Chain Monte Carlo (MCMC) and the nested sampling.

\begin{figure}[t]
\begin{center}
\includegraphics[width=7cm]{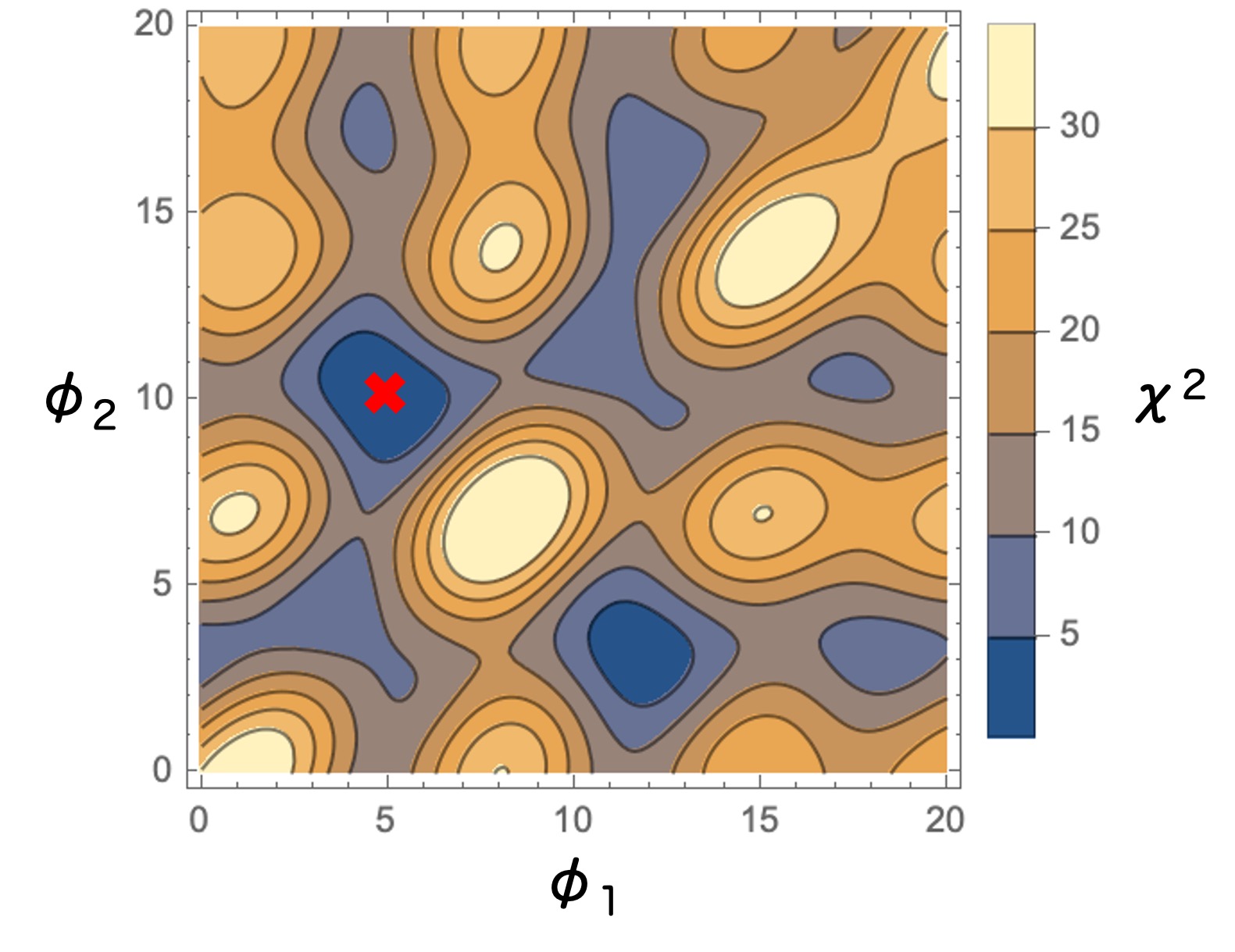} 
\end{center}
\begin{center}
\includegraphics[width=7cm]{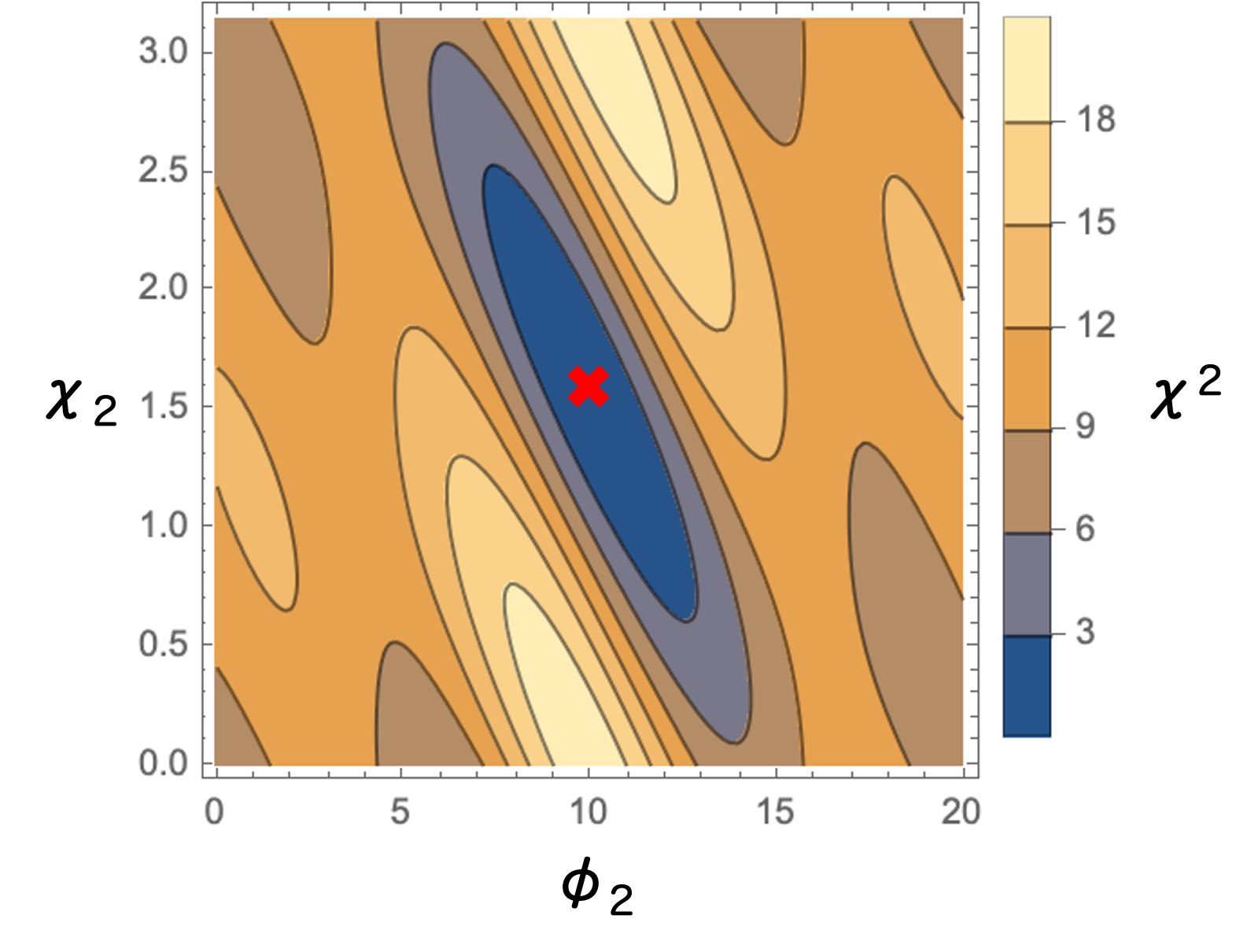} 
\end{center}
\begin{center}
\includegraphics[width=7cm]{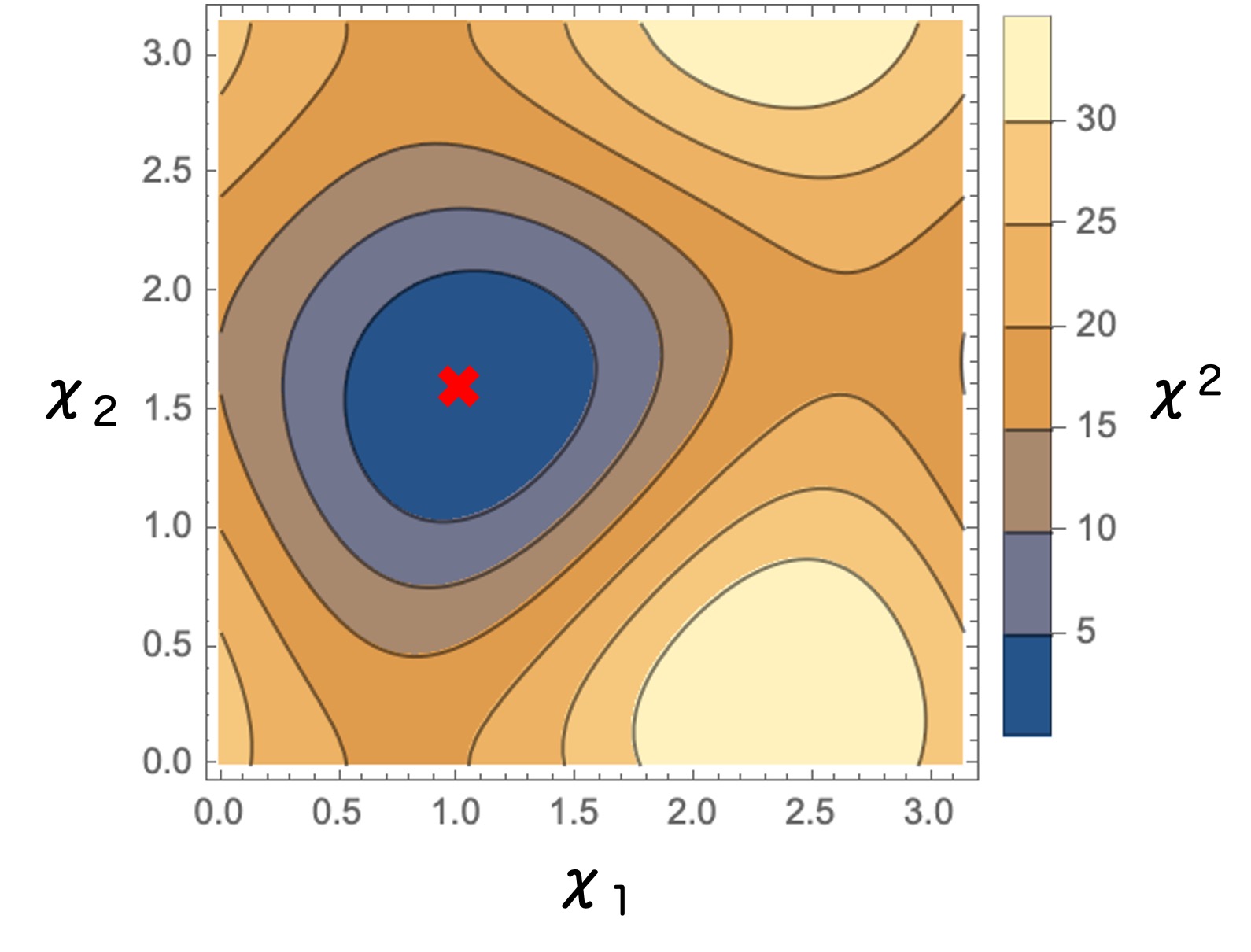} 
\end{center}
\caption{Contour of chi-square in $(\phi_1, \phi_2)$, $(\phi_2, \chi_2)$ and $(\chi_1, \chi_2)$ planes, assuming an observation of two delta-function sources with a band of $0.1~{\rm m^2} \leq \lambda^2 \leq 0.5~{\rm m^2}$. The two sources have the same polarization intensity and other parameters are set as $\phi_1 = 5~{\rm rad/m^2}, \phi_2 = 10~{\rm rad/m^2}, \chi_1 = \pi/3~{\rm rad}$ and  $\chi_2 = \pi/2~{\rm rad}$. Observation errors are not taken into account so that $\chi^2 = 0$ at the correct parameter set shown as a cross.}
\label{fig:chi2}
\end{figure}

Another difficulty lies in the chi-square structure in the parameter space. In general, the further away from the correct parameter set in the parameter space, the larger the chi-square value becomes. However, depending on the nature of the problem, the chi-square can take a local minimum at a parameter set away from the correct one. If there are many such local minima, it becomes difficult to find the global minimum by the gradient method or MCMC. In the case of Faraday tomography, periodic functions are often involved due to the nature of Fourier transform and many local minima of chi-square are considered to exist in the parameter space.

As an example to show this, let us consider an FDF which consists of two delta functions. They are assumed to have the same polarization intensity and other parameters are set as $\phi_1 = 5~{\rm rad/m^2}, \phi_2 = 10~{\rm rad/m^2}, \chi_1 = \pi/3~{\rm rad}$ and  $\chi_2 = \pi/2~{\rm rad}$. Fig.~\ref{fig:chi2} shows the contour of chi-square in $(\phi_1, \phi_2)$, $(\phi_2, \chi_2)$ and $(\chi_1, \chi_2)$ planes, assuming an observation band of $0.1~{\rm m^2} \leq \lambda^2 \leq 0.5~{\rm m^2}$. Here, parameters other than the two are set to the correct values. Observation errors are not taken into account so that $\chi^2 = 0$ at the correct parameter set shown as a cross. While the chi-square takes the global minimum here, several local minima can be seen in $(\phi_1, \phi_2)$ plane.

\citet{2019MNRAS.482.2739M} investigated the performance of QU fit by the MCMC method in model selection and parameter estimation, by simulating observation of two Gaussian-function sources. It was found that, compared to RM CLEAN, QU fit can resolve two sources in many cases even if the dirty FDF has only one peak. While QU fit can resolve two sources more often for a larger separation in the $\phi$ space, due to the interference in the $\phi$ space of the two sources, it may not work even if the separation is larger than the FWHM of the RMSF.

\begin{figure}[t]
\begin{center}
\includegraphics[width=6cm]{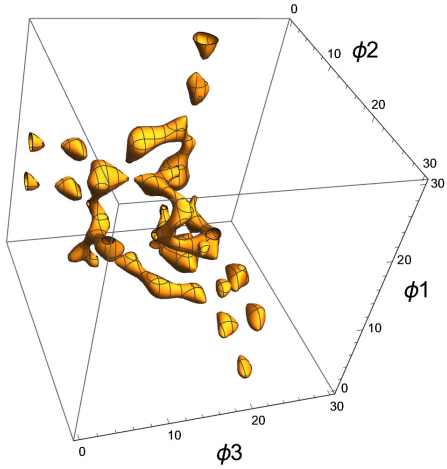} 
\end{center}
\caption{Isosurface of chi-square in $(\phi_1, \phi_2, \phi_3)$ space in the case with three delta-function sources assuming an observation band of $0.1~{\rm m^2} \leq \lambda^2 \leq 0.5~{\rm m^2}$. The three sources have the same polarization intensity and other parameters are set as $\phi_1 = 5~{\rm rad/m^2}, \chi_1 = \pi/3~{\rm rad}, \phi_2 = 10~{\rm rad/m^2}, \chi_2 = \pi/2~{\rm rad}, \phi_3 = 18~{\rm rad/m^2}$ and $\chi_3 = \pi/6~{\rm rad}$. Parameters other than $(\phi_1, \phi_2, \phi_3)$ are set to the correct values to draw the isosurface.}
\label{fig:chi2_3D}
\end{figure}

The structure of chi-square in the parameter space becomes more complicated as the number of parameters increases. Fig.~\ref{fig:chi2_3D} shows an isosurface of chi-square in $(\phi_1, \phi_2, \phi_3)$ space in the case with three delta-function sources. The polarization intensity is assumed to be the same for the three sources and other parameters are set as $\phi_1 = 5~{\rm rad/m^2}, \chi_1 = \pi/3~{\rm rad}, \phi_2 = 10~{\rm rad/m^2}, \chi_2 = \pi/2~{\rm rad}, \phi_3 = 18~{\rm rad/m^2}$ and $\chi_3 = \pi/6~{\rm rad}$. Parameters other than $(\phi_1, \phi_2, \phi_3)$ are set to the correct values to draw the isosurface. We can see many local minima in $(\phi_1, \phi_2, \phi_3)$ space and finding the global minimum is much harder in this case compared to the previous case with two sources.

In this way, an FDF model with more parameters is expected to have more complicated structure of chi-square in the parameter space. More efficient algorithm of QU fit is necessary to search a huge and complicated parameter space and to find the global minimum. Computational cost is a serious problem, especially in an era where millions of polarization sources can be obtained with wide-field surveys.

\subsection{Sparse modelling}
\label{subsection:sparse}

As we saw before, the fundamental difficulty of Faraday tomography, that is, the procedure to obtain the FDF from observed polarization spectrum, is that polarization spectrum can be measured only in a finite range of $\lambda^2$, while complete Fourier transform need it for $-\infty < \lambda^2 < \infty$. There are infinite number of FDFs which can explain observed polarization spectrum of a finite range. Sparse modelling, or compressive sampling, assumes that the FDF is sparse and search for the sparsest possible solution while reproducing the observation data. It is an idea similar to Occam's razor which selects the simplest model that can explain observation data. In astronomy, sparse modelling has been successful in image synthesis of radio interferometry and application to Faraday tomography was initiated in \citet{2011A&A...531A.126L,2012AJ....143...33A,2018arXiv181110610A} and recently further developed in \citet{2023MNRAS.518.1955C}.

To explain the principle of sparse modelling, we discretize the relation between polarization spectrum and the FDF, Eq.~(\ref{eq:P}).
\begin{equation}
P(\lambda_j^2) = \sum_k e^{2 i \phi_k \lambda_j^2} F(\phi_k)
\end{equation}
This equation can be expressed as,
\begin{eqnarray}
&& P_j = \sum_k M_{jk} F_k \\
&& P_j \equiv P(\lambda_j^2), ~~~ j=1,2,\cdots,J \\
&& F_k \equiv F(\phi_k), ~~~ k=1,2,\cdots,K \\
&& M_{jk} \equiv e^{2 i \phi_k \lambda_j^2},
\end{eqnarray}
where $J$ is the number of data, $K$ is the number of grid in the $\phi$ space and we generally have $J < K$. We simplify the expression further as,
\begin{equation}
{\mathbf P} = {\mathbf M} {\mathbf F}.
\label{eq:discrete}
\end{equation}
Eq.~(\ref{eq:discrete}) is a set of linear equations, but since the number of unknowns $F_k$ ($K$) is larger than the number of equations ($J$), there are an infinite number of solutions that satisfy this. Then, we assume that the true solution has a sparsity. Most simply, a sparsity is that the number of non-zero $F_k$ is the fewest of all solutions. Denoting $L_0$ norm, the number of non-zero component, as $||{\mathbf F}||_0$, we consider the following minimization problem.
\begin{equation}
{\rm min}_{\mathbf F} ||{\mathbf F}||_0 ~~ {\rm subject ~ to} ~~  {\mathbf P} = {\mathbf M} {\mathbf F}
\end{equation}
Here, "${\rm min}_{\mathbf F}$" represents minimization with respect to ${\mathbf F}$. Therefore, this problem is to find a solution with the least value of $L_0$ norm among infinite number of solutions which explain the observation data (${\mathbf P} = {\mathbf M} {\mathbf F}$). This is the basic idea of sparse modelling, but solving this problem actually requires trying all combinations of which $F_k$ should be zero (combinatorial optimization) and is practically impossible to solve when $K$ is large due to the computational cost.

We therefore define a problem that is equivalent to $L_0$-norm minimization and feasible. First, let us define $L_p$ norm with $p \geq 1$ as,
\begin{equation}
||{\mathbf F}||_p \equiv \left( \sum_j |F_j|^p \right)^{1/p}.
\end{equation}
We consider minimization of $L_1$ norm, instead of $L_0$ norm. $L_1$ norm is the sum of absolute values of all $F_j$, and a solution with minimum $L_1$ norm often has minimum $L_0$ norm. Minimization of $L_1$ norm is convex optimization, rather than combinatorial optimization,  and can be solved with a reasonable computational cost. Therefore, a problem to solve can be written as,
\begin{equation}
{\rm min}_{\mathbf F} ||{\mathbf F}||_1 ~~ {\rm subject ~ to} ~~  {\mathbf P} = {\mathbf M} {\mathbf F}
\end{equation}
In a practical situation, due to observational errors, ${\mathbf P} = {\mathbf M} {\mathbf F}$ should not be satisfied exactly and a solution with a small value of $||{\mathbf P} - {\mathbf M} {\mathbf F}||_2^2$ can be considered to explain observation data well. Denoting a typical observation error as $\sigma$, the minimization problem which takes observation errors into account is written as,
\begin{equation}
{\rm min}_{\mathbf F} ||{\mathbf F}||_1 ~~ {\rm subject ~ to} ~~  ||{\mathbf P} - {\mathbf M} {\mathbf F}||_2^2 < \sqrt{J} \sigma.
\label{eq:L1}
\end{equation}
In fact, since it is necessary to make both $L_1$ norm and the deviation from the observed data as small as possible, the problem of Eq.~(\ref{eq:L1}) is equivalent to the following problem.
\begin{equation}
{\rm min}_{\mathbf F} \left( ||{\mathbf P} - {\mathbf M} {\mathbf F}||_2^2 + \Lambda ||{\mathbf F}||_1 \right)
\label{eq:LASSO}
\end{equation}
Here, the first term is the chi-square that represents the goodness of fit of the observed data with the model, and the second term is the $L_1$ norm. $\Lambda$ is a hyperparameter that determines the balance between data fit and sparsity, and needs to be given in advance. Such a problem setting is called LASSO (the least absolute shrinkage and selection operator).

Two methods to extend Eq.~(\ref{eq:LASSO}) have been studied to improve the performance. One is to change the basis. So far, sparsity has been considered to be that $F$ is non-zero with as few $\phi_k$ as possible, but this is not the case with sources spread in the Faraday depth space. Denoting a new basis of FDFs as $\psi_\ell(\phi)$, we can write as,
\begin{equation}
F(\phi) = \sum_\ell \xi_\ell \psi_\ell(\phi),
\end{equation}
where $\xi_\ell$ are coefficients. It should be noted that the basis of the original FDF is delta functions as $\psi_\ell(\phi) = \delta(\phi - \phi_\ell)$ and $\xi_\ell = F(\phi_\ell)$. Other bases often used are Gaussian functions, top-hat functions and wavelets. If we take Gaussian functions as the basis, the sparsity means that $F(\phi)$ represented by the sum of as few Gaussian functions as possible.

In order to rewrite the LASSO with a new basis, we write the new basis with delta functions.
\begin{equation}
\psi_\ell(\phi) = \sum_k V_{k \ell} \delta(\phi - \phi_k)
\end{equation}
Then, the FDF can be written as,
\begin{equation}
F(\phi) = \sum_\ell \xi_\ell \psi_\ell(\phi) = \sum_{k, \ell} \xi_\ell V_{k \ell} \delta(\phi - \phi_k),
\end{equation}
and we have,
\begin{equation}
\xi_\ell = \sum_k V_{k \ell}^{-1} F_k \equiv \sum_k W_{k \ell} F_k,
\end{equation}
where $W_{k \ell}$ is the inverse matrix of $V_{k \ell}$. Therefore, the LASSO in the new basis can be written as,
\begin{equation}
{\rm min}_{\mathbf F} \left( ||{\mathbf P} - {\mathbf M} {\mathbf F}||_2^2 + \Lambda ||{\mathbf W}{\mathbf F}||_1 \right)
\end{equation}
In order to choose good basis, we need to consider in what sense the true FDF is sparse. \citet{2011A&A...531A.126L,2012AJ....143...33A} used wavelet basis to reconstruct Faraday-thin and Faraday-thick sources through a series of simulations.

\begin{figure}[t]
\begin{center}
\includegraphics[width=8cm]{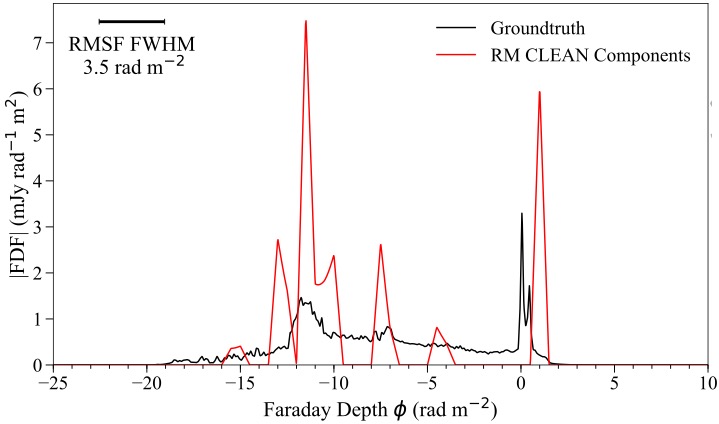} 
\end{center}
\begin{center}
\includegraphics[width=8cm]{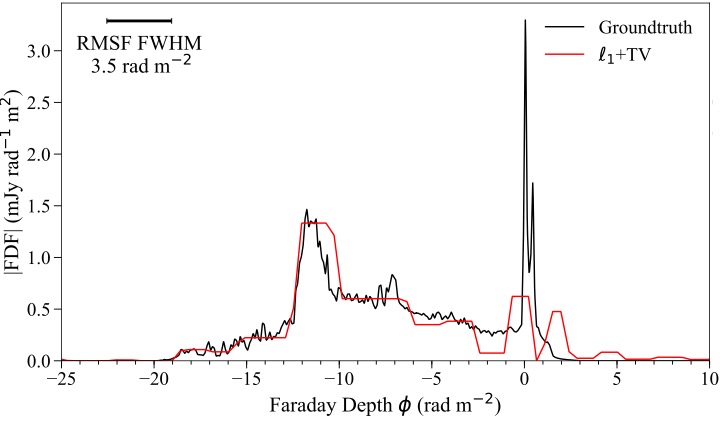} 
\end{center}
\begin{center}
\includegraphics[width=8cm]{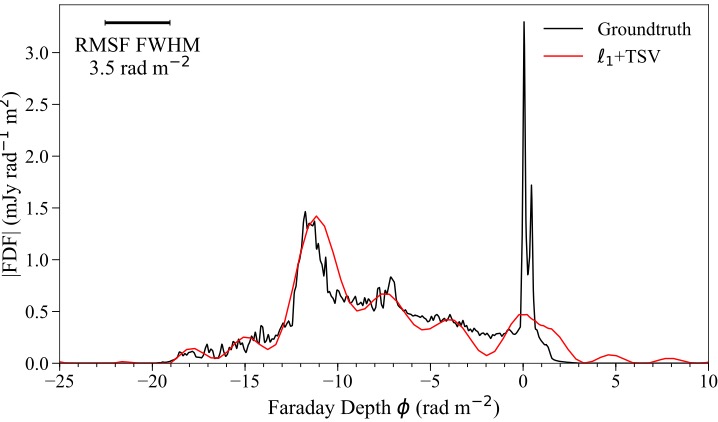} 
\end{center}
\caption{Simulations of reconstruction of an FDF through sparse modelling \citep{2018arXiv181110610A}. A model of galaxy developed in \citet{2014ApJ...792...51I} was used as the ground truth and an observation band of $300~{\rm MHz} - 3~{\rm GHz}$ was assumed the results of RM CLEAN (top), $L_1 + {\rm TV}$ and $L_1 + {\rm TSV}$ are compared.}
\label{fig:Akiyama}
\end{figure}

Sparse modelling with delta-function basis tends to be too sparse and not to give a reasonable reconstruction of FDFs. Then, \citet{2018arXiv181110610A} adopted constraints on total variation (TV) and total squared variation (TSV) to obtain sparse and smooth FDFs. This is the second extension of the LASSO.
\begin{itemize}
\item total variation (TV)
\begin{equation}
|| {\mathbf F} ||_{\rm TV} = \sum_k | F_{k+1} - F_k |
\end{equation}
\item total squared variation (TSV)
\begin{equation}
|| {\mathbf F} ||_{\rm TSV} = \sum_k | F_{k+1} - F_k |^2
\end{equation}
\end{itemize}
These are constraints on the difference between adjacent $F_k$. If we adopt the constraint on total variation, the optimization problem can be written as follows.
\begin{equation}
{\rm min}_{\mathbf F} \left( ||{\mathbf P} - {\mathbf M} {\mathbf F}||_2^2 + \Lambda_\ell ||{\mathbf F}||_1 + \Lambda_t || {\mathbf F} ||_{\rm TV} \right)
\end{equation}
Here, $\Lambda_\ell$ and $\Lambda_t$ are hyperparameters to determine the weights of constraints of $L_1$ norm and total variation. Fig.~\ref{fig:Akiyama} represents the comparison of results from RM CLEAN, $L_1 + {\rm TV}$ and $L_1 + {\rm TSV}$, through mock observation data of $300~{\rm MHz} - 3~{\rm GHz}$. For the FDF model, a galactic model of \citep{2014ApJ...792...51I} shown in Fig.~\ref{fig:Ideguchi-2014} was adopted, which is expected to express the complexity of realistic FDFs. The hyperparameters are determined through cross validation and $(\Lambda_\ell, \Lambda_t) = (10, 1)$ and $(\Lambda_\ell, \Lambda_t) = (10, 10^3)$ were chosen for $L_1 + {\rm TV}$ and $L_1 + {\rm TSV}$, respectively. We can see that RM CLEAN picks up major peaks, but hardly reproduces the overall shape. On the other hand, $L_1 + {\rm TV}$ and $L_1 + {\rm TSV}$ do not reproduce small-scale structure, especially the sharp peak at $\phi = 0$, while overall shape is well reproduced. It should be noted that the peak at $\phi \sim -11~{\rm rad/m^2}$ is resolved with a slightly higher resolution compared with the FWHM of the RMSF of this band, which is about $3.5~{\rm rad/m^2}$.

Sparse modelling has been very successful in image synthesis for radio interferometry, but its application to Faraday tomography is still immature and needs further study. In particular, because it is expected that FDFs of some polarization sources are Faraday thick and have a complicated structure, it is necessary to select a basis that fits well with such FDFs. Since the amount of calculation for sparse modelling is generally large compared to other methods such as RM CLEAN, it is also necessary to develop an algorithm that can reduce the computational cost.

\subsection{CRAFT}
\label{subsection:CRAFT}

\citet{2021MNRAS.500.5129C} proposed a new method, called CRAFT (Constraining and Restoring iterative Algorithm for Faraday Tomography), to estimate the FDF from observed polarization spectrum. This is an application of an algorithm \citep{2020PASJ...72...61C} which estimate unobserved signal by imposing reasonable assumptions about the properties of signals in Fourier space.

\begin{figure}[t]
\begin{center}
\includegraphics[width=7cm]{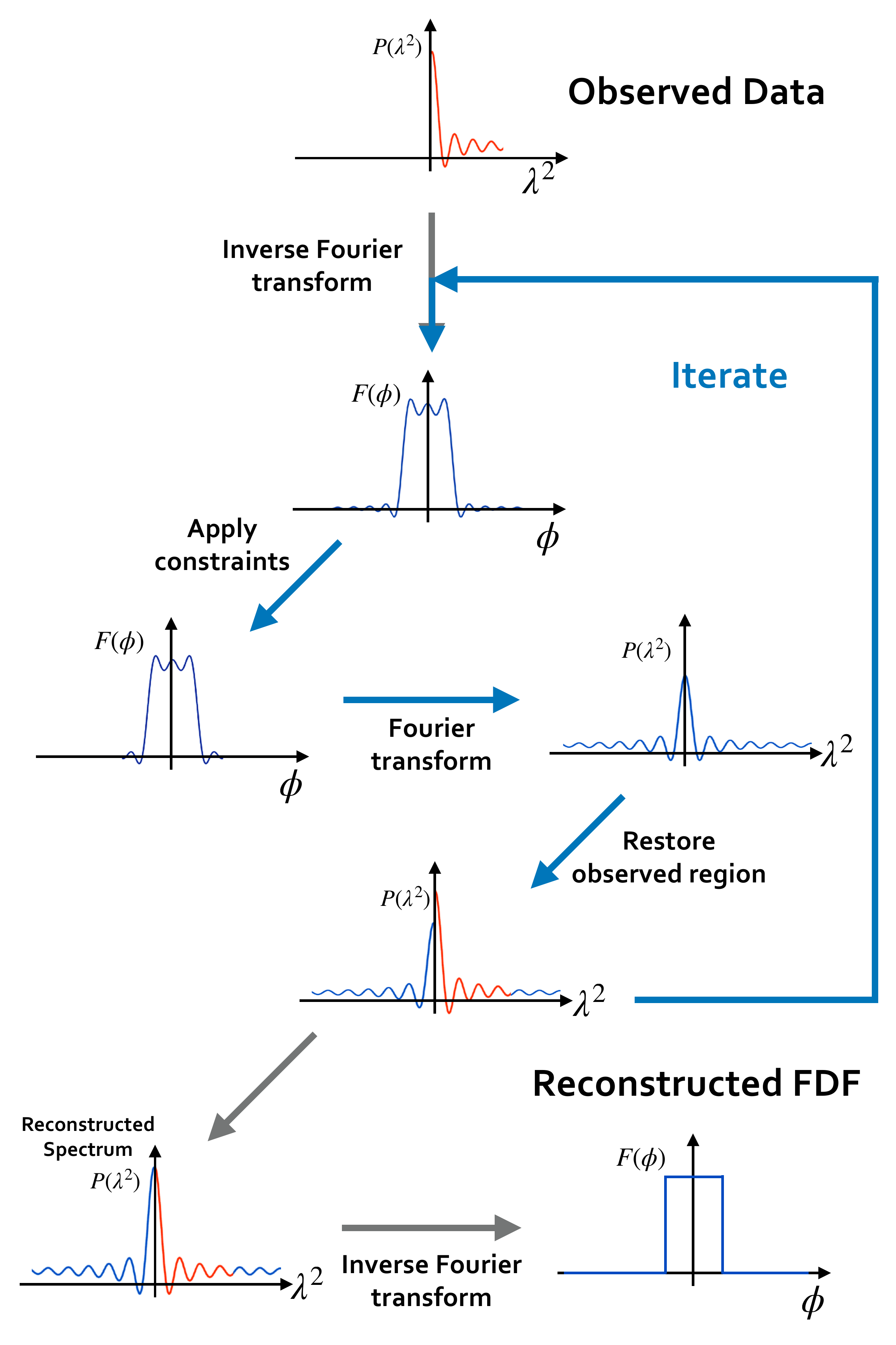} 
\end{center}
\caption{Conceptual diagram of CRAFT \citep{2022PASJ..tmp...82C}. Here, "Fourier transform" and "inverse Fourier transform" correspond to the inverse RM synthesis and RM synthesis, respectively.}
\label{fig:Cooray-2021_diagram}
\end{figure}

A conceptual diagram of CRAFT is shown in Fig.~\ref{fig:Cooray-2021_diagram}. The basic idea is to limit the $\phi$-space range where non-zero polarization intensity exist and to estimate the polarization spectrum of unobserved band including negative $\lambda^2$. This restriction on the FDF comes from the fact that $\phi$ does become too large within a realistic source and an assumption of sparsity of the FDF. The specific algorithm is as follows.
\begin{enumerate}
\item Compute the dirty FDF from Fourier transform of observed polarization spectrum $\tilde{P}(\lambda^2) = W(\lambda^2) P(\lambda^2)$ and denote it as $F_0(\phi)$. As we saw in section \ref{subsection:RMSF}, the dirty FDF has sidelobes and polarization intensity spreads over a wide range of Faraday-depth space.
\item Compute a new FDF $F'_{i}(\phi)$ from $F_i(\phi)$ in the following way.
\begin{itemize}
\item Consider a physically reasonable limit on the range in Faraday-depth space and set the polarization intensity outside that range to zero.
\item Set the polarization intensity at $\phi$ with $|F(\phi)| < \mu$ to zero assuming the sparsity of the FDF. Here $\mu$ is a threshold we give in advance. Further replace $|F(\phi)|$ with $|F(\phi)| - \mu$ for $\phi$ range with $|F(\phi)| > \mu$.
\end{itemize}
\item Perform Fourier transform of $F'_i(\phi)$ to obtain polarization spectrum $P'_{i+1}(\lambda^2)$. This polarization spectrum have non-zero values in unobserved band including $\lambda^2 < 0$ as well.
\item Replace $P'_{i+1}(\lambda^2)$ of the observed band with the observation data $\tilde{P}(\lambda^2)$. This new spectrum is denoted as $P_{i+1}(\lambda^2)$.
\item Compute a new FDF $F_{i+1}(\phi)$ from Fourier transform of $P_{i+1}(\lambda^2)$.
\item Repeat the steps 2-5 above. $P_i(\lambda^2)$ coincides with the observation data in the observed band and is expected to approach the true value outside the observed band through the iteration. Thus, $F_i(\phi)$ is also expected to approach the true FDF. Terminate the iteration when $F_i(\phi)$ sufficiently converges.
\end{enumerate}

\begin{figure}[t]
\begin{center}
\includegraphics[width=8cm]{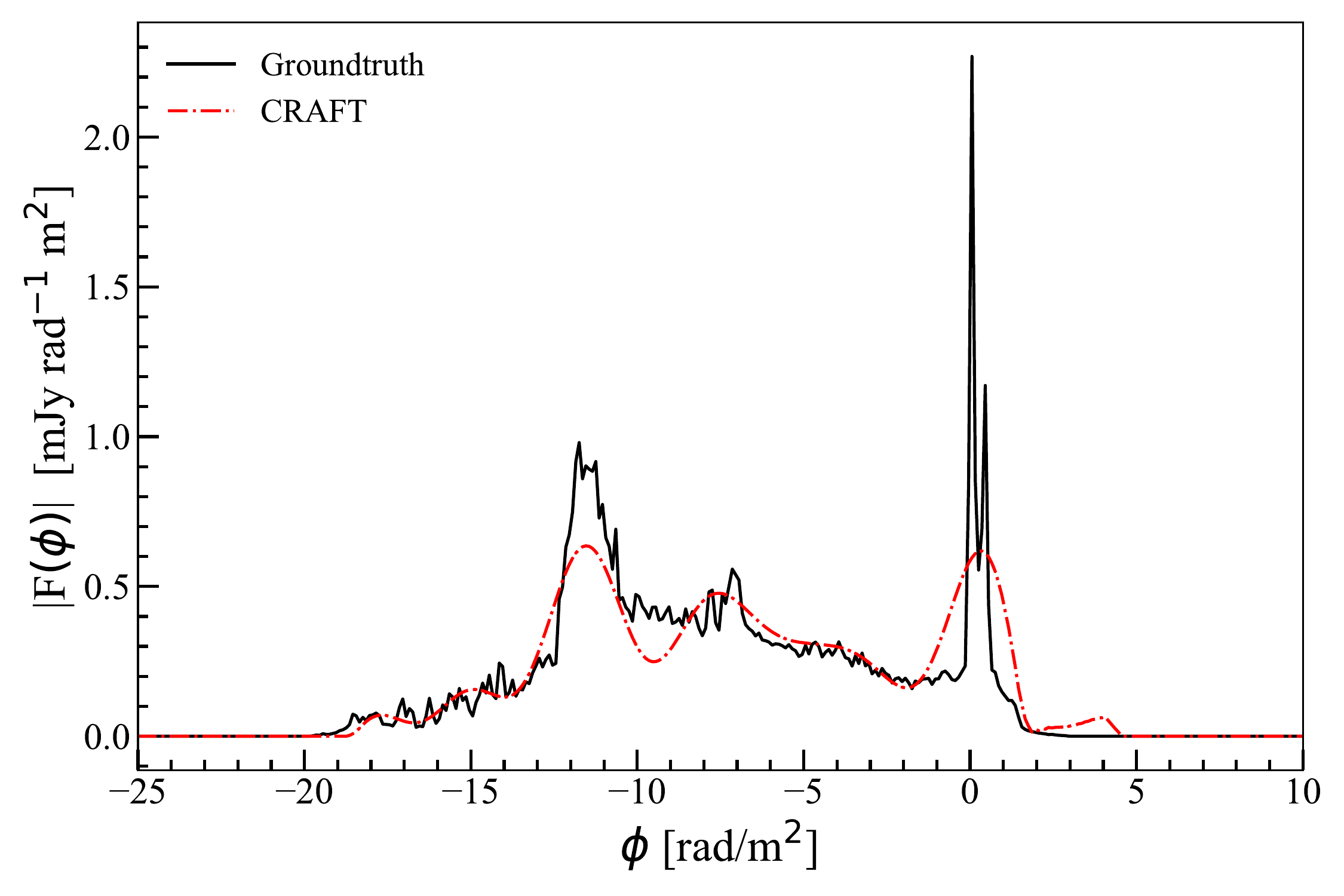} 
\end{center}
\caption{Reconstruction of a galactic FDF model \citep{2014ApJ...792...51I} with CRAFT \citep{2021MNRAS.500.5129C}. An observation band of $300~{\rm MHz} - 3~{\rm GHz}$ was assumed.}
\label{fig:Cooray-2021_FDF}
\end{figure}

Fig.~\ref{fig:Cooray-2021_FDF} is an example of reconstruction of the FDF by CRAFT with a mock polarization spectrum. The galactic model of \citet{2014ApJ...792...51I} was used as the true FDF and an observation band of $300~{\rm MHz} - 3~{\rm GHz}$ was assumed. Compared to the result of sparse modelling in Fig.~\ref{fig:Akiyama}, the overall shape is reproduced with about the same quality and the reproduction of the sharp peak at $\phi = 0$ is slightly better. Again, It should be noted that the peaks are resolved with a slightly higher resolution compared with the FWHM of the RMSF of this band ($3.5~{\rm rad/m^2}$). In this example, 532 iterations were required to converge the results, but the computation time was comparable to RM CLEAN and significantly faster than sparse modelling. \citet{2021MNRAS.500.5129C} notes that the assumption applied in step 2 can be flexible. \citet{2022PASJ..tmp...82C} has extended the method to use wavelets as the basis of sparsity with different sets of assumptions, further improving the reproducibility of the FDF. 

\section{Application of Faraday tomography}
\label{section:applications}

In recent years, wide-band polarization observation has become possible and Faraday tomography has been applied to real data. In this section, we will see some examples of application of Faraday tomography. For other important applications, see \cite{2017NatAs...1..621M,2019ApJ...871..106D,2021A&A...654A...5T,2021ApJ...923L...5P,2022MNRAS.512..945C,2022MNRAS.513.3289I,2022arXiv220801336P}, for example.

\subsection{Resolution in Faraday-depth space}
\label{subsection:resolution}

\citet{2012MNRAS.421.3300O} performed observations of 4 quasars with ATCA (Australia Telescope Compact Array) with a wide observational band of $1.1-3.1~{\rm GHz}$ and frequency resolution of $1~{\rm MHz}$. The beam size was $\sim 10'' \times 10''$ and all quasars were not resolved in image and observed as point sources. However, the polarization spectra of two of four quasars (PKS B1610-771 and PKS B1039-47) could not be fitted well with a single delta-function model and multiple sources were necessary to explain the data. Therefore, Faraday tomography could resolve the components of the two sources which could not be resolved by imaging.

\begin{figure}
\begin{center}
\includegraphics[width=8cm]{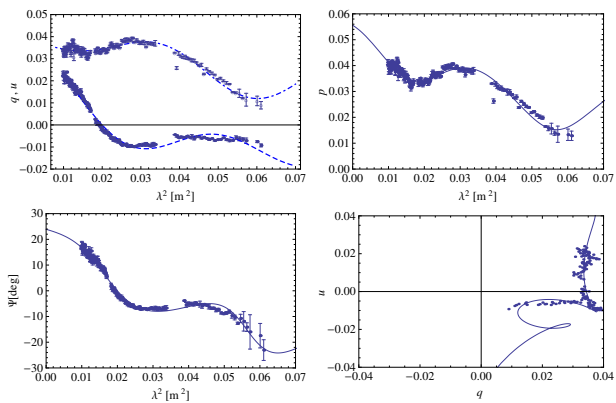}
\end{center}
\caption{Polarization spectrum of PKS B1039-47 obtained by ATCA observation \citep{2012MNRAS.421.3300O}. $Q$ and $U$ (top-left), polarization fraction (top-right) and polarization angle (bottom-left) are shown as functions of $\lambda^2$. Bottom-right panel represents $(Q,U)$ plane. Curves are the results of QU-fitting with 3-delta-function model.}
\label{fig:OSullivan_QU}
\end{figure}

Fig.~\ref{fig:OSullivan_QU} shows the polarization spectrum of PKS B1039-47 and the result of QU-fitting with 3 delta functions. The bottom left panel is polarization angle as a function of $\lambda^2$ and it is evident that it is not linear and cannot be fitted by a single delta function. The polarization fraction shown in the top right panel oscillates with $\lambda^2$ and indicate that multiple sources are interfering. The bottom-left panel also implies the existence of interfering sources because it should be circular in the case of a single source. This polarization spectrum was fitted with one, two and three delta-function models and the last model was selected by the BIC. By this model, the complicated behavior of both $Q$ and $U$ for the whole observational band are well explained as shown in the top left panel.

\begin{figure}
\begin{center}
\includegraphics[width=8cm]{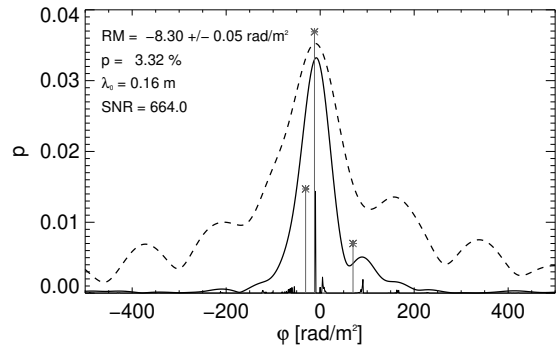}
\end{center}
\caption{FDFs from Faraday tomography of polarization spectrum in Fig.~\ref{fig:OSullivan_QU} \citep{2012MNRAS.421.3300O}. The dirty FDF (dashed), cleaned FDF (solid) and QU-fit with 3 delta functions (asterisks) are compared.}
\label{fig:OSullivan_FDF}
\end{figure}

The obtained FDF of PKS B1039-47 is shown in Fig.~\ref{fig:OSullivan_FDF}. The dirty FDF, cleaned FDF and the result of QU-fit are compared. The cleaned FDF also implies the existence of multiple sources but QU-fit resolves the main peak of the cleaned FDF into two delta functions, whose Faraday depths are about $-13~{\rm rad/m^2}$ and $-30~{\rm rad/m^2}$, respectively. The gap of the two sources in Faraday-depth space is smaller than the FWHM of the RMSF of the observational band which is about $60~{\rm rad/m^2}$. Thus, QU-fit succeeded in resolving two sources which are closer than the resolution.

In fact, PKS B1039-47 was resolved in image with VLBI observation by the LBA (Australian Long Baseline Array) and a jet structure extending about 20 masec ($\sim 160~{\rm pc}$) was found. In the jet, there are 3 spots which are bright in Stokes I and it is possible that the 3 polarization sources found by QU-fit correspond to these spots. If this is the case, the 3 delta functions are independent sources. In fact, it is also possible that there is only one polarized source with a complicated FDF and it is necessary to perform observations with a wider band and Faraday tomography with more advanced methods to verify the possibility.

\subsection{Correspondence between physical space and Faraday-depth space}
\label{subsection:real-Faraday}

As stated before, because there is generally no one-to-one correspondence between physical space and Faraday-depth space, distribution of polarization intensity in physical space cannot be directly known from the FDF, even if we could reconstruct the FDF perfectly. However, a combination with other kind of observation may allow us to infer the distribution in physical space. Here we show an example of such studies.

\begin{figure}
\begin{center}
\includegraphics[width=8cm]{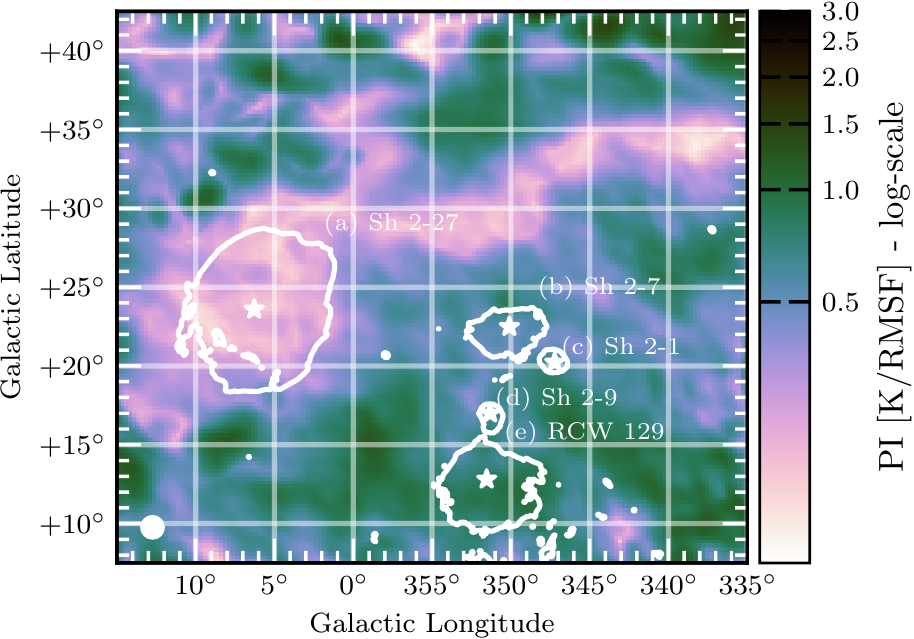}
\end{center}
\caption{Peak heights of the FDF around Sh 2-27 region \citep{2019MNRAS.487.4751T}. White curve represents HII regions and the corresponding central stars are represented by white stars.}
\label{fig:Thomson-2019_peak}
\end{figure}

\citet{2019MNRAS.487.4751T}  observed Sharpless 2-27 (Sh 2027) with a band of $300 - 480~{\rm MHz}$ with the Parkes radio telescope. "Sharpeless" is the name of a catalogue of HII regions and Sh 2-27 is an HII region with an angular size of $5.5$ degree around Zeta Ophiuchi. In this paper, they first obtained the Faraday cube of a region around Sh 2-27, which is 3-dimensional data which consists of the FDF at each point in the sky, by Faraday tomography of the polarization spectra. Fig.~\ref{fig:Thomson-2019_peak} represents the peak height of the FDF. The peak height within Sh 2-27 region is relatively low compared with the outside but a finite value is detected that is significantly larger than the error.

It is considered that Sh 2-27 itself does not emit synchrotron radiation. According to an RM catalogue of \citet{2009ApJ...702.1230T}, there is a large dispersion in the Faraday depths of extragalactic objects behind Sh 2-27 and a part of the dispersion, $\sigma \approx 74 \pm 1~{\rm rad/m^2}$, is estimated to be contributed from Sh 2-27. At low frequencies such as 300-480MHz, the polarization from behind Sh 2-27 is considered to be completely depolarized due to beam depolarization. Then, the polarized emission coming from the direction of Sh 2-27 is produced in front of Sh 2-27. Noting the distance of Zeta Ophiuchi is $182^{+53}_{-30}~{\rm pc}$ and assuming the HII region is spherical, the near-side boundary of the HII region is located at the distance of $164^{+48}_{-33}~{\rm pc}$. Thus, the FDF toward Sh 2-27 is contributed from a nearby region around $160~{\rm pc}$ from the Sun.

\begin{figure}
\begin{center}
\includegraphics[width=7cm]{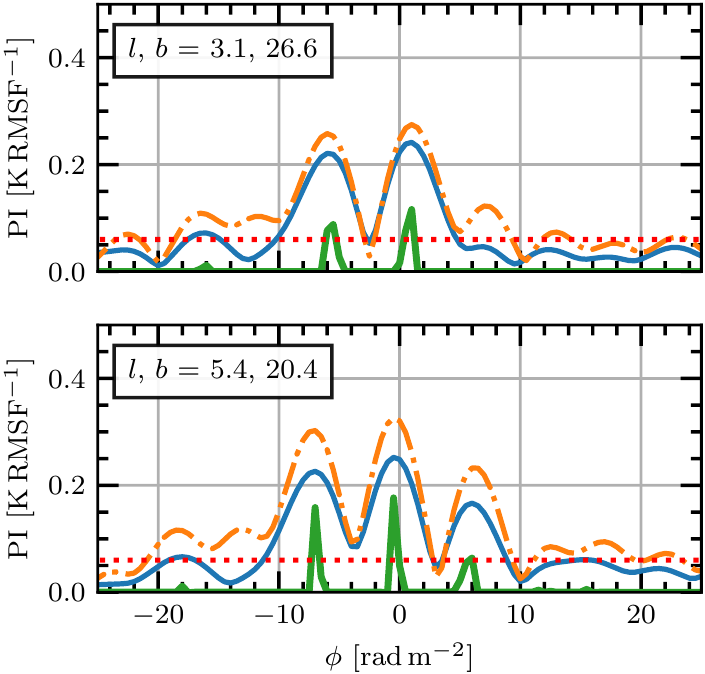}
\end{center}
\caption{FDFs toward two directions within Sh 2-27 \citep{2019MNRAS.487.4751T}. The top and bottom panels correspond to LOSs which intersect one and two neutral clouds, respectively. Orange long-dashed curve, blue solid curve and green solid curve represent dirty FDF, cleaned FDF and CLEAN components, respectively. Horizontal dotted line is the threshold for RMCLEAN ($\epsilon$ in section \ref{subsection:RMCLEAN}).}
\label{fig:Thomson-2019_FDF}
\end{figure}

Fig.~\ref{fig:Thomson-2019_FDF} is the result of Faraday tomography toward Sh 2-27. The bottom panel is the FDF at the center of Sh 2-27 and three peaks are seen there. On the other hand, the upper panel corresponds to an off-center LOS and the peak at $\phi \sim 6~{\rm rad/m^2}$ seen in the bottom panel is absent. It was argued that, considering the typical gas density, ionization rate and strength of magnetic fields, Cold Neutral Medium (CNM) is a reasonable candidate of the polarization sources appearing in the FDF. In fact, from the 3-dimensional dust distribution map obtained by the STILISM project (STructuring by Inversion the Local Interstellar Medium) \citep{2018A&A...616A.132L}, it was found that two neutral clouds exist in front of Sh 2-27.

\begin{figure}
\begin{center}
\includegraphics[width=8cm]{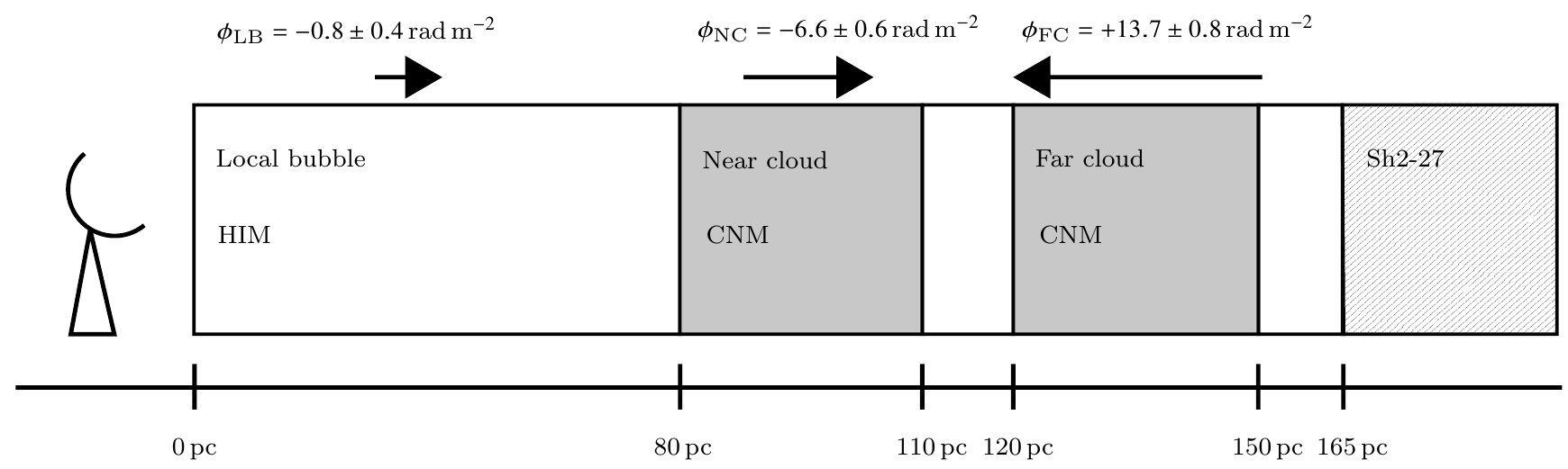}
\end{center}
\caption{Distribution of the LOS component of magnetic field and polarization intensity in physical space which was reconstructed by Faraday tomography and observation of neutral clouds \citep{2019MNRAS.487.4751T}. There are the Local Bubble and two neutral clouds in front of Sh 2-27.}
\label{fig:Thomson-2019_LOS}
\end{figure}

As an interpretation of the FDFs, considering the Local Bubble near the Solar System, which is Hot Ionized Medium (HIM), and these two neutral gas clouds to be the main sources of synchrotron radiation, \citet{2019MNRAS.487.4751T} assigned them the three peaks. Fig.~\ref{fig:Thomson-2019_LOS} is a schematic view of the polarization distribution in physical space. Further, assuming the typical thermal-electron density of each phase, the LOS components of magnetic field of the Local Bubble, the near neutral cloud and the far neutral cloud are estimated to be $2.5~{\rm \mu G}$, $15~{\rm \mu G}$ and $30~{\rm \mu G}$, respectively. On the other hand, the top panel of Fig.~\ref{fig:Thomson-2019_FDF} is interpreted as an LOS which intersects only the Local Bubble and the near neutral cloud, which is consistent with the FDF with only two peaks.

As described above, by combining Faraday tomography of polarization spectra with other observations, it is possible to reconstruct the LOS distributions of emission region, magnetic fields and other phases of gases. There have been many other attempts in this direction \citep{2018PASJ...70...27S,2018Galax...6..137S,2020A&A...644L...3B,2021A&A...654A...5T} and physical interpretation through MHD simulations is also presented in \citet{2022A&A...663A..37B}. In the future, as surveys at other wavelengths progress, this kind of reconstruction will become possible on various scales more precisely.

\section{Conclusion}
\label{section:conclusion}

This review described Faraday tomography from its basic principles to physical interpretation, algorithms and applications. Synchrotron radiation, Faraday rotation, and depolarized waves, which have been major tools for the research of cosmic magnetic fields, form the basis for Faraday tomography, and it can be considered to be an advanced version of them. In other words, Faraday rotation and depolarization generally depend on the wavelength reflecting the the line-of-sight structure of the polarization source, and Faraday tomography reconstructs it from the polarization spectrum over a wide band. More specifically, it yields the Faraday dispersion function, which contains information about the cosmic-ray electron density and the component of magnetic fields perpendicular to the line of sight, which are associated with synchrotron radiation, and the thermal-electron density and the line-of-sight component of magnetic fields, which are associated with Faraday rotation. By combining 2-dimensional imaging and Faraday tomography, we can investigate the 3-dimensional structure of polarization sources.

However, we saw that there are two major problems in applying Faraday tomography. One is the reconstruction of the Faraday dispersion function from the observed polarization spectrum. The Faraday dispersion function is mathematically equal to the Fourier transform of the polarization spectrum, and if the polarization spectrum can be obtained from negative infinity to positive infinity with respect to the square of the wavelength, the Fourier transform can be performed to obtain the complete Faraday dispersion function. However, the polarization spectrum is physically meaningful only in the region where the square of the wavelength is positive, and the observation is limited to a finite range there. Therefore, it is necessary to reconstruct the Faraday dispersion function from imperfect information as accurately as possible. We saw that, in addition to standard methods such as RM CLEAN and QU-fit, more advanced algorithms have been proposed. Considering the future development of observation facilities, the computational cost is also an important factor, and it is necessary to develop a method to reconstruct a physically reasonable Faraday dispersion function while suppressing the computational cost.

Another problem is the physical interpretation of the Faraday dispersion function. The Faraday dispersion function is a function of the Faraday depth, which generally does not have one-to-one correspondence with the physical space so that the distribution of physical quantities in the physical space is not immediately deduced from the polarization intensity distribution in the Faraday depth space. In this review, we started with a simple model with a uniform magnetic field and polarization intensity and examined a toy model of a galaxy including turbulent magnetic fields. Then, it was shown that the Faraday dispersion functions can have complex shapes even for simple models. The Faraday dispersion functions of more realistic galaxy models are expected to be further complicated, and even if it could be reconstructed accurately from the observed polarization spectrum, its physical interpretation is not straightforward. In the future, if higher-sensitivity observations in a wider band including long wavelengths become possible, the resolution and maximum width in the Faraday depth space will improve, and the complex structure of the Faraday dispersion function will become visible. In preparation for this, both the approach of understanding the physical meaning of the Faraday dispersion function from a simple model and the approach of predicting it with a realistic model would be necessary.

Faraday tomography has already been applied to various polarization sources and has produced fruitful results. It is useful to compare the obtained Faraday dispersion function with the theoretical model and to combine it with other observables. In the future, large-scale surveys with various wavelengths are scheduled to proceed, and Faraday tomography is expected to play a major role in it because it has information on the line-of-sight distribution of sources.

\section*{Acknowledgements}

KT is partially supported by JSPS KAKENHI Grant Numbers 20H00180, 21H01130 and 21H04467, Bilateral Joint Research Projects of JSPS, and the ISM Cooperative Research Program (2021- ISMCRP-2017).

\end{document}